\documentclass[fleqn,usenatbib]{mnras}

\usepackage{newtxtext,newtxmath}

\usepackage[T1]{fontenc}

\DeclareRobustCommand{\VAN}[3]{#2}
\let\VANthebibliography\thebibliography
\def\thebibliography{\DeclareRobustCommand{\VAN}[3]{##3}\VANthebibliography}

\usepackage{graphicx}	
\usepackage{amsmath}	
\usepackage{ulem}

\title[Collision fragments \& water transport to Earth]{
The evolution of collision debris near the $\nu_6$ secular resonance and its role in the origin 
of terrestrial water}

\author[Á. Süli and E. Forgács-Dajka]{
Á. Süli$^{1,2}$,\thanks{E-mail: suli.aron@ttk.elte.hu (ÁS)}
Forgács-Dajka E.$^{1,3,4}$
\\
$^{1}$ELTE E\"otv\"os Lor\'and University, Institute of Physics and Astronomy, Department of Astronomy, H-1117 Budapest, P\'azm\'any P\'eter s\'et\'any 1/A, Hungary\\
$^{2}$Konkoly Observatory, Research Centre for Astronomy and Earth Sciences, H-1121 Budapest, Konkoly Thege Mikl\'os \'ut 15-17, Hungary\\
$^{3}$HUN-REN–SZTE Stellar Astrophysics Research Group, H-6500 Baja, Szegedi út, Kt. 766, Hungary\\
$^{4}$Wigner Research Centre for Physics, H-1525 Budapest, P.O. Box 49, Hungary\\
}

\date{Accepted 2023 October 15. Received 2023 October 13; in original form 2023 June 01}

\pubyear{2015}

\begin{document}
\label{firstpage}
\pagerange{\pageref{firstpage}--\pageref{lastpage}}
\maketitle

\begin{abstract}This work presents novel findings that broadens our understanding of the amount 
of water that can be transported to Earth. The key innovation lies in the combined usage 
of Smoothed Particle Hydrodynamics (SPH) and $N$-body codes to assess the role of collision 
fragments in water delivery. We also present a method for generating initial conditions 
that enables the projectile to impact at the designated location on the target's surface
with the specified velocity. The primary objective of this study is to simulate giant 
collisions between two Ceres-sized bodies by SPH near the $\nu_6$ secular resonance  
and follow the evolution of the ejected debris by numerical $N$-body code. With our 
method 6 different initial conditions for the collision were determined and the 
corresponding impacts were simulated by SPH. Examining the orbital evolution of the 
debris ejected after collisions, we measured the amount of water delivered to Earth, 
which is broadly 0.001 ocean equivalents of water, except in one case where one large 
body transported 7\% oceans of water to the planet. Based on this, and taking into 
account the frequency of collisions, the amount of delivered water varies between 1.2 
and 8.3 ocean's worth of water, depending on the primordial disk mass. According to our 
results, the prevailing external pollution model effectively accounts for the assumed 
water content on Earth, whether it's estimated at 1 or 10 ocean's worth of water.
\end{abstract}




\section{Introduction}

The inner Solar System planets are generally dry \citep{Abe2000,Raymond2022}. However, 
there is some evidence of water on certain inner rocky planets. Earth is the only planet 
in the inner Solar System known to have liquid water on its surface. About 70\% of Earth's 
surface is covered in oceans. All of Earth's oceans contain roughly 
$2.5\cdot 10^{-4} \mathrm{M_\oplus}$ water, where $\mathrm{M_\oplus}$ denotes the mass of the Earth. 
This amount is often referred to as one 'ocean' of water, hereafter its mass is denoted by 
$\mathrm{M_\mathrm{o}}$. However, the amount of water in Earth's interior is much more uncertain, 
with different studies suggesting vastly different quantities of water trapped in hydrated 
silicates and other minerals. It is generally assumed that the amount of water 
in the interior and on the surface of Earth is between 1 and 10 ‘oceans’ 
\citep[e.g.][]{Abe2000,Hirschmann2006,Marty2012}.

\par
Overall, studies suggest that the origin of Earth's water is a complex process that 
involves the accretion of water from multiple sources over a long period of time.
Below, we list these potential sources of water on rocky planets. Cosmochemical tracers 
such as isotopic ratios, mainly the D/H ratio, may constrain the origin of Earth’s water 
\citep[for an overview of the D/H measurements, see e.g.][]{Alexander2012,Morbidelli2000,Marty2006}. 
There are six potential models to explain how water was delivered to Earth. They can broadly be 
divided into two categories based on whether the sources of water are local or external. 
The local source models include (a) the adsorption of water vapour onto silicate grains and (b) 
the oxidation of H-rich primordial envelope. The external source models include (i) the "pebble" 
snow model, (ii) the wide feeding zone, (iii) the external pollution, and (iv) the inward 
migration model. For a detailed description of these models see e.g. \cite{Raymond2022}.

\par
According to recent research by \cite{Raymond2022,Morbidelli2000}, the prevailing theory 
suggests that the main source of water on Earth is external pollution from the outer regions 
of the asteroid belt, rather than local sources. It is believed that this water originates 
from a few planetary embryos located in the outer part of the asteroid belt, as proposed 
by \cite{Morbidelli2000}. According to the external pollution model, the presence of giant 
planets strongly influences the orbits of the remaining planetesimals, especially during 
the phase of rapid gas accretion. In the course of this period, the mass of the giant planets
can increase rapidly, from $\sim\,10 - 20\, \mathrm{M_\oplus}$ to hundreds of Earth masses on a few $10^5$ 
years timescale. The rapid mass increase excites the orbits of the nearby planetesimals that 
managed to avoid accretion by the growing giants. A multitude of planetesimals experience 
close encounters with the growing planet, resulting in scattering in all directions. Bodies 
that are on elliptical orbits that cross the orbit of the giant planet at apostar distance 
may experience significant energy loss due to gas drag. As a result, the planetesimal's 
farthest point moves inward and the body will not suffer any more close encounters 
with the proto-Jupiter.

\par
The most significant secular resonance in our solar system is the $\nu_6$ resonance 
\citep{Ito2006,Minton2011}, which has been in the focus of research of resonant dynamics for several
decades and thus has an abundant literature. 
The $\nu_6$ secular resonance involves the apsidal precession of asteroids 
denoted by $g$ and Saturn denoted by $g_6$. The 
resonance condition is defined by $|g - g_6| < \delta$, where $g_6 \approx 28.2455$ arcsec/year 
(derived from the LONGSTOP 1B numerical integration \citep{Nobili1989}), and the half-width of the resonant 
strip, $\delta$, is usually taken as 1 arcsec/yr.

\par
\cite{Froeschle1986} conducted numerical integrations over 1 million years to study the orbital 
evolution in the $\nu_6$ at 2.05 au. The results revealed that significant variations in 
eccentricity led to the presence of direct Earth-crossers originating from this secular 
resonance. The authors suggest that secular resonances should be considered as potential 
sources of meteorites. Later on
numerical simulations by \cite{Scholl1991} demonstrated that the $\nu_6$ secular resonance 
often in synergy with the 4/1 mean motion resonance with Jupiter, creates a wide region 
where Earth-crossing chaotic orbits are produced. The 4/1 resonance rapidly increases 
eccentricity to 0.5 earth-crosser, whereas remaining in the $\nu_6$ resonance requires at 
least several hundred thousand years to become an earth-crosser.
The analytical model of \cite{Morbidelli1991} is suitable for a global description of 
dynamics in secular resonances of order 1. The model is applied to investigate the secular 
resonances $\nu_6$, $\nu_5$, and $\nu_{16}$, and illustrations of the secular motion are 
provided.
The $\nu_6$ resonance sets the inner boundary of the asteroid belt at approximately 2 au. A significant 
number of near-Earth asteroids originate from the inner main belt rather than the middle or outer belt 
\citep{Bottke2002}. Many of these asteroids within the inner main belt are affected by the $\nu_6$ 
resonance and eventually acquire Earth-crossing orbits \citep{Morbidelli1994,Bottke2000}.

\par
In a recent paper by \cite{Smallwood2018} the authors conducted $N$-body simulations of a planetary 
system with an asteroid belt to study how the architecture of the system affects the asteroid impact rate 
on Earth. The results revealed the significant role of the $\nu_6$ resonance in the frequency of asteroid 
collisions with Earth. In further investigations, they replaced Earth with a 10 $\mathrm{M_\oplus}$
super-Earth and 
measured how the asteroid collision rate with Earth changes. If the super-Earth orbited beyond 0.7 au, a 
significant increase in the number of collisions was observed. However, when it orbited between 1 and 1.5 
au, the collision rate decreased. Additionally, altering Saturn's semi-major axis substantially reduced the 
asteroid collision rate, while increasing its mass raised it.

\par
The formation of the Hungaria family provided further motivation for selecting the location 
of the collision. The inner edge of the main asteroid belt close to $\nu_6$ is home to the Hungaria 
asteroids, including the Hungaria family, which originated from the catastrophic collision of the 
(434) Hungaria asteroid \citep{Warner2009,Forgacs2022}. This is a direct evidence that collisions 
occurred in the past within this region.

\par
In a recent paper by \cite{Martin2021} (hereafter ML) the researchers compared the impact 
efficiencies of different locations in the asteroid belt, namely the $\nu_6$ secular resonance
at $a \approx 2.05$ au, at the 2:1 mean-motion resonance with Jupiter at $a \approx 3.32$ au and 
the outer asteroid belt at $a \approx 4.05$ au, where $a$ denotes the semi-major axis. 
They estimated the potential amount of water 
delivered to Earth and found that the majority of asteroids that collided with Earth originate 
from the $\nu_6$ resonance. About 2\% of asteroids from $\nu_6$ collide with Earth, while 
the collision probabilities from other resonances are orders of magnitude smaller.
If the majority of asteroids in the primordial asteroid belt were displaced to the $\nu_6$ 
secular resonance, e.g. by chaotic diffusion \citep{Kovari2023}, gravitational scattering, 
gas drag or the Yarkovsky effect, it is possible that up 
to $\sim 8$ oceans' worth of water could have been transported to Earth. Thus the delivery 
of at least 1 $\mathrm{M_\mathrm{o}}$ of water from the asteroid belt is possible, 
but to explain the delivery of 10 $\mathrm{M_\mathrm{o}}$ is more challenging.

\par
In this study, we perform an analysis similar to that of ML
but specifically focus on tracking the evolution 
of debris ejected during collisions of large protoplanets.
Following the oligarchic growth model proposed by \cite{Kokubo2000}, we assume 
that the terrestrial planet formation has been almost finished. At this stage, protoplanets in 
the inner disk may have 
grown to Ceres size and potentially larger beyond the snowline.
The impact of two Ceres-sized bodies 
occurs near the $\nu_6$ secular resonance, where the ejected debris instantly acquires 
significant eccentricity and inclination. The resonance further increases the 
eccentricity and inclination of a portion of the debris. As a result, the majority of 
the debris is rapidly transported to chaotic orbits and can reach the inner planets. 
The water content of the debris is determined by the original composition of the colliding 
bodies, 
and the smoothed particle hydrodynamics (SPH) code consistently accounts for the composition 
of individual fragments. Thus, in this aspect, we take a step closer to simulating reality. 
In this study, the simulations of collisions were conducted using the SPH code, which is 
described in detail in the paper of \cite{Schaefer2020}.

\par
In this research, we take a step forward in the simulation of the origin of Earth water. We
focus on estimating the amount of water that could have been transported 
to Earth via collision fragments which were created by a giant impact between two protoplanets. 
To this end, the dynamical evolution of all fragments will be followed until they impact a 
large body or the simulation time ends. This scenario is an extension of the external pollution 
model, since the sudden eccentricity and inclination growth is principally due to collision and
not to gravitational interactions. In Section 2, we present the definition of the collision
geometry, a short description of the applied SPH and $N$-body simulations and a detailed 
description of how the initial conditions were constructed from the parameters defining the
impact. Section 3 presents the results of the simulations of both the SPH and the $N$-body codes,
and we present estimations of how much water could have been transported to the Earth via 
fragments resulted form a giant collision. We discuss the consequences of water delivery 
to Earth and draw our conclusions in Section 4.

\section{Collision geometry, SPH and $N$-body simulations}

In this study the Runge--Kutta--Nystrom 7(6) (RKN) \cite{Dormand1978} algorithm was 
utilized with adaptive step size control. The method has an acceptable speed and 
accuracy in most situations. The integrator was used to model the evolution of a 
population of asteroids and test particles under the gravitational attraction of 
Venus, Earth, Mars, Jupiter and Saturn. Uranus and Neptune are not considered in the 
simulation due to their negligible impact on the dynamics of Jupiter, Saturn and 
the inner Solar System objects. In the $N$-body code based on the RKN algorithm,
all massive bodies interact with all others. Test particles do not influence the motion 
of the other bodies. In the model we have assumed that all planets have reached
their current mass and move on their current orbits.
In the simulations, the gas disc had dissipated by this stage and thus gas friction in our 
simulations is ignored. 
In the $N$-body simulations, all collisions between bodies are treated as follows: when a close 
approach with a separation less than the sum of the two radii is detected, the two colliding protoplanets 
merge to form a new body with a mass equal to the sum of their masses. The orbital elements of the newly 
formed body are determined based on the position and velocity of the center of mass of the two colliding 
bodies at the moment of impact. In each collision the composition of the resulting body is 
determined based on the material composition of the colliding bodies. Most of the previous work on 
planet formation used this oversimplified treatment of collisions. In the $N$-body runs in order to 
determine the effect of 
collisions on the composition of massive bodies, we track the collision history of each individual 
fragment with mass.

\subsection{Initial condition}

The process to create collision fragments that originate from impacting bodies with given orbital 
elements and collision parameters is complex. A description of this procedure is given below and 
can be divided into the following four main steps:
\begin{enumerate}
    \item Calculate the radii of the target $R_\mathrm{t}$ and the projectile $R_\mathrm{p}$ depending 
    on the composition of the objects.
    \item Generate the position and velocity vectors for the planets as well as for the 
    projectile and target. Using the specified orbital elements of the target and 
    the collision-defining parameters one has to calculate the initial conditions
    for projectile. With these initial conditions, the system is 
    integrated backwards in time for approximately $T_\mathrm{back} \approx 10^4$ 
    seconds.
    \item Compute the collision with the SPH code. From the system obtained at the time 
    $t = T_\mathrm{back}$, the initial conditions for the SPH code have to be calculated. 
    With these initial conditions defining the barycentric position and velocity of the 
    Sun, the 5 planets, and the target and projectile, the SPH code is used to simulate 
    the collision for a duration of approximately $T_\mathrm{SPH} \approx 8\cdot 10^4$ 
    seconds. The end state of the SPH run defines the initial state for the next step.
    \item Calculate the initial conditions for the $N$-body code from the SPH run's end 
    state. In this step, it is necessary to identify the cohesive debris fragments, 
    which are bodies composed of one or more SPH particles. Once this is accomplished, 
    their position and velocity vectors must be computed which in conjunction with the data 
    of the major planets constitute the initial conditions for the N-body simulation.
    The $N$-body simulator is then integrates the system for a duration of $T=10^6$ 
    years.
\end{enumerate}
Typically, it takes about 1 million years for these fragments to be transported 
into Earth-crossing orbits when their eccentricity is directly increased to around $\approx 
0.6$ by the resonance \citep{Morbidelli1994,Smallwood2018}. Below, we provide a detailed 
description of the above four steps.

\begin{figure}
\includegraphics[width=0.95\columnwidth,keepaspectratio=true]{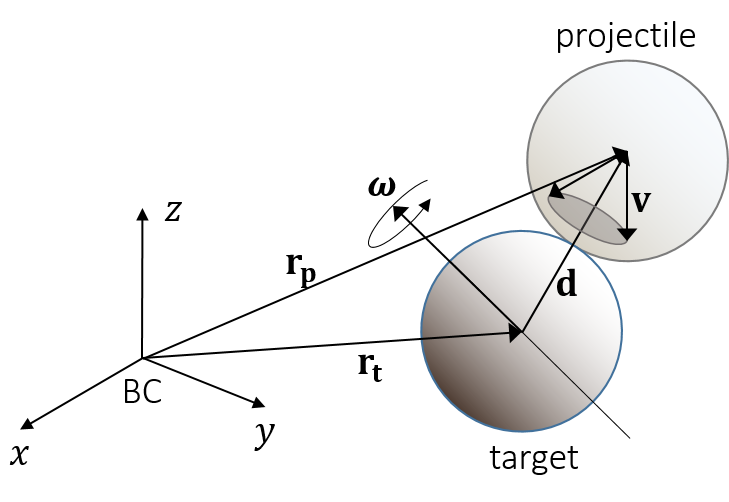}
\caption{The collision of two bodies in the barycentric coordinate system. The geometry of 
the collision is defined by the $\mathbf{d}$ relative position and $\mathbf{v}$ relative 
velocity vectors.}
\label{Fig_01}
\end{figure}

\begin{figure}
\includegraphics[width=0.95\columnwidth,keepaspectratio=true]{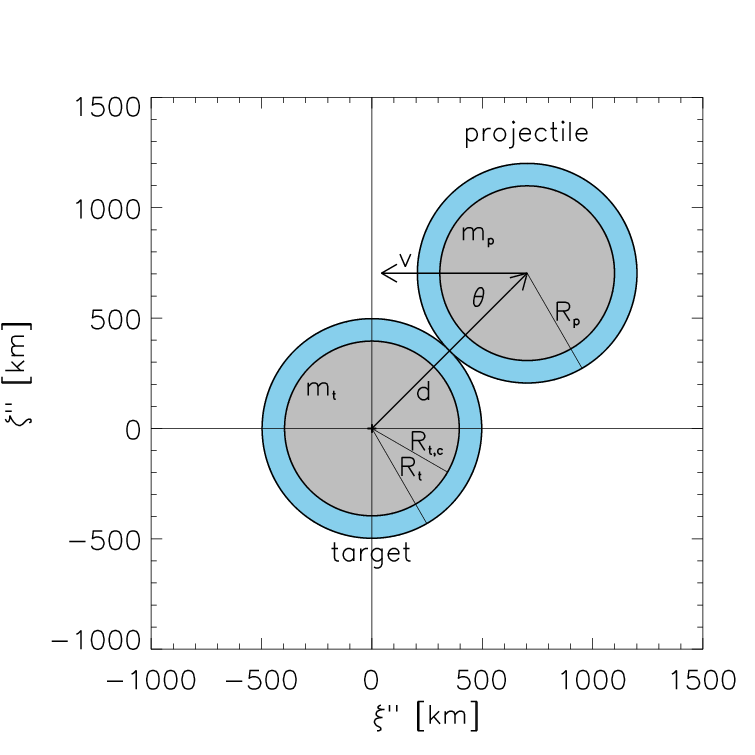}
\caption{The plane of $\xi'' - \zeta''$ is defined by the $\mathbf{d}$ and $\mathbf{v}$ vectors. 
The grey-coloured circle denotes the basalt core, while the sky blue denotes the ice mantel.}
\label{Fig_02}
\end{figure}

\par
To complete the first step, we based our assumptions on the observations of \cite{Thomas2005}. 
The authors have analyzed the Ceres' spectral features in reflected sunlight 
and demonstrated that the presence of a rocky core and a water-ice mantle aligns with the 
data assuming the densest materials for the core with a mean density of 2.077 kg m$^{-3}$. They 
estimated that the ice mantles have a thickness ranging from 110 to 124 km, accounting for 24-26\% 
of the total mass of the object. In accordance with these results, in the present paper, the 
composition of the impacting bodies consists of a basaltic core and a mantle composed of water ice.
This structural composition of bodies were utilized by \cite{Maindl2014b} when studying the 
fragmentation of icy objects.

\par
Therefore, our scenarios include two Ceres-mass $M_\mathrm{Ceres} = 9.38\cdot 10^{20}$ kg 
protoplanets, where both bodies consist of a basalt core (75\% of mass) and a mantle/shell of water 
ice (25\% of mass).
The two protoplanets together contain approximately 0.31 $\mathrm{M_\mathrm{o}}$ of water.
In Fig.~\ref{Fig_01}, we depict the moment of collision between two bodies in the $\mathrm{BC}xyz$ 
barycentric coordinate system. The target is the larger of the two colliding bodies, and if the 
masses are equal, then the object with the smaller identifier number becomes the target.
The barycentric position vectors of the target and projectile are denoted as 
$\mathbf{r}_t$ and $\mathbf{r}_p$, respectively. The relative position vector of the 
projectile with respect to the center of the target is $\mathbf{d}$, given by 
$\mathbf{d} = \mathbf{r}_p - \mathbf{r}_t$, 
and the relative velocity vector $\mathbf{v}$ is defined as 
$\mathbf{v} = \mathbf{v}_p - \mathbf{v}_t$, 
where $\mathbf{v}_t$ and $\mathbf{v}_p$ represent the barycentric velocity vectors of the target 
and projectile, respectively. The $\mathbf{\omega}$ vector corresponds to the angular velocity
vector of the target.

\par
In Fig.~\ref{Fig_02} the two impacting bodies are shown in the plane of the collision defined
by the $\mathbf{d}$ and $\mathbf{v}$ vectors. The basalt core is represented by a light grey
coloured circle, while the ice mantle is denoted by a sky blue coloured one. 
The spatial arrangement of the SPH particles composing the target and projectile is governed by a 
hexagonal lattice structure. This lattice structure was determined based on the given core and shell 
masses. The radii of the target and the projectile are equal and denoted by
$R_\mathrm{t} = R_\mathrm{p} = 4.974 \cdot 10^5$ m, resulting in a bulk density of 
$\rho = 1.82$ g cm$^{-3}$. The radii of the target's and projectile's core are equal too
and denoted by $R_\mathrm{t,c} = R_\mathrm{p,c} = 3.956 \cdot 10^5$ m.

\par
\begin{figure}
\includegraphics[width=0.95\columnwidth,keepaspectratio=true]{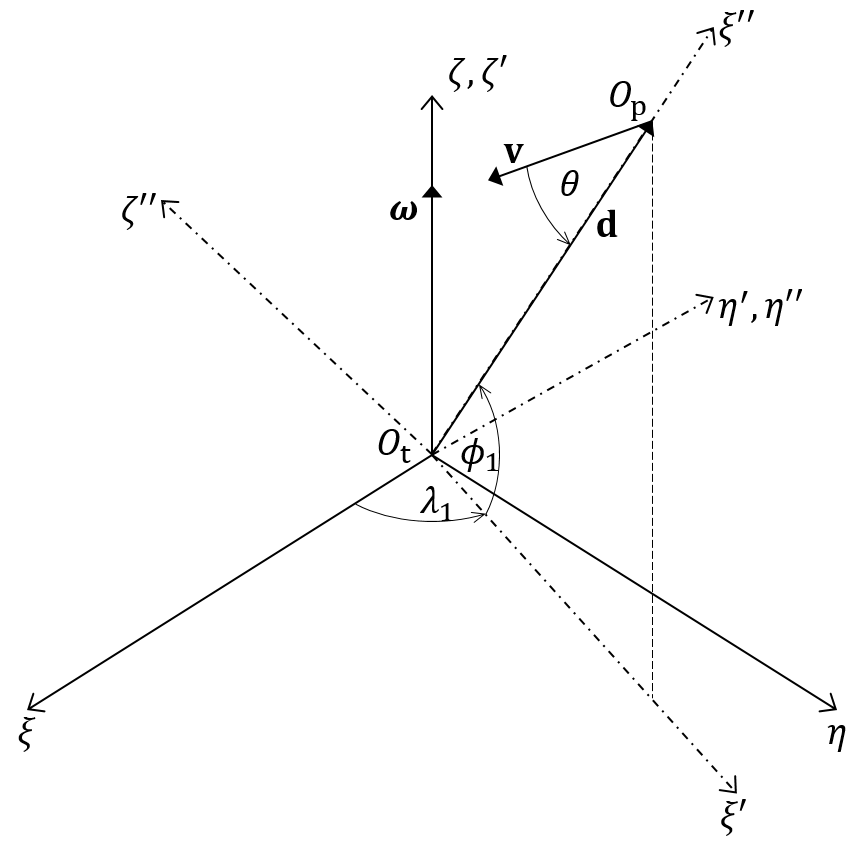}
\caption{The $O_\mathrm{t}\xi\eta\zeta$ targetgraphic coordinate system, where 
$\mathbf{\omega}$ is the angular velocity of the target, 
$\lambda$ denotes the longitude and $\phi$ the latitude, $\mathbf{d}$ is the position vector 
of the projectile, $\mathbf{v}$ is the impact velocity vector. The $\mathbf{d}$ and 
$\mathbf{v}$ vectors determine the plane of collision, $\theta$ is the impact angle.}
\label{Fig_03}
\end{figure}

To execute the second step, multiple consecutive rotational transformations need to be 
performed. It is convenient to perform these rotational transformations in the coordinate 
system fixed to the target's center denoted as $O_\mathrm{t}$. We denote this coordinate
system as $O_\mathrm{t}\xi\eta\zeta$, where the $O_\mathrm{t}\zeta$ axis corresponds to the
rotation axis $\mathbf{\omega}$ of the target, and the plane perpendicular to the 
$O_\mathrm{t}\zeta$ axis and passing through the centre is the $\xi-\eta$ plane, 
see Fig.~\ref{Fig_03}.

\par
The geometry of the collision is completely determined by the impact position 
vector $\mathbf{d}$ and the velocity vector $\mathbf{v}$ relative to the origin $O_\mathrm{t}$. 
In previous studies, the $\mathbf{d}$ vector is usually not provided, only the collision
velocity $v=|\mathbf{v}|$ and the impact angle $\theta$ are given, where $\theta$ is defined 
as the smaller angle between the vectors $\mathbf{d}$ and $\mathbf{v}$ at the moment of first
contact, see Fig.~\ref{Fig_02}. Therefore, only a "collision cone" is defined, with the 
apex at the centre of 
the target $O_\mathrm{t}$ and the lateral surface composed of possible collision velocity
vectors. The impact velocity is given in mutual escape velocity units which is defined
as follows \citep{Genda2012}:
\begin{equation}
v_{\mathrm{esc}} = k_\mathrm{G}\sqrt{2\frac{m_{\mathrm t} + m_{\mathrm p}}{R_{\mathrm t} + R_{\mathrm p}}},
\label{Eq_mutualescspeed}
\end{equation}
where $m_{\mathrm t}$ and $m_{\mathrm p}$ denote the mass of the target and projectile, 
respectively and $k_\mathrm{G}$ is the Gaussian constant of gravity.

\par
When $\theta = 0^\circ$, the collision is referred to as head-on, while $\theta = 90^\circ$ 
corresponds to a grazing collision. In terms of geometric shapes, $\theta$ represents the half-angle 
of the cone. For the sake of generality, in the following discussion, we also take into account 
the $\mathbf{d}$ vector. Given $\mathbf{r}_\mathrm{t}$ and $\mathbf{v}_\mathrm{t}$ 
of the target, as well as the $O_\mathrm{t}\zeta$ axis, $\theta$, 
and $v$, we aim to determine the barycentric position vector $\mathbf{r}_\mathrm{p}$ and 
velocity vector $\mathbf{v}_\mathrm{p}$ of the projectile, see Fig.~\ref{Fig_01}. Let us 
express the components of the $\mathbf{d}$ and $\mathbf{v}$ vectors in the 
$O_\mathrm{t}\xi\eta\zeta$ "targetgraphic" coordinate system
\begin{equation}
\mathbf{d} = d \left( \begin{array}{r}
   \cos \lambda_1 \cos \phi_1 \\
   \sin \lambda_1 \cos \phi_1 \\
   \sin \phi_1
    \end{array} \right), \quad 
   \mathbf{v} = v \left( \begin{array}{r}
   \cos \lambda_2 \cos \phi_2 \\
   \sin \lambda_2 \cos \phi_2 \\
   \sin \phi_2
   \end{array} \right)
\label{Eq_rho_component}
\end{equation}
where $\lambda_1$ is the longitude and $\phi_1$ is the latitude of vector $\mathbf{d}$,
and similarly $\lambda_2$ is the longitude and $\phi_2$ is the latitude of vector $\mathbf{v}$
(since $\mathbf{v}$ is a free vector, thus we have imaginarily shifted it to the origin 
$O_\mathrm{t}$).

\par
Let $\mathbf{f^{(1)}}$ and $\mathbf{f^{(2)}}$ denote the unit vectors in the directions 
of $\mathbf{d}$ and $\mathbf{v}$, respectively
\begin{equation}
\mathbf{f}^{(1)} = \frac{\mathbf{d}}{d}, \quad \mathbf{f}^{(2)} = 
\frac{\mathbf{v}}{v}.
\label{Eq_f1_f2}
\end{equation}
Rotate the $O_\mathrm{t}\xi\eta\zeta$ coordinate system around the $O_\mathrm{t}\zeta$ axis with 
$\lambda_1$ angle and then the resulting the $O_\mathrm{t}\xi'\eta'\zeta'$ coordinate system 
around the $O_\mathrm{t}\eta'$ axis with $\phi_1$. Let us denote the rotation matrix around 
axis $O_\mathrm{t}\zeta$ by $\mathbf{R}_\mathrm{\zeta}(\lambda_1)$, and around the 
$O_\mathrm{t}\eta'$ axis by $\mathbf{R}_\mathrm{\eta'}(\phi_1)$. They are defined as
\begin{eqnarray}
\mathbf{R}_\mathrm{\zeta}(\lambda_1) = \left( \begin{array}{rrr}
   \cos \lambda_1 & \sin \lambda_1 & 0 \\
  -\sin \lambda_1 & \cos \lambda_1 & 0 \\
   0 & 0 & 1 
    \end{array} \right), \\
   \mathbf{R}_\mathrm{\eta'}(\phi_1) = \left( \begin{array}{rrr}
   \cos \phi_1 & 0 & \sin \phi_1 \\
   0 & 1 & 0 \\
   -\sin \phi_1 & 0 & \cos \phi_1
   \end{array} \right).
\label{Eq_R_matrix}
\end{eqnarray}

\begin{figure}
\includegraphics[width=0.95\columnwidth,keepaspectratio=true]{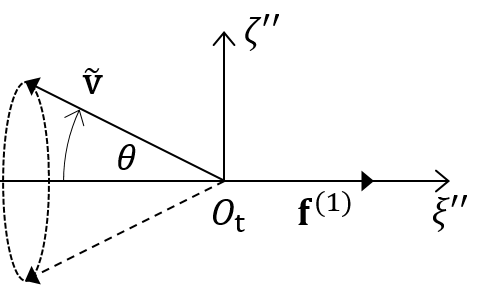}
\caption{The projection of the collision cone onto the plane $O_\mathrm{t}\xi''\zeta''$. 
The $\mathbf{\Tilde{v}}$ is a possible impact vector with the right magnitude $v$ and 
collision angle $\theta$, which is the half-angle of the cone.}
\label{Fig_04}
\end{figure}

The rotated system is shown in Fig \ref{Fig_04}, denoted by $O_\mathrm{t}\xi''\eta''\zeta''$.
The intersection of the collision cone and the $O_\mathrm{t}\xi''\zeta''$ plane specifies a 
possible collision vector, which we denoted as $\mathbf{\Tilde{v}}$ 
in Fig.~\ref{Fig_04}. An $\alpha$-angle rotation about the 
$O_\mathrm{t}\xi''$ axis connects the original $\mathbf{v}$ and $\mathbf{\Tilde{v}}$. 
Then if we denote by $\mathbf{\Tilde{f}^{(2)}}$ the 
unit vector in the direction of $\mathbf{\Tilde{v}}$:
\begin{equation}
\mathbf{\Tilde{f}^{(2)}} = \frac{\mathbf{\Tilde{v}}}{\Tilde{v}} =
    \left( \begin{array}{r}
    -\cos \theta \\
    0 \\
    \sin \theta
    \end{array} \right)
\label{tilde_f2}
\end{equation}
and then rotate it around the axis $O_\mathrm{t}\xi''$ by $\alpha$ it yields
\begin{equation}
\mathbf{{f}^{(2)}} = \left( \begin{array}{rrr}
   1 & 0 & 0 \\
   0 & \cos \alpha & -\sin \alpha \\
   0 & \sin \alpha & \cos \alpha 
    \end{array} \right) \mathbf{\Tilde{f}^{(2)}} 
= \left( \begin{array}{r}
   -\cos \theta \\
   -\sin \alpha \sin \theta \\
    \cos \alpha \sin \theta
    \end{array} \right).
\label{Eq_f2}
\end{equation}

\par
Let us transform $\mathbf{f}^{(2)}$ back to the original $O_\mathrm{t}\xi\eta\zeta$ 
coordinate system. Applying $\mathbf{R}_\mathrm{\eta'}(-\phi_1)$ and then 
$\mathbf{R}_\mathrm{\zeta}(-\lambda_1)$, we get
\begin{equation}
\mathbf{v} = v \left( \begin{array}{rrr}
\cos \lambda_1\cos \phi_1 & -\sin \lambda_1 & -\cos \lambda_1 \sin \phi_1 \\
\sin \lambda_1\cos \phi_1 &  \cos \lambda_1 & -\sin \lambda_1 \sin \phi_1 \\
\sin \phi_1 &  0 & \cos \phi_1 \end{array} \right) \left( \begin{array}{r}
    f^{(2)}_x \\
    f^{(2)}_y \\
    f^{(2)}_z 
    \end{array} \right),
\label{Eq_f2_ppppp}  
\end{equation}
where the components of $\mathbf{{f}^{(2)}}$ are given in Eq. (\ref{Eq_f2}).

As a last step, one has to transform the vectors into the inertial barycentric coordinate 
system $\mathrm{BC}xyz$ as shown in Fig.~\ref{Fig_01}. The connection between 
$\mathrm{BC}xyz$ and $O_\mathrm{t}\xi\eta\zeta$ systems is defined by the 
$\mathbf{\omega}$ angular velocity vector of the target. The components 
of $\mathbf{\omega}$ in the $\mathrm{BC}xyz$ coordinate system may be given as
\begin{equation}
\mathbf{\omega} = \omega \left( \begin{array}{r}
   \cos \lambda \cos \phi \\
   \sin \lambda \cos \phi \\
   \sin \phi
    \end{array} \right),
\label{Eq_omega_component}
\end{equation}
where $\lambda$ is the longitude and $\phi$ is the latitude in the "barygraphic" coordinate system. 
The transformation is defined by two subsequent rotations by applying $\mathbf{R}_\mathrm{x}(-\lambda)$ 
and then $\mathbf{R}_\mathrm{y'}(-\phi)$. The computation of these matrices is similarly done as before.

\par
After performing the aforementioned transformations, we obtain the barycentric coordinates 
and velocities of the projectile. From these, one can calculate the orbital elements associated 
with the given collision. These computed orbital elements together with the those of the
other bodies are listed in Table \ref{tab_01} for all scenarios and serve as the initial
conditions for the backward integration. For all the scenarios the initial orbital elements 
of the bodies are the same except for the projectile, which are given in the table for
all the six scenarios.

\par 
According to the last objective in step 2, the backward integration is performed for 
$T_\mathrm{back} \approx 10^4$ seconds. As a result, the separation between the colliding 
bodies will be at least 5 times the sum of their radii. The SPH code requires this 
initial separation distance in order for the SPH particles to arrange themselves properly
due to mutual tidal and gravitational forces. 
The end state of this backward simulation is used to generate the initial condition for 
the SPH code. In this article, for simplicity, we assume that initially neither the 
target nor the projectile rotates around its axis, and that the collision 
occurs in a plane parallel to the $\mathrm{BC}xy$ plane, thus $\lambda_1 = 0$
and $\phi_1 = 0$ therefore $\mathbf{d} = (d, 0, 0)$.

\begin{table*}
\centering
\begin{tabular}{ccccrrrrrrr}
\hline
Sim. & $v [v_\mathrm{esc}]$ & $\theta$ [deg]   &  Body    & $m\, [\mathrm{M_\oplus}]$  &      $a$ [au]   &    $e$        &      $i$ [rad]  & $\omega$ [rad]&   $\Omega$ [rad] &  $M$ [rad]  \\
\hline
     &     &    &  Venus      &   2.44757e-06  &  0.72332  &  6.76124e-3 &  5.92446e-2  &  0.95363 &   1.33761 &  0.6327  \\
     &     &    &  Earth      &   3.00244e-06  &  0.99994  &  1.70615e-2 &  4.94382e-5  &  4.61622 &   3.47543 &  2.17442 \\
     &     &    &  Mars       &   3.22611e-07  &  1.52358  &  9.34559e-2 &  3.22607e-2  &  5.00092 &   0.86416 &  1.36914 \\
     &     &    &  Jupiter    &   9.54296e-04  &  5.20147  &  4.89592e-2 &  2.27578e-2  &  4.77563 &   1.75444 &  2.20341 \\
     &     &    &  Saturn     &   2.85725e-04  &  9.55384  &  5.37045e-2 &  4.33936e-2  &  5.93821 &   1.98306 &  2.50830 \\
     &     &    &  Target     &   4.71882e-10  &  2.08000  &  5.00000e-2 &  8.72664e-2  &        0 &         0 &        0 \\
\hline                                                                                                                  
\hline                                                                                                                  
run1 & 3.0 & 30 &  Projectile &   4.71882e-10  &  2.10526  &  8.55781e-2 &  1.21687e-1  &  0.83226 &         0 &  5.57213 \\
\hline                                                                                                                  
run2 & 3.0 & 45 &  Projectile &   4.71882e-10  &  2.11117  &  8.03866e-2 &  1.35865e-1  &  0.69963 &         0 &  5.68244 \\
\hline                                                                                                                  
run3 & 5.0 & 30 &  Projectile &   4.71882e-10  &  2.13530  &  1.23688e-1 &  1.44479e-1  &  1.02614 &         0 &  5.45849 \\
\hline                                                                                                                  
run4 & 5.0 & 45 &  Projectile &   4.71882e-10  &  2.14546  &  1.12679e-1 &  1.67923e-1  &  0.87794 &         0 &  5.56950 \\
\hline                                                                                                                  
run5 & 7.0 & 30 &  Projectile &   4.71882e-10  &  2.17693  &  1.65018e-1 &  1.67121e-1  &  1.12143 &         0 &  5.44273 \\
\hline                                                                                                                  
run6 & 7.0 & 45 &  Projectile &   4.71882e-10  &  2.19173  &  1.48589e-1 &  1.99636e-1  &  0.96437 &         0 &  5.54775 \\
\hline
\end{tabular}
\caption{Initial conditions for the $N$-body simulation scenarios for backward integration. The first 
column contains the name of the simulation. Columns 2 and 3 show the impact velocity $v$ and angle 
$\theta$, respectively. Column 4 displays the name of the body, Column 5 represents its mass in 
Earth-mass units. Columns 6-11 show the orbital elements of the bodies, $a$ is the semi-major 
axis in au, $e$ is the eccentricity, $i$ is the inclination, $\omega$ is the argument of 
pericenter, $\Omega$ is the longitude of the ascending node, and $M$ is the mean anomaly. 
The orbital elements of planets listed above the double line are the same for all six runs.}
\label{tab_01}
\end{table*}

\par
In step 3, the SPH code developed by \cite{Schaefer2020} was utilized. This code has proven to be highly 
effective in simulating high-resolution surface impacts \citep{Maindl2014a,Maindl2015} as well as 
investigating collisions between planetesimals and protoplanets \citep{Maindl2014b,Maindl2014c}. 
It has been successfully utilized in both scenarios, yielding valuable insights and accurate results. 
In SPH-based simulations, continuous bodies are represented by discrete SPH particles 
\citep[e.g.][]{Gingold1977,Benz1995,Maindl2013,Schaefer2005}. These particles carry the 
specific physical properties of interest and contribute to the overall physical 
properties of their respective volume elements. In this study, we employ a relatively 
low resolution of $N_\mathrm{SPH} \approx 20 \cdot 10^3$ SPH particles, which means that
each SPH particle has a mass of $m_\mathrm{SPH} = M_\mathrm{t}/N_\mathrm{SPH} = 4.695 \cdot 10^{16}$ kg. 

\par
Due to a collision, a large amount of debris is generated and scattered. If we were to 
consider every debris as a massive body in the $N$-body code, the simulation would take a 
very long time to run. In order to strike a balance between the statistically significant number 
of bodies and the runtime of the simulation, we only considered debris with masses denoted 
as $m_\mathrm{frag}$ above a certain threshold as massive bodies and treated the rest as 
test particles. A fragment is considered as protoplanet if 
$m_\mathrm{frag}\in [10^{-10}, 10^{-7}]\,\mathrm{M_\oplus}$, if 
$m_\mathrm{frag}\in [10^{-13}, 10^{-10})\,\mathrm{M_\oplus}$ then it was treated as planetesimal, 
and any remaining mass below $10^{-13}\,\mathrm{M_\oplus} = 1.87\cdot 10^{17}$ kg as test particle. 
This lower mass limit 
corresponds to approximately 4 SPH particles. This compromise ensures to get statistically 
meaningful results and to keep the number of massive bodies manageable while guaranteeing 
that the debris is not only present in the form of test particles in the $N$-body simulator. 

\begin{figure*}
\includegraphics[width=0.9\columnwidth,keepaspectratio=true]{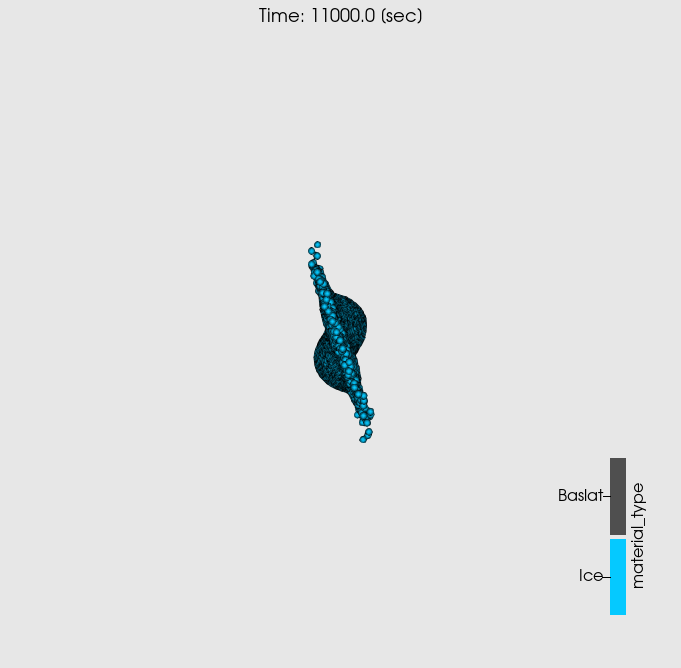}
\includegraphics[width=0.9\columnwidth,keepaspectratio=true]{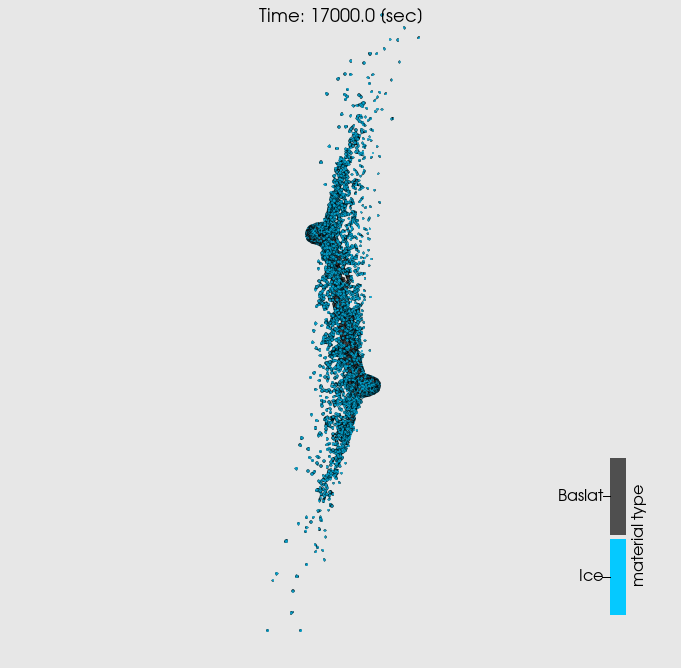}
\includegraphics[width=0.9\columnwidth,keepaspectratio=true]{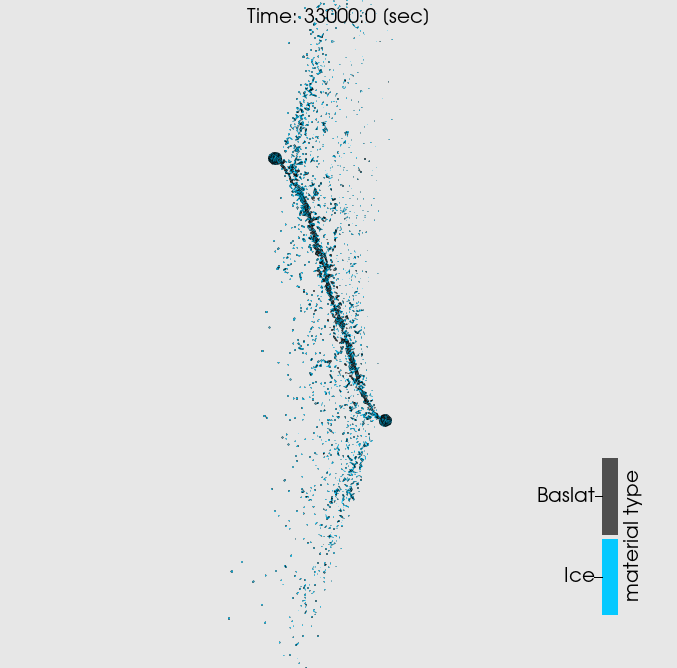}
\includegraphics[width=0.9\columnwidth,keepaspectratio=true]{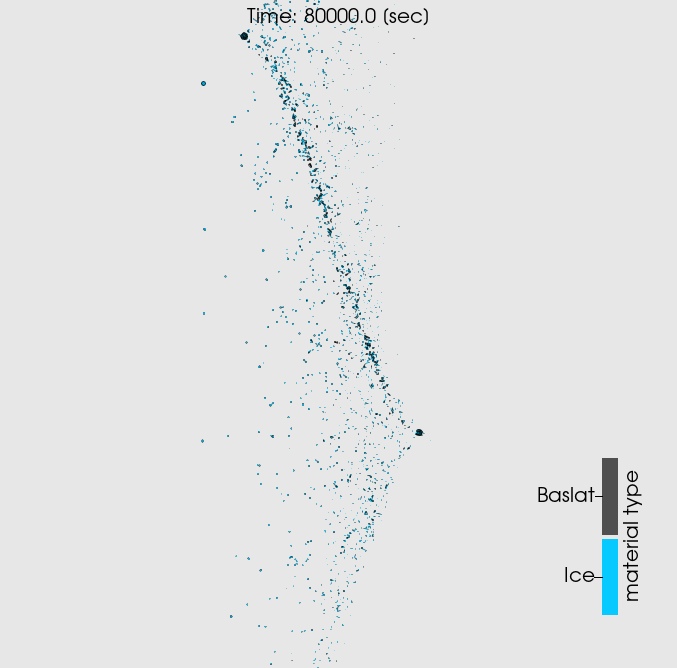}
\caption{Collision snapshots for $v = 3.0\, [v_\mathrm{esc}]$ and $\theta = 30^\circ$. 
The simulation time for each frame is indicated at the top. The actual collision takes 
place at $t \approx 10^4$ seconds. Basalt material is represented in grey, while 
water ice is depicted in sky blue. Both bodies have a layer of water ice 
(25 percent of its total mass). We note that the scale increases on successive snapshots. 
In the top left panel, it is a few times $\sim 10^3$ km, while in the bottom right panel, 
it is a few times $\sim 10^5$ km kilometers.}
\label{Fig_05}
\end{figure*}

In Fig.~\ref{Fig_05}, a series of four snapshots are presented for the collision with 
$\theta = 30^\circ$ and $v = 3.0\, [v_\mathrm{esc}]$. The spherical objects are composed 
of basaltic rock represented by grey and water ice represented by sky blue. Neither 
of the bodies is rotating at the start of the simulation. The collision happens at $t\approx 10^4$ 
seconds, and the evolution of the colliding bodies is followed via the SPH code for 
$t = 8\cdot 10^4$ seconds. The lower right frame in Fig.~\ref{Fig_05} illustrates the 
final outcome of the SPH simulation, which will serve as the input for the subsequent 
$N$-body simulation. The region occupied by the fragments in the final snapshot is 
$x\in [1.981,\,1.982]$ au, $y\in [8.638,\,10.61] \times 10^{-3}$ and 
$z\in [-4.737,\,182.6]\times 10^{-5}$ au, which fits in a rectangular cuboid with sizes 
of $\Delta x = 10^{-3},\, \Delta y = 2\cdot 10^{-3}$, and $\Delta z = 2 \cdot 10^{-3}$ au. 

\par
\begin{figure*}
\includegraphics[width=0.27\paperwidth,keepaspectratio=true]{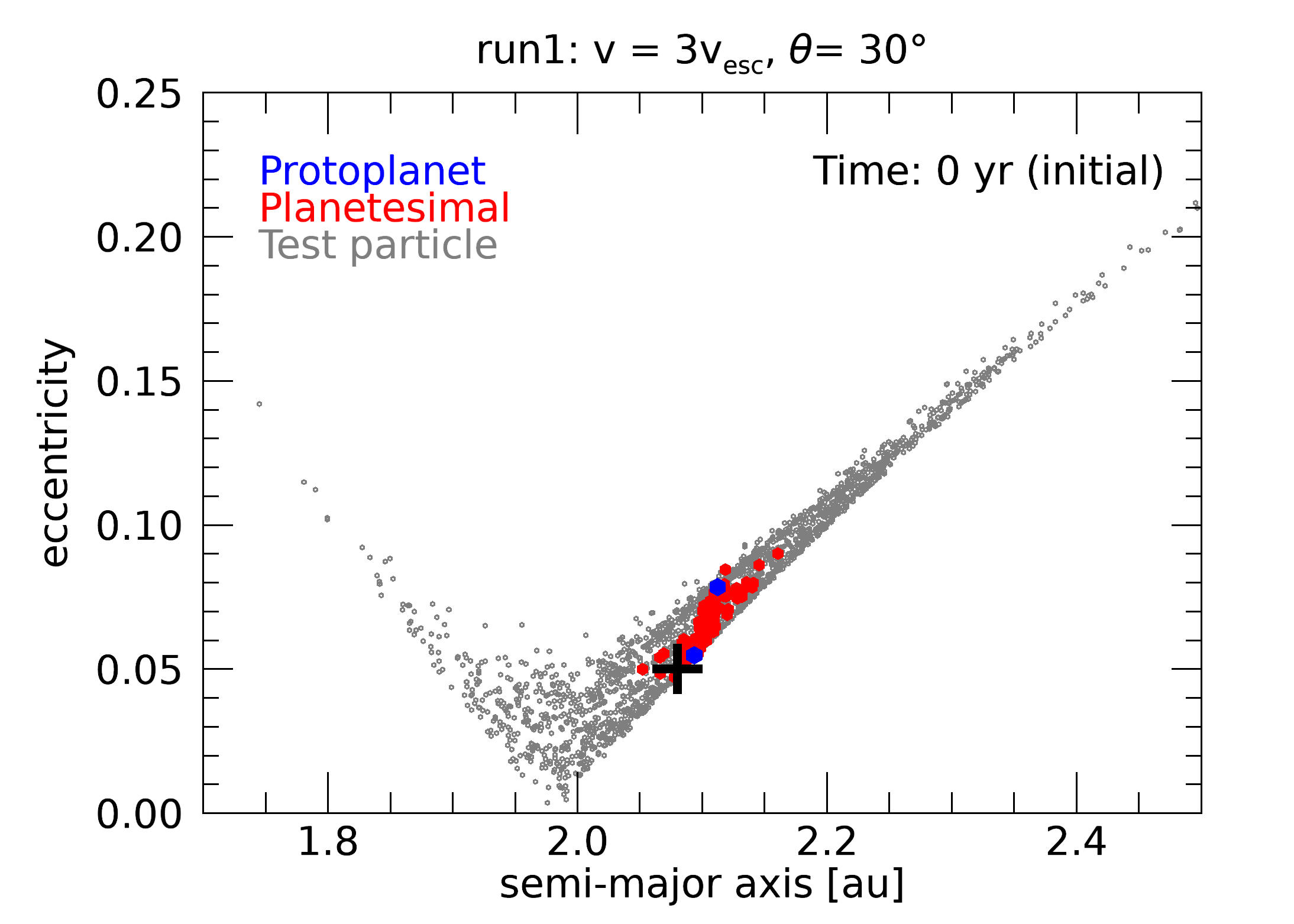}
\includegraphics[width=0.27\paperwidth,keepaspectratio=true]{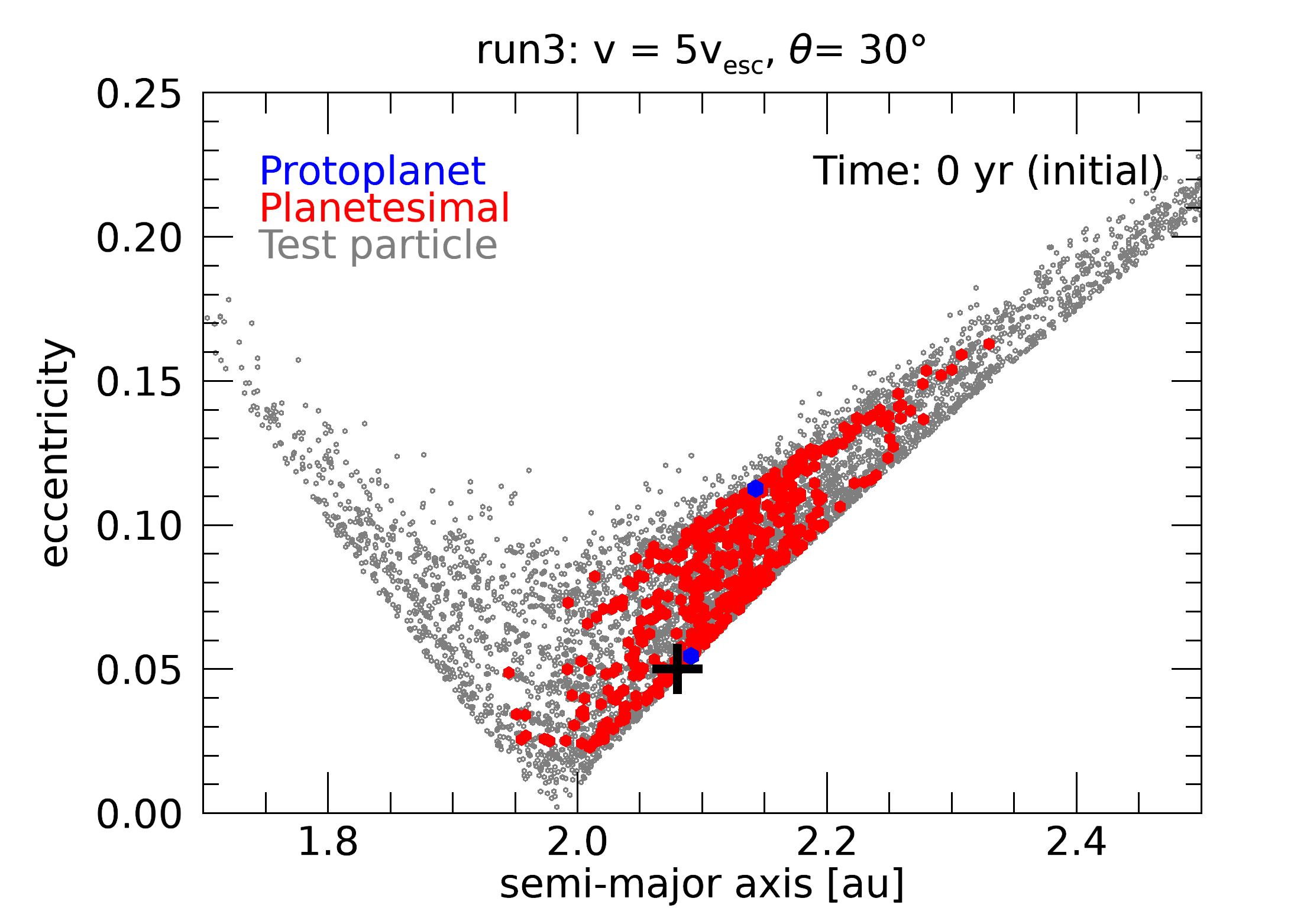}
\includegraphics[width=0.27\paperwidth,keepaspectratio=true]{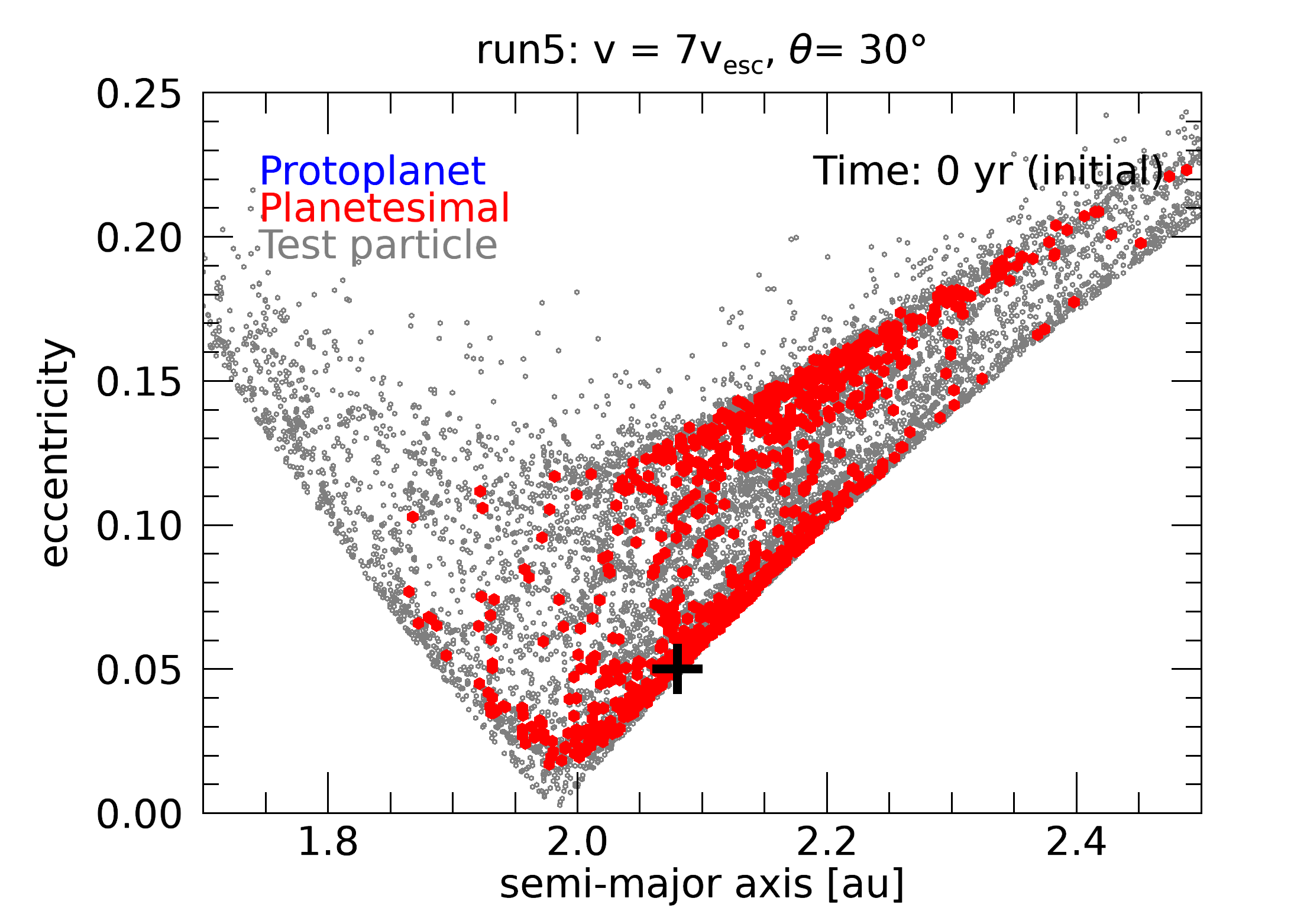} \\
\includegraphics[width=0.27\paperwidth,keepaspectratio=true]{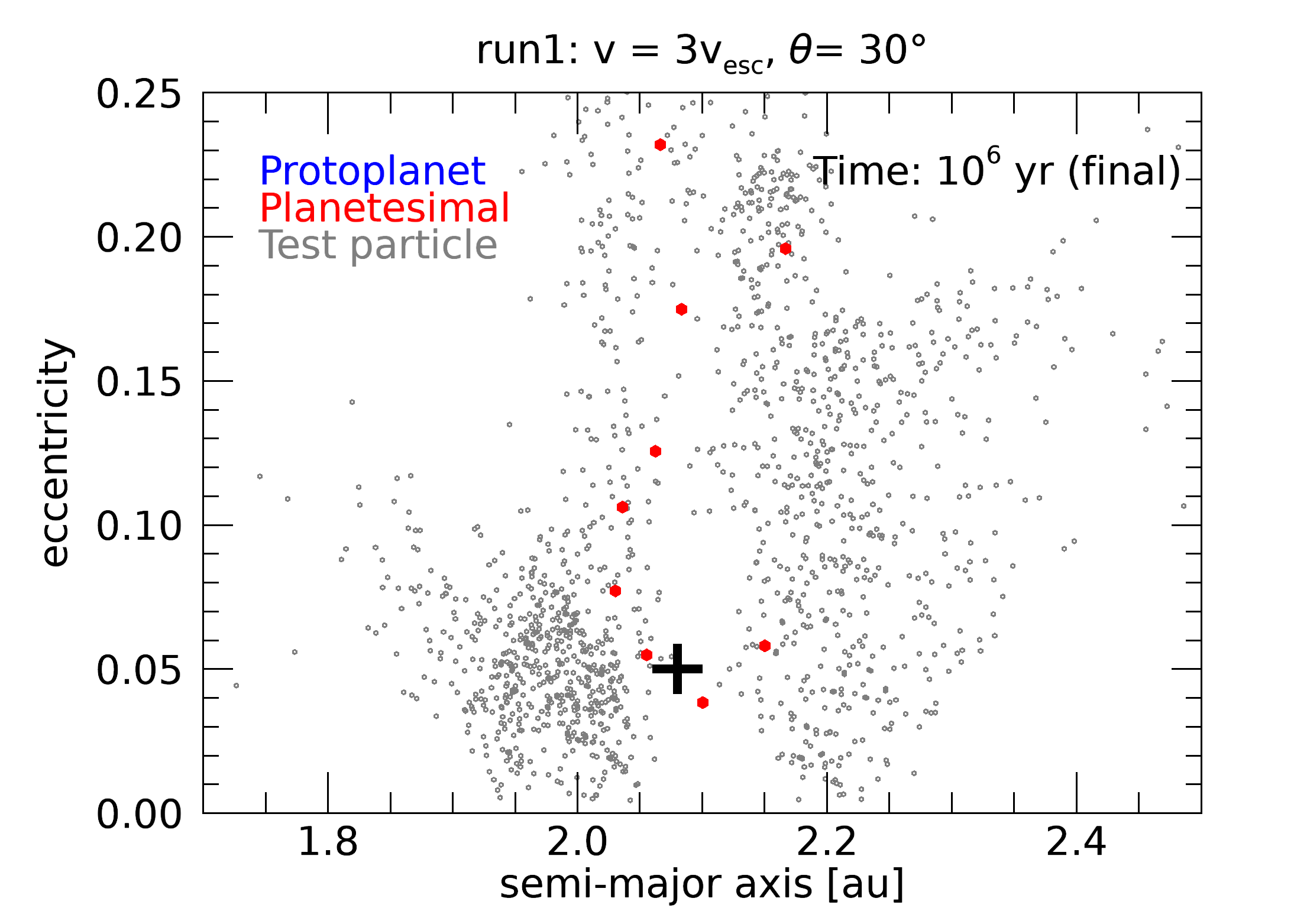}
\includegraphics[width=0.27\paperwidth,keepaspectratio=true]{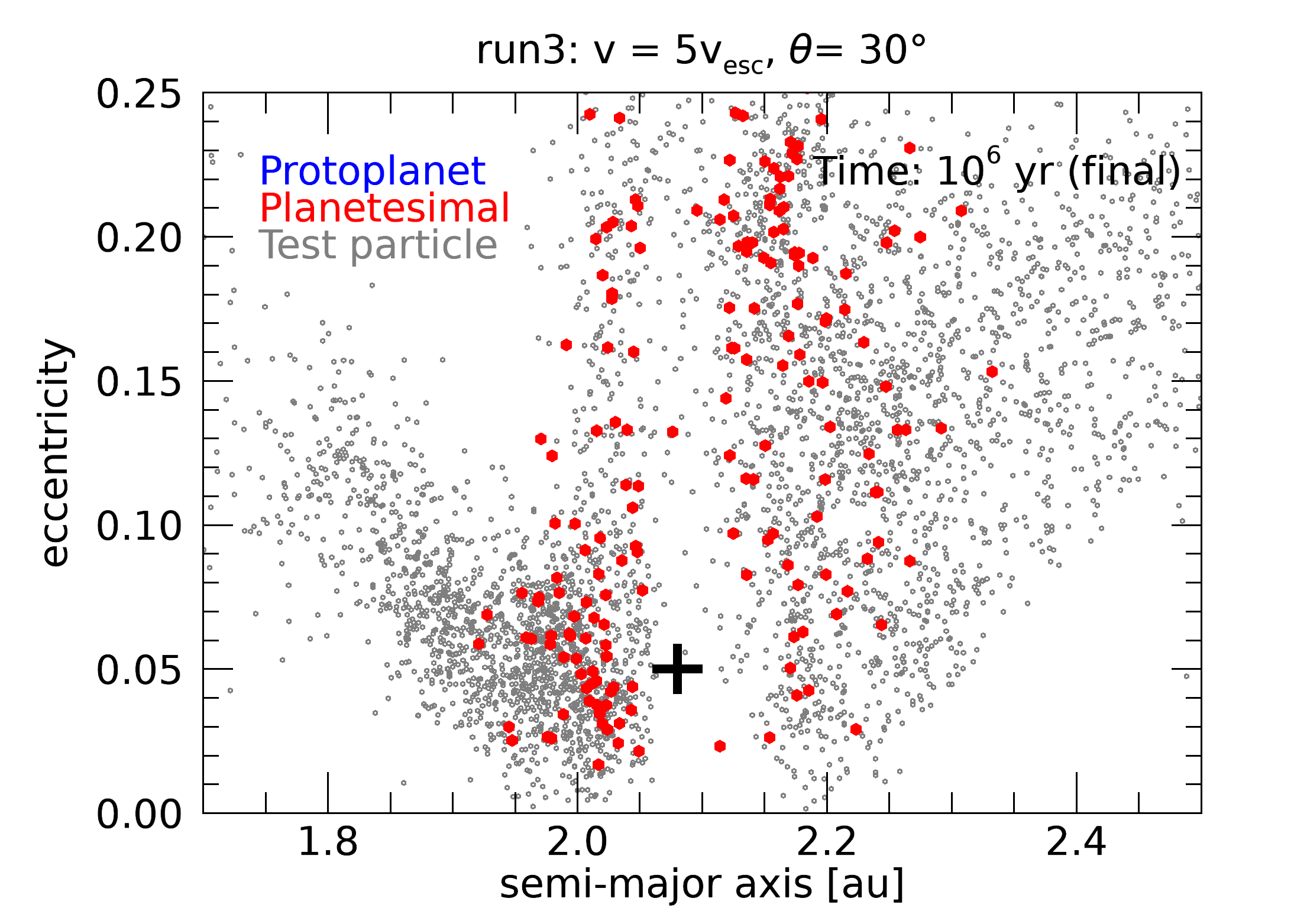}
\includegraphics[width=0.27\paperwidth,keepaspectratio=true]{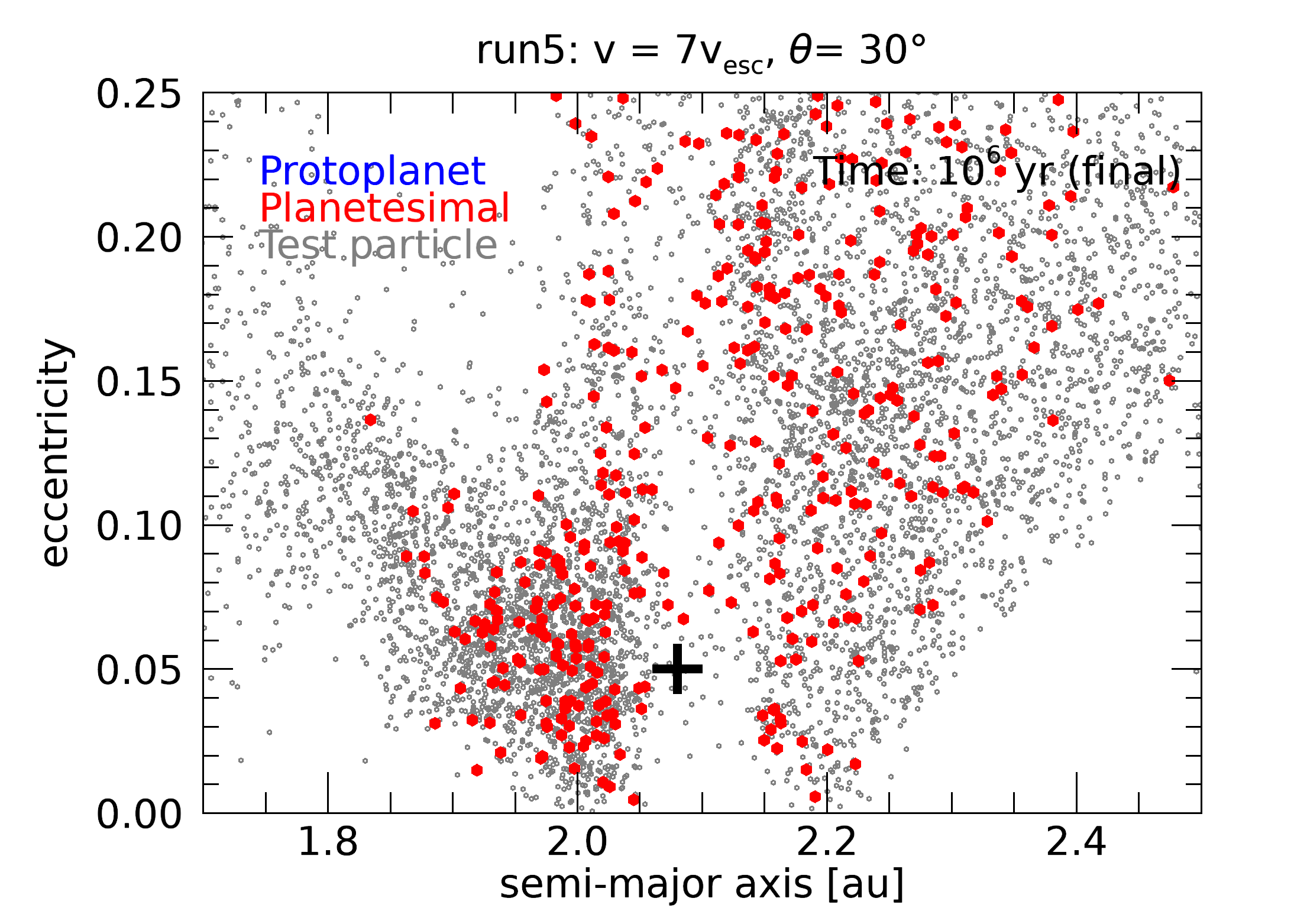} \\
\includegraphics[width=0.27\paperwidth,keepaspectratio=true]{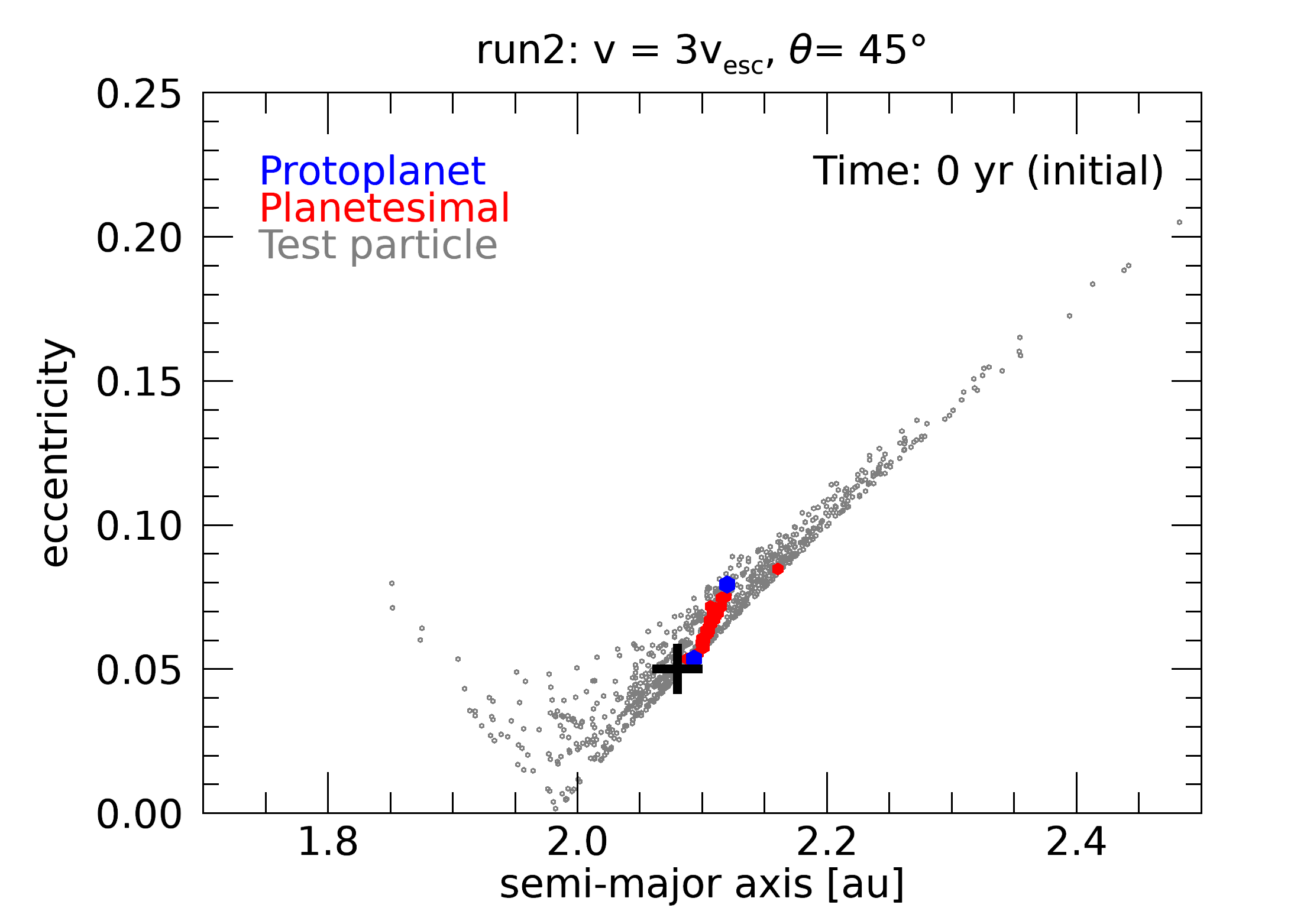}
\includegraphics[width=0.27\paperwidth,keepaspectratio=true]{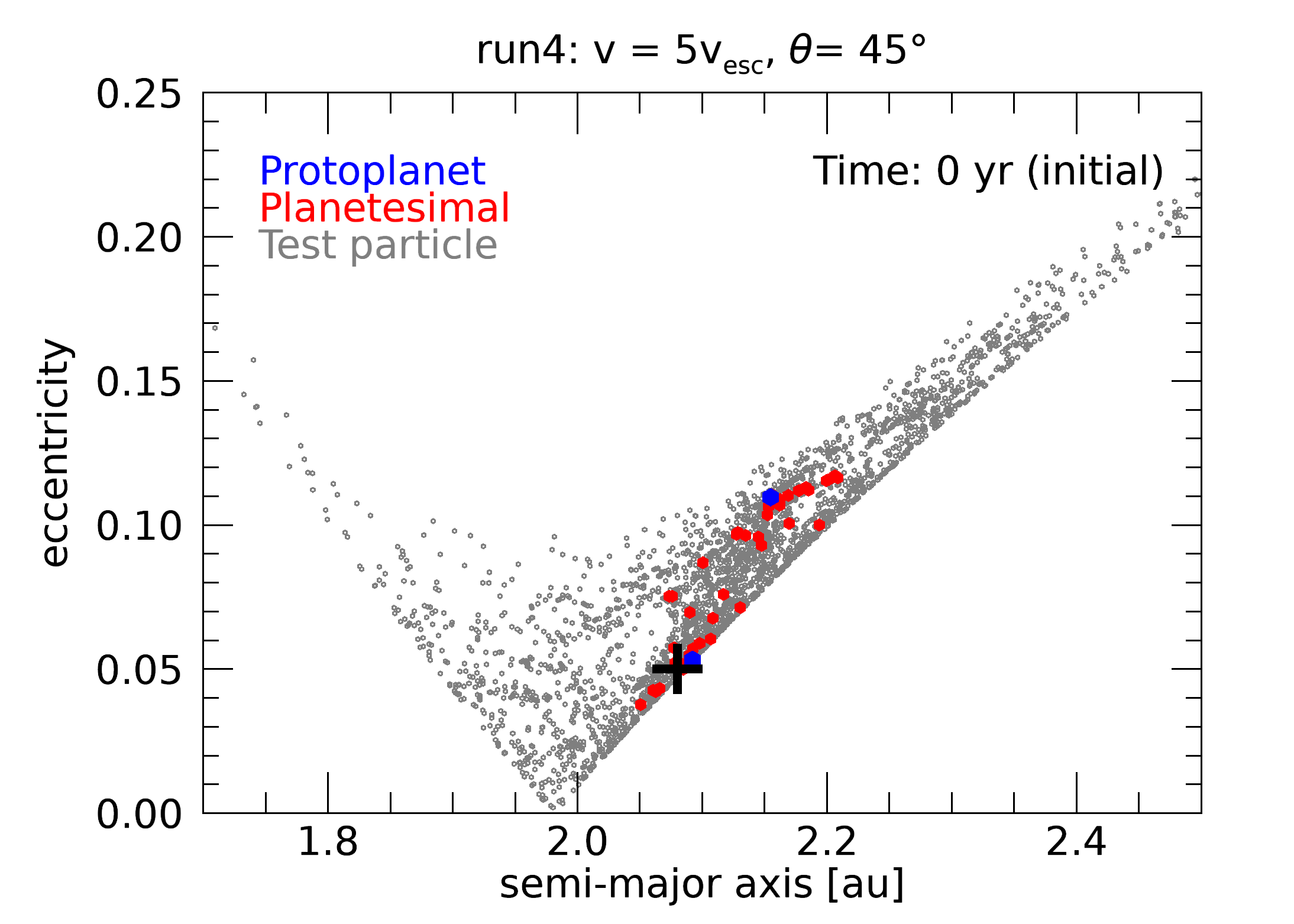}
\includegraphics[width=0.27\paperwidth,keepaspectratio=true]{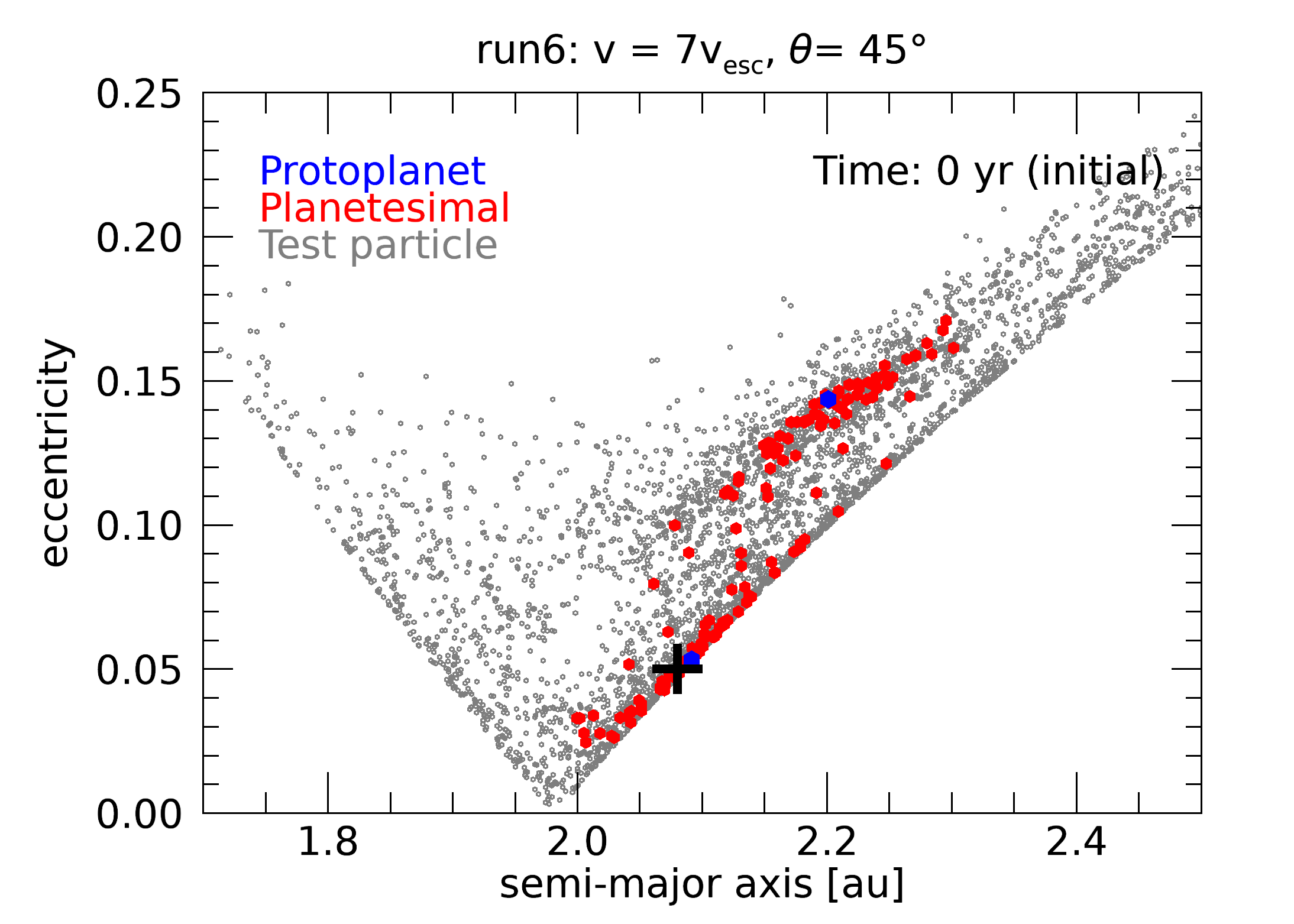} \\
\includegraphics[width=0.27\paperwidth,keepaspectratio=true]{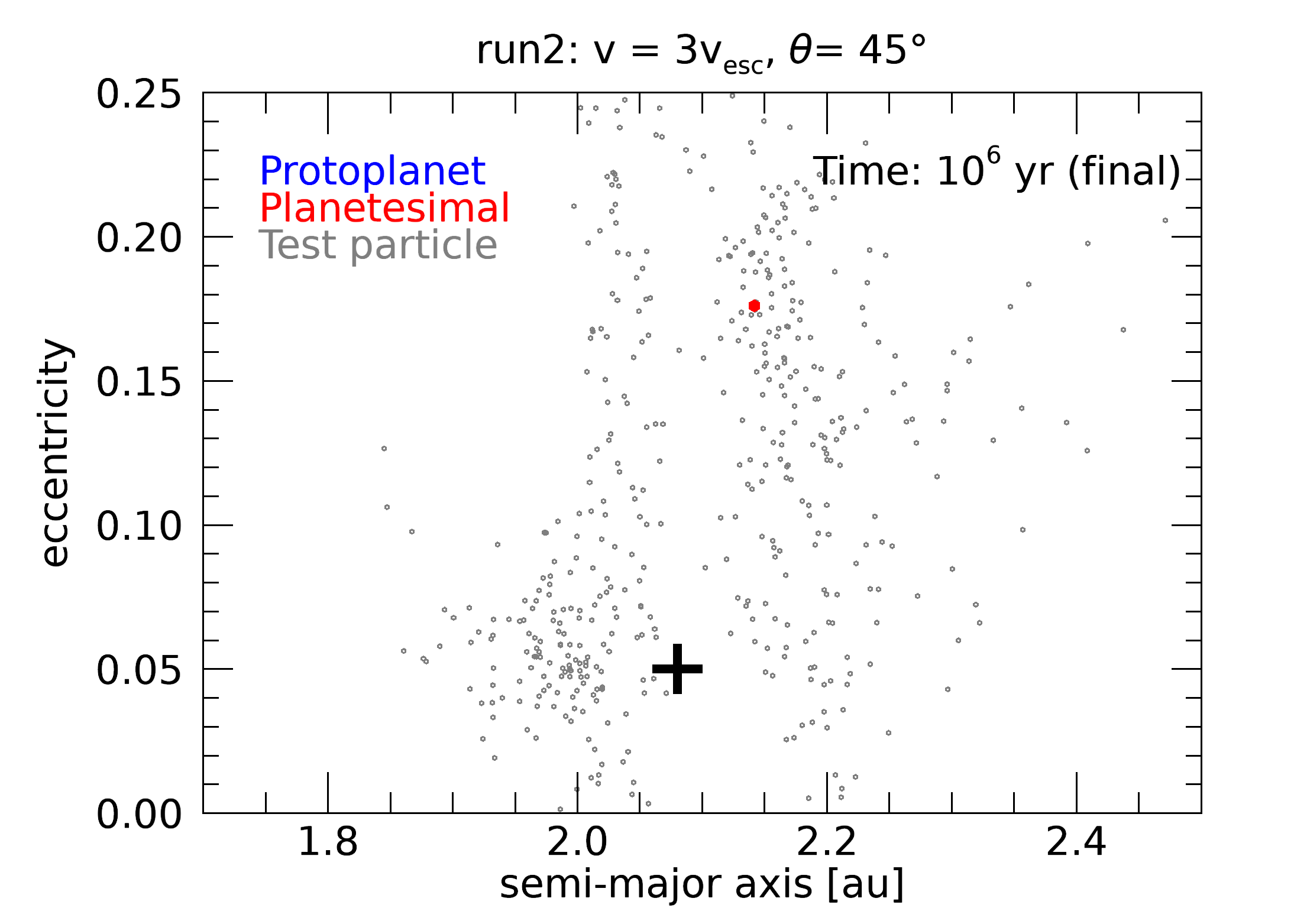}
\includegraphics[width=0.27\paperwidth,keepaspectratio=true]{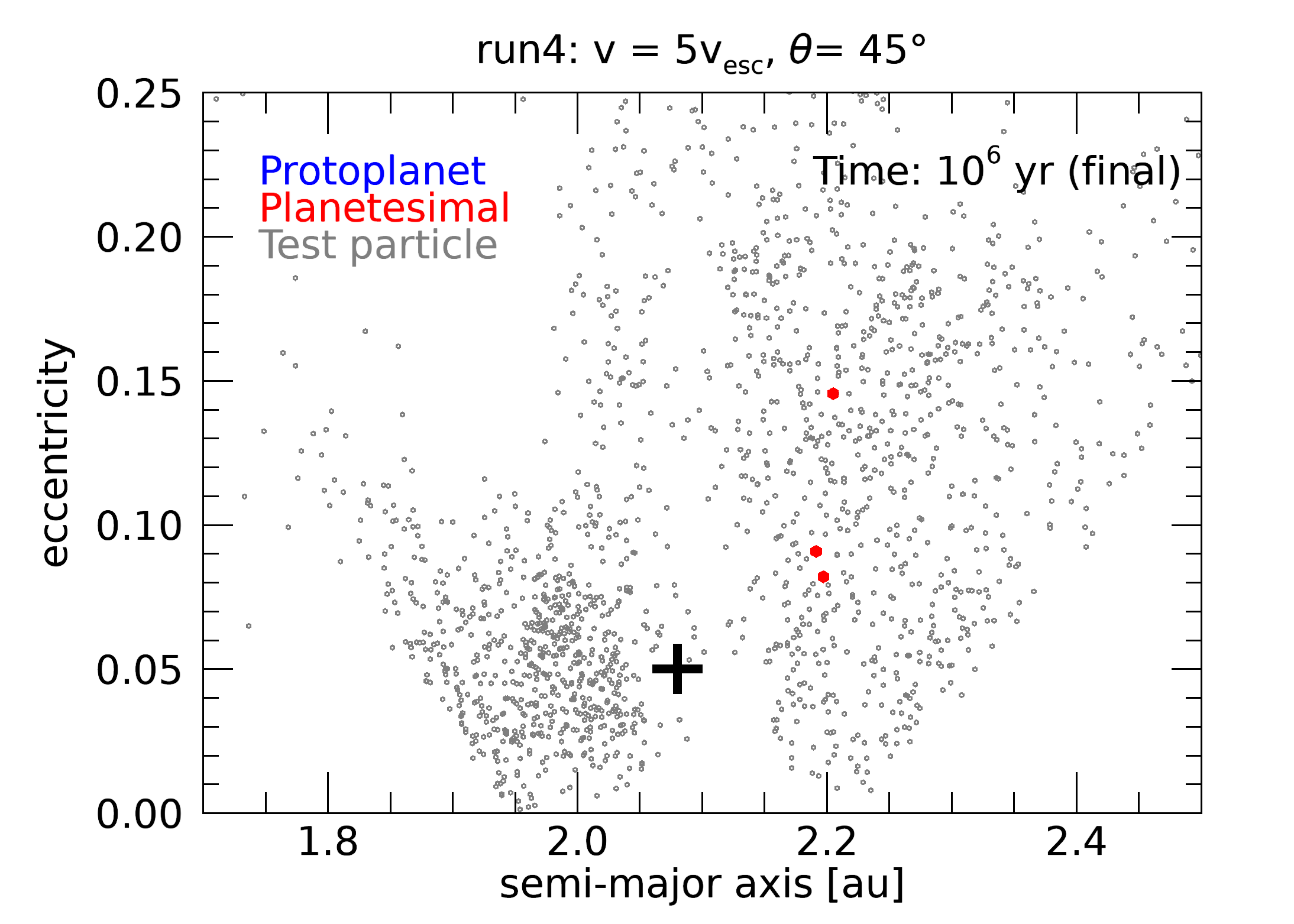}
\includegraphics[width=0.27\paperwidth,keepaspectratio=true]{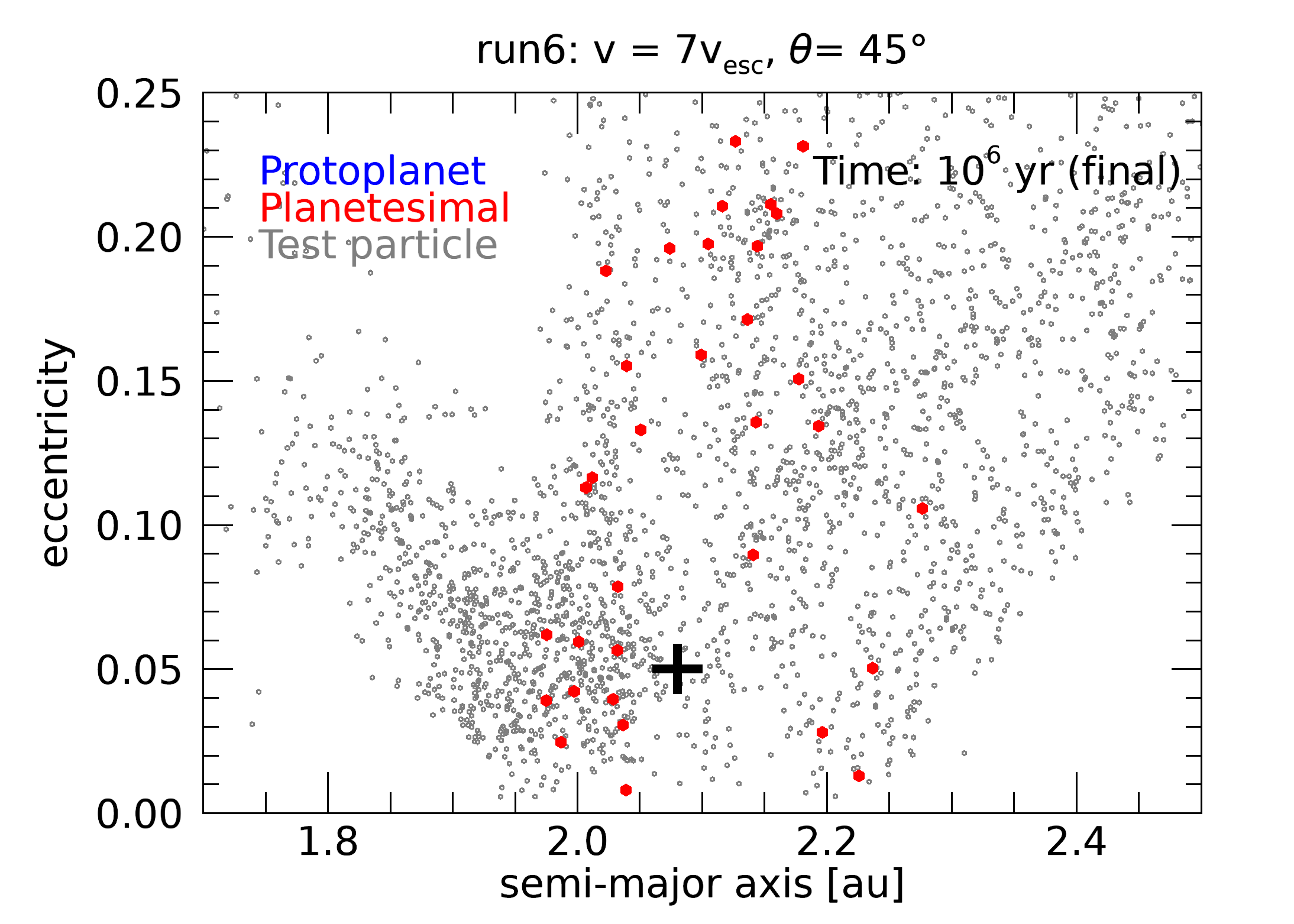} 
\caption{The ($a-e$) plane of the fragments in the $N$-body simulations. 
The top two rows show the initial (first row) and final (second row) eccentricities of the fragments with 
$\theta = 30^\circ$ for different impact velocities, while the third and fourth rows for $\theta = 45^\circ$ 
(see Table \ref{tab_02}).
The initial eccentricities were computed using the barycentric position and velocity vectors taken 
from the last frame of the SPH simulation (see Fig.~\ref{Fig_05}). The big black plus denotes the 
position of the impact. Blue circles represent protoplanets, red ones planetesimals while grey dots 
correspond to test particles.}
\label{Fig_06}
\end{figure*}
\begin{figure*}
\includegraphics[width=0.27\paperwidth,keepaspectratio=true]{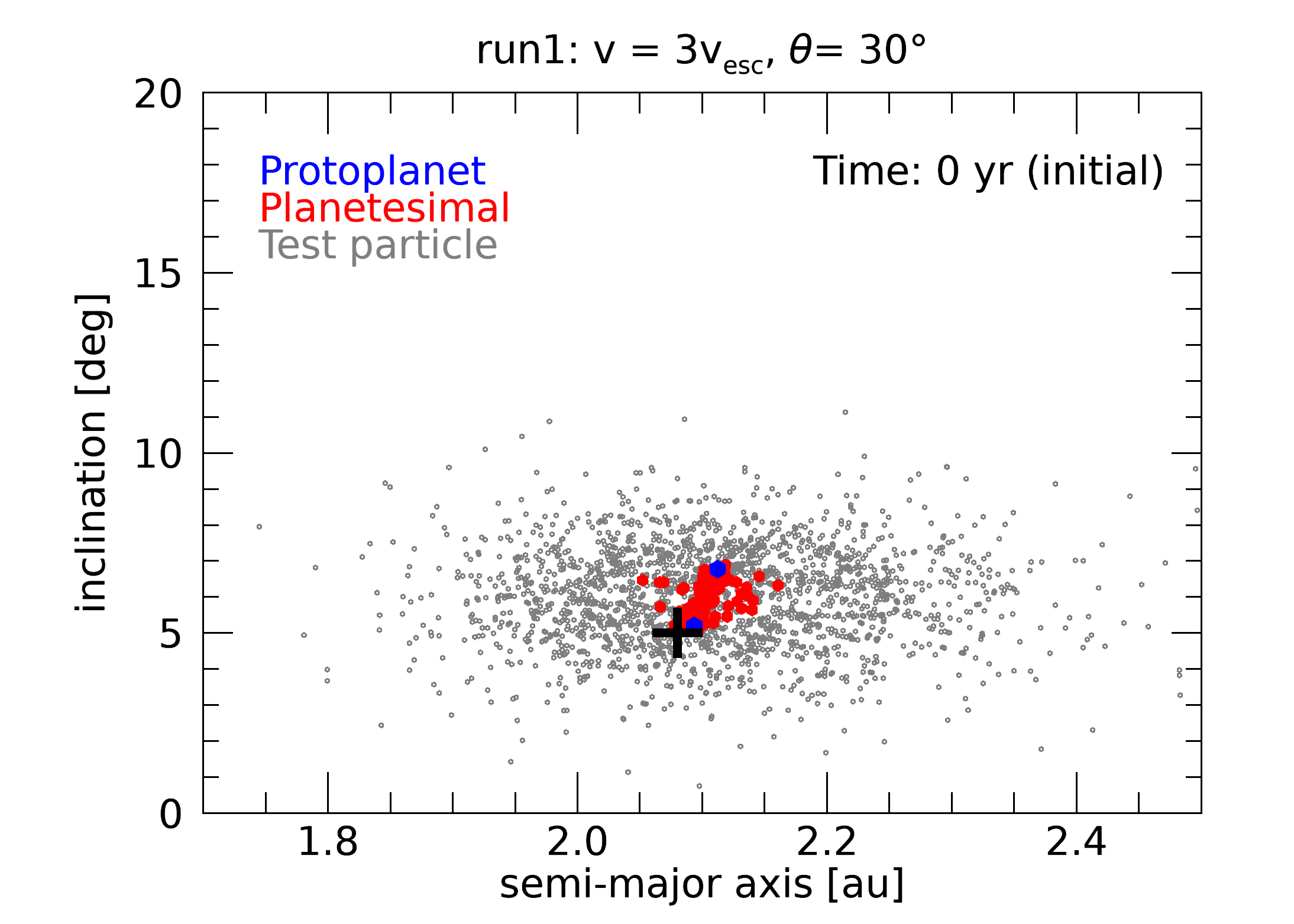}
\includegraphics[width=0.27\paperwidth,keepaspectratio=true]{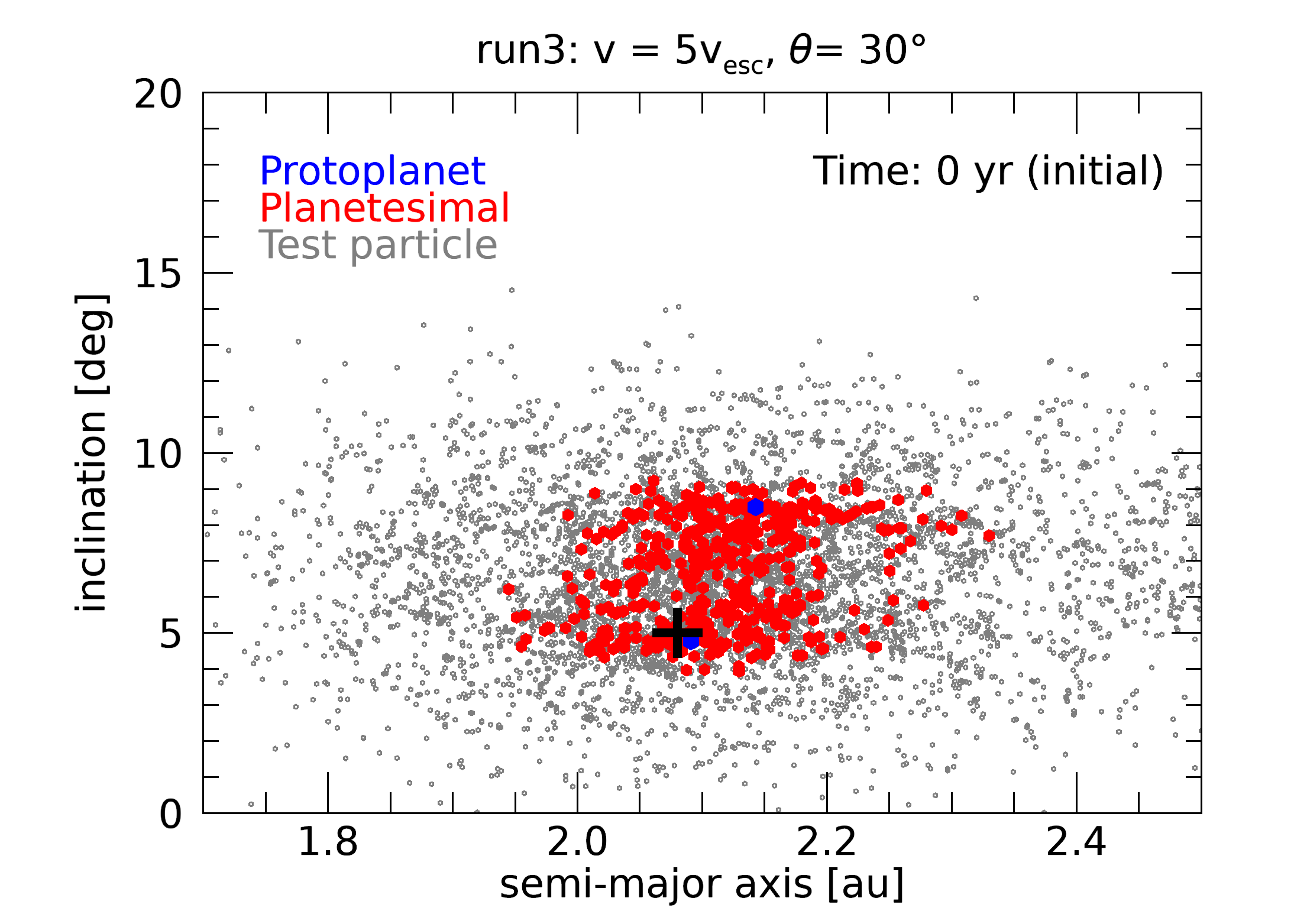}
\includegraphics[width=0.27\paperwidth,keepaspectratio=true]{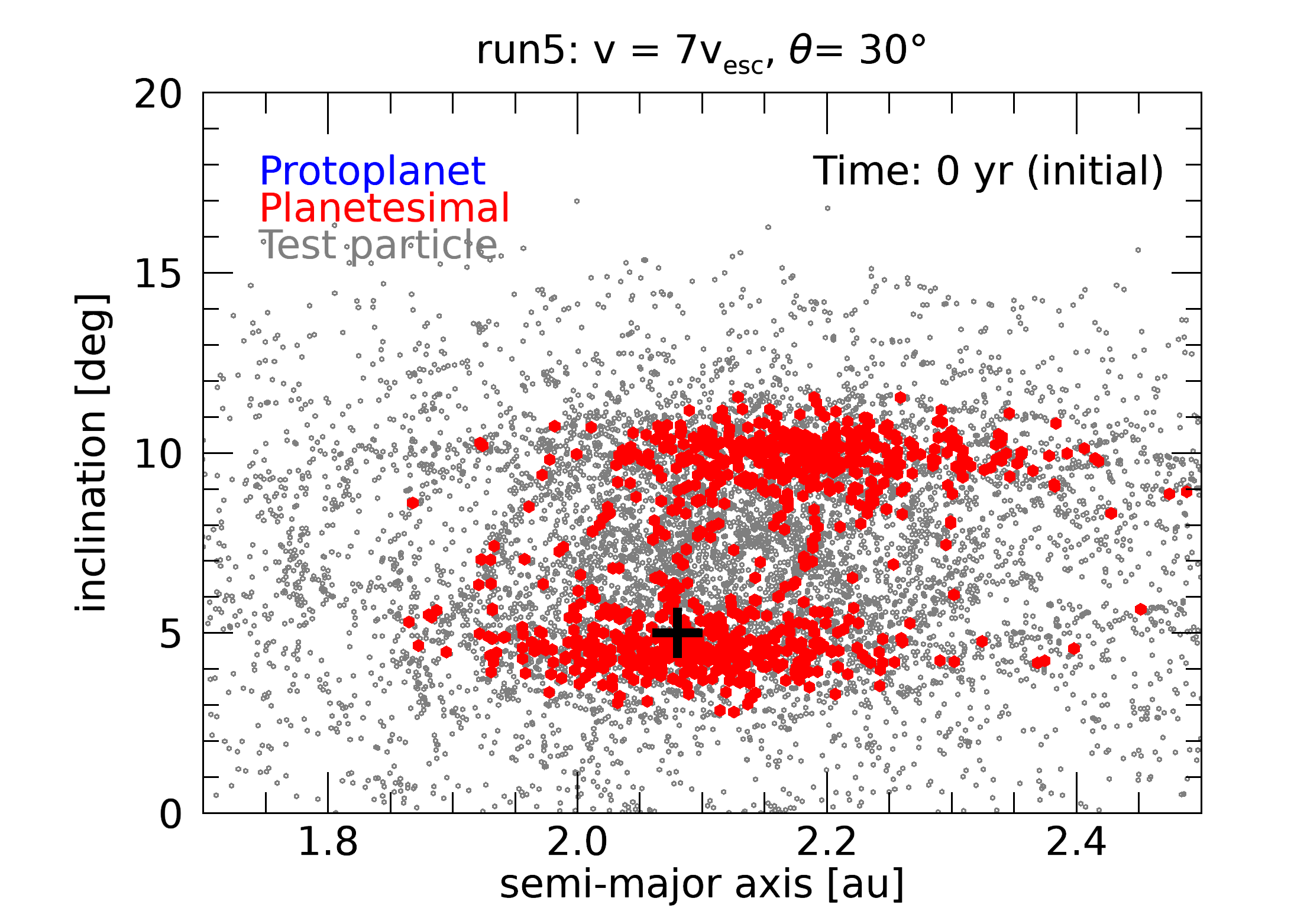} \\
\includegraphics[width=0.27\paperwidth,keepaspectratio=true]{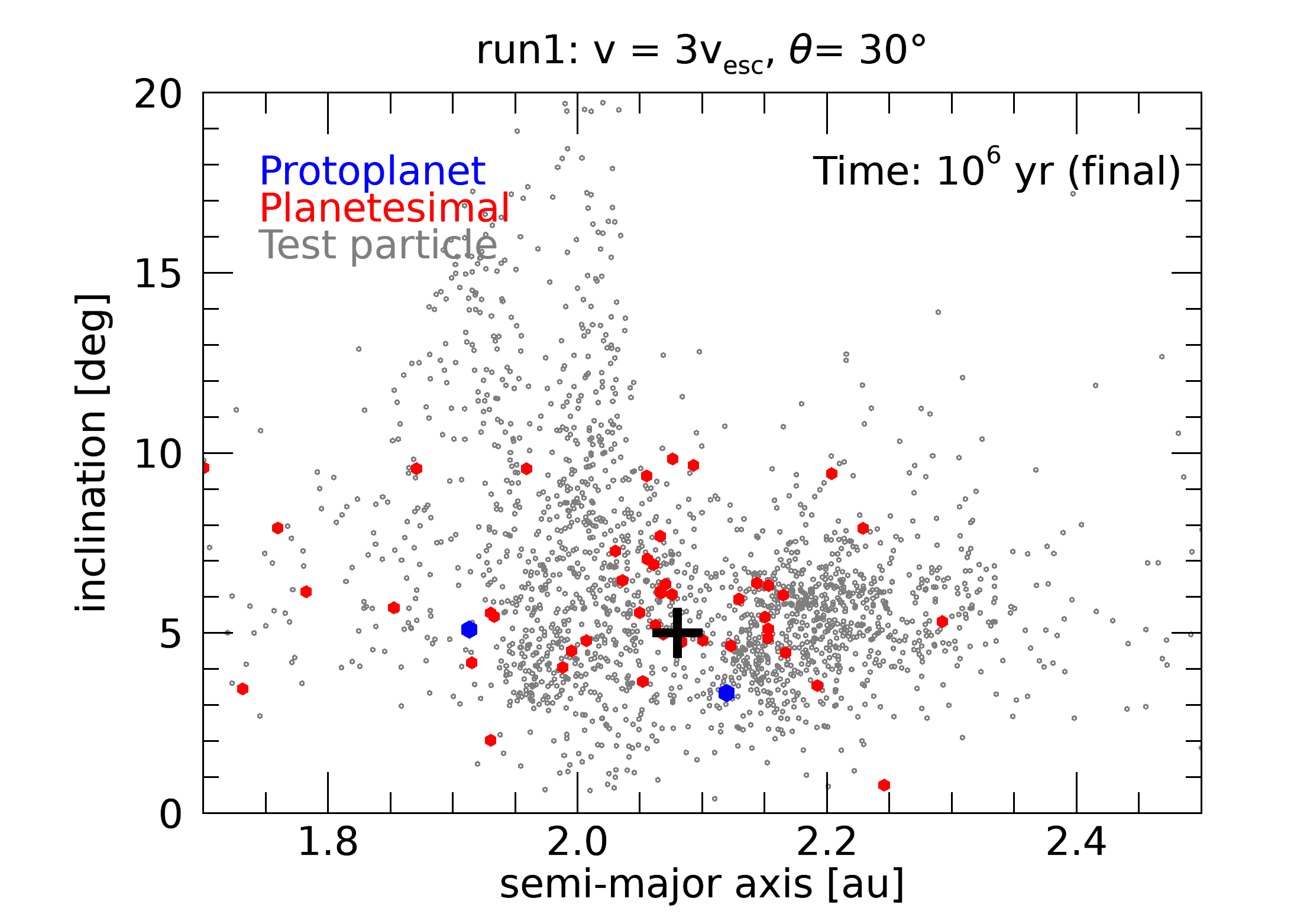}
\includegraphics[width=0.27\paperwidth,keepaspectratio=true]{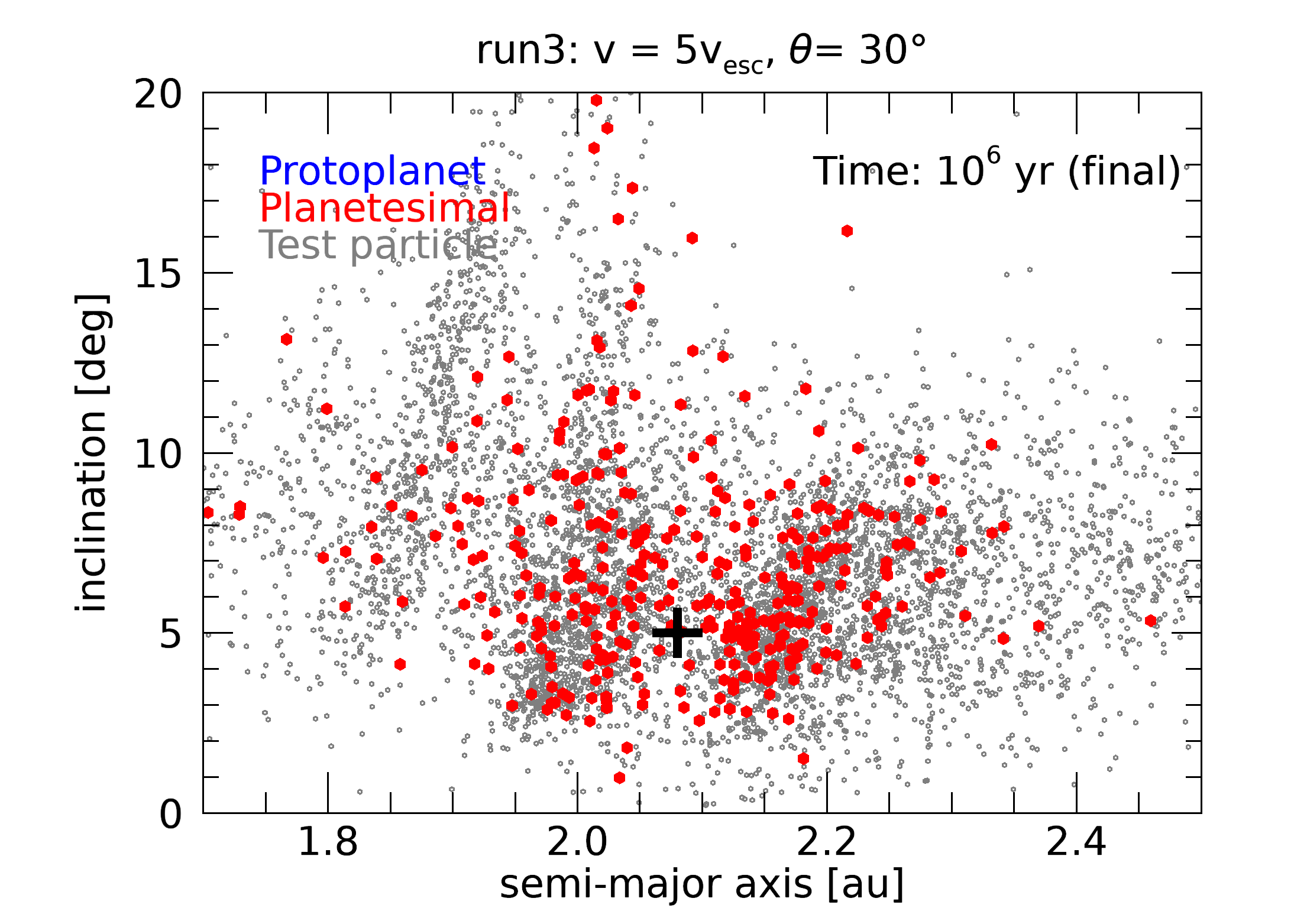}
\includegraphics[width=0.27\paperwidth,keepaspectratio=true]{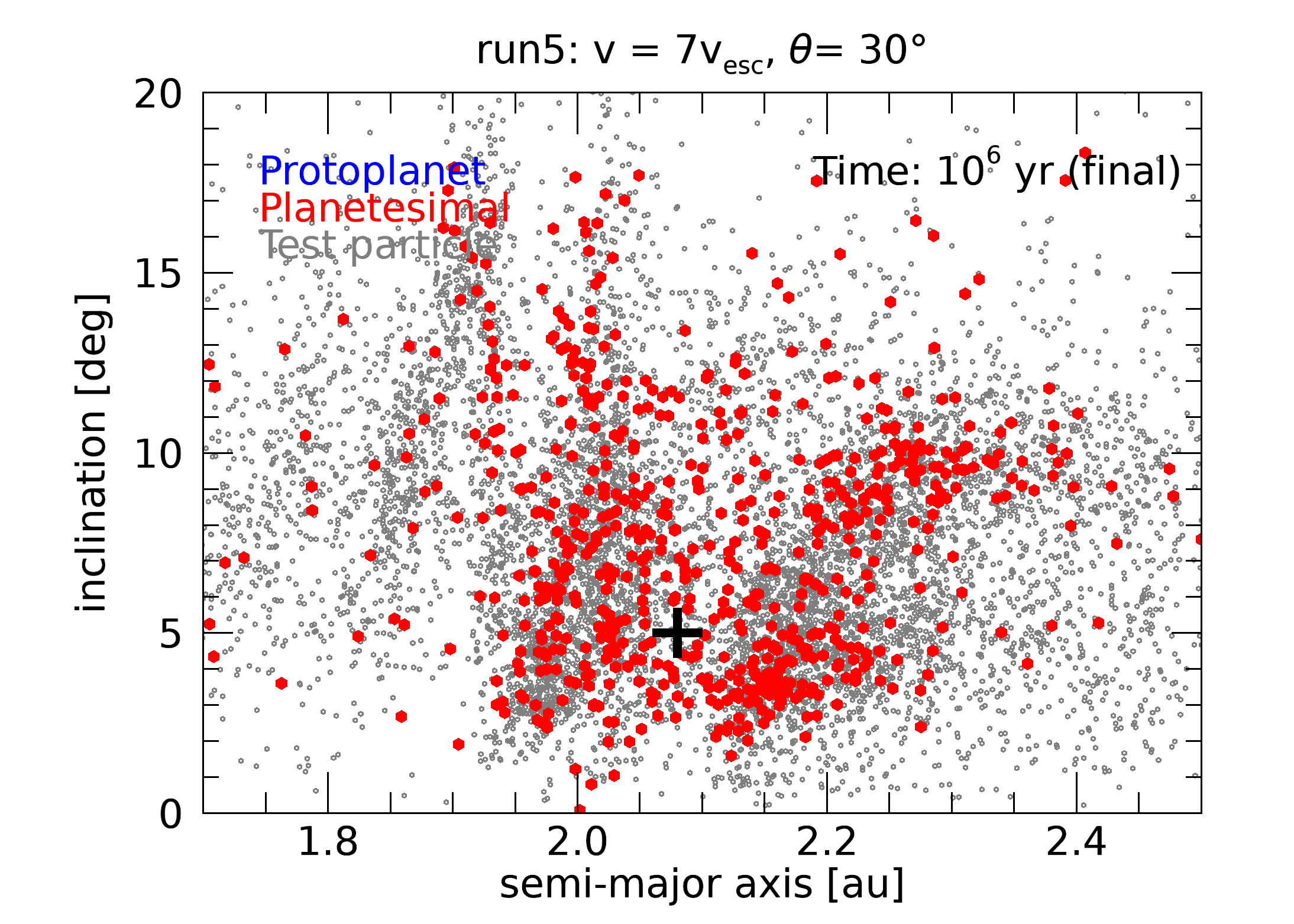} \\
\includegraphics[width=0.27\paperwidth,keepaspectratio=true]{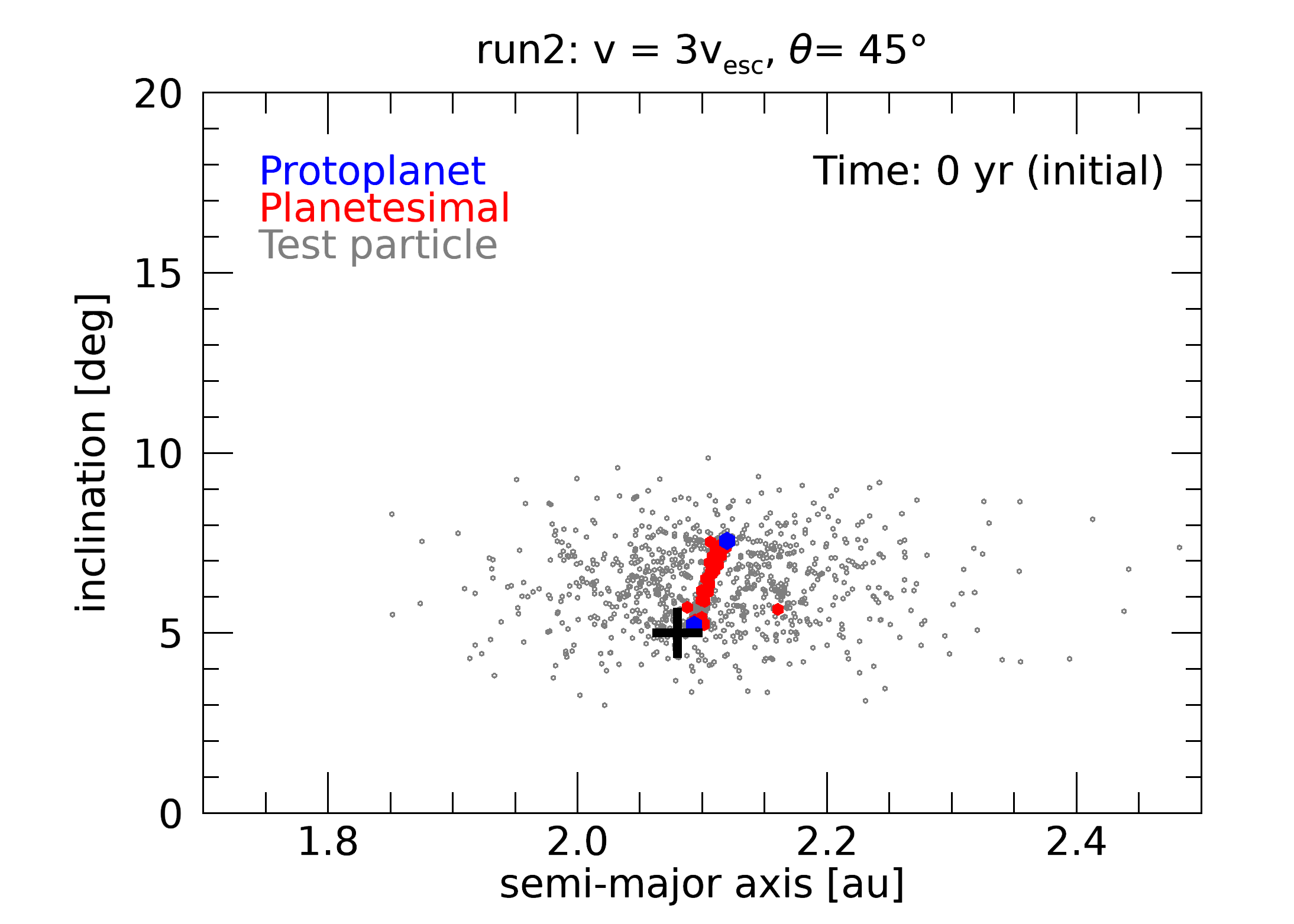}
\includegraphics[width=0.27\paperwidth,keepaspectratio=true]{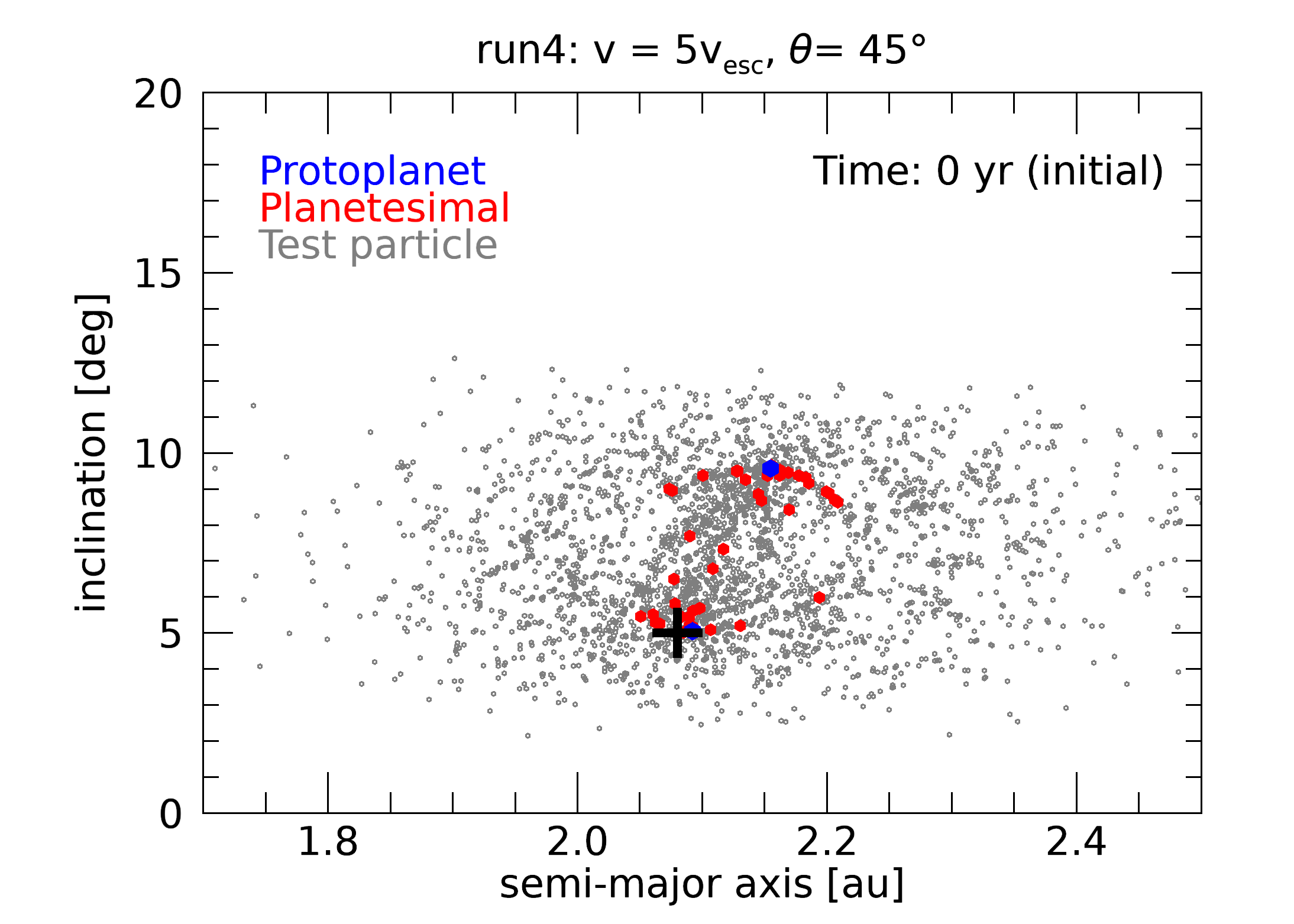}
\includegraphics[width=0.27\paperwidth,keepaspectratio=true]{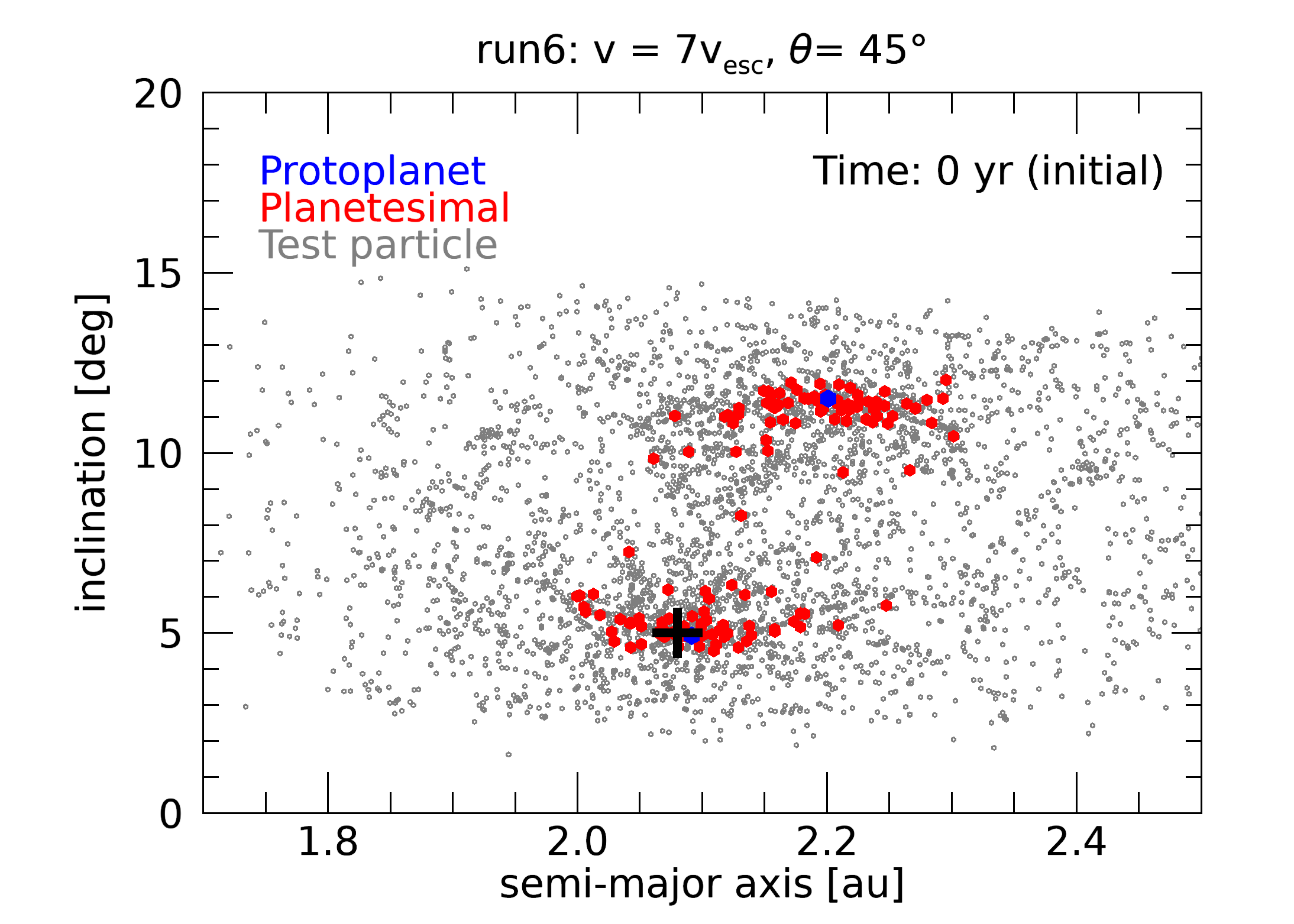} \\
\includegraphics[width=0.27\paperwidth,keepaspectratio=true]{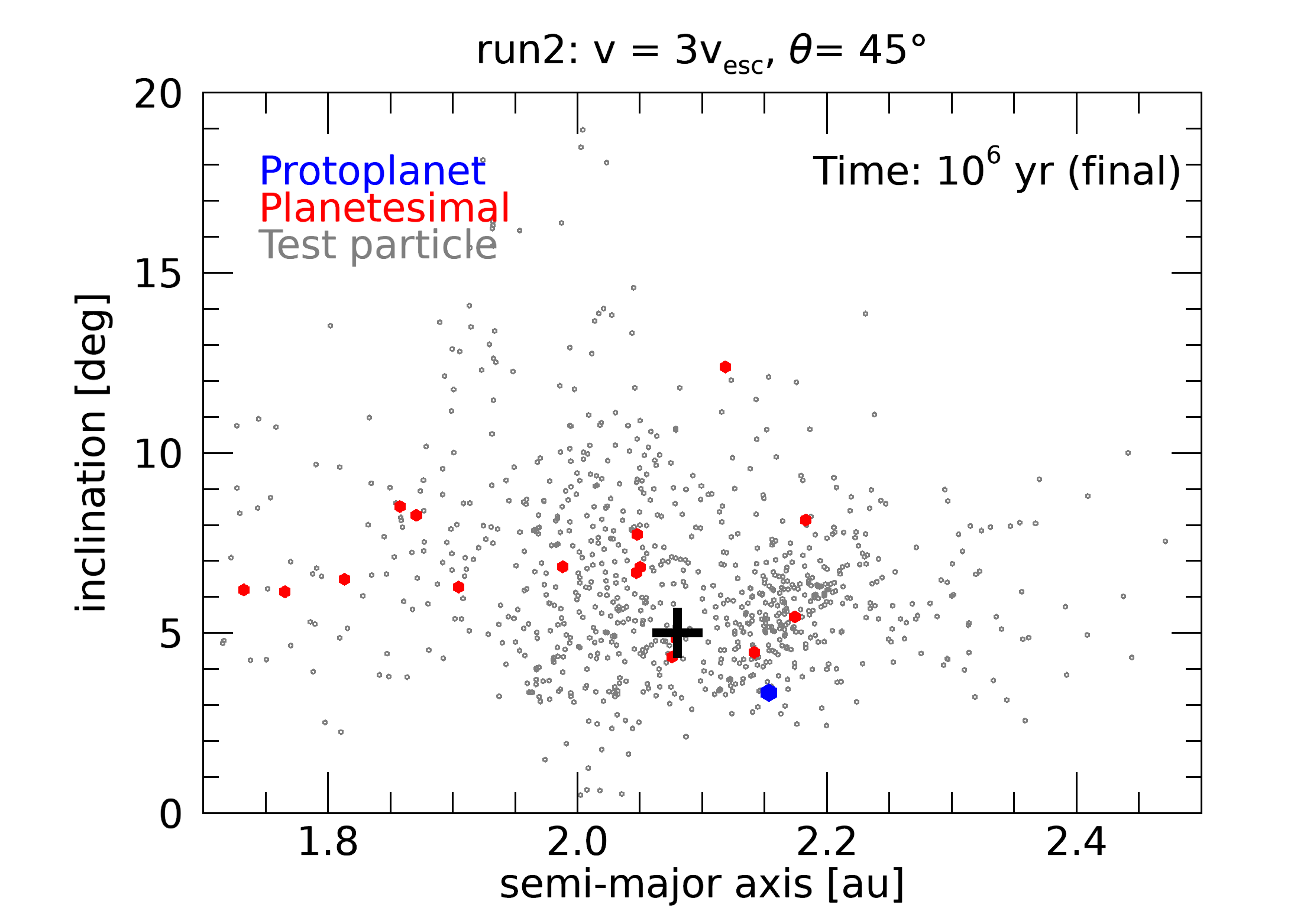}
\includegraphics[width=0.27\paperwidth,keepaspectratio=true]{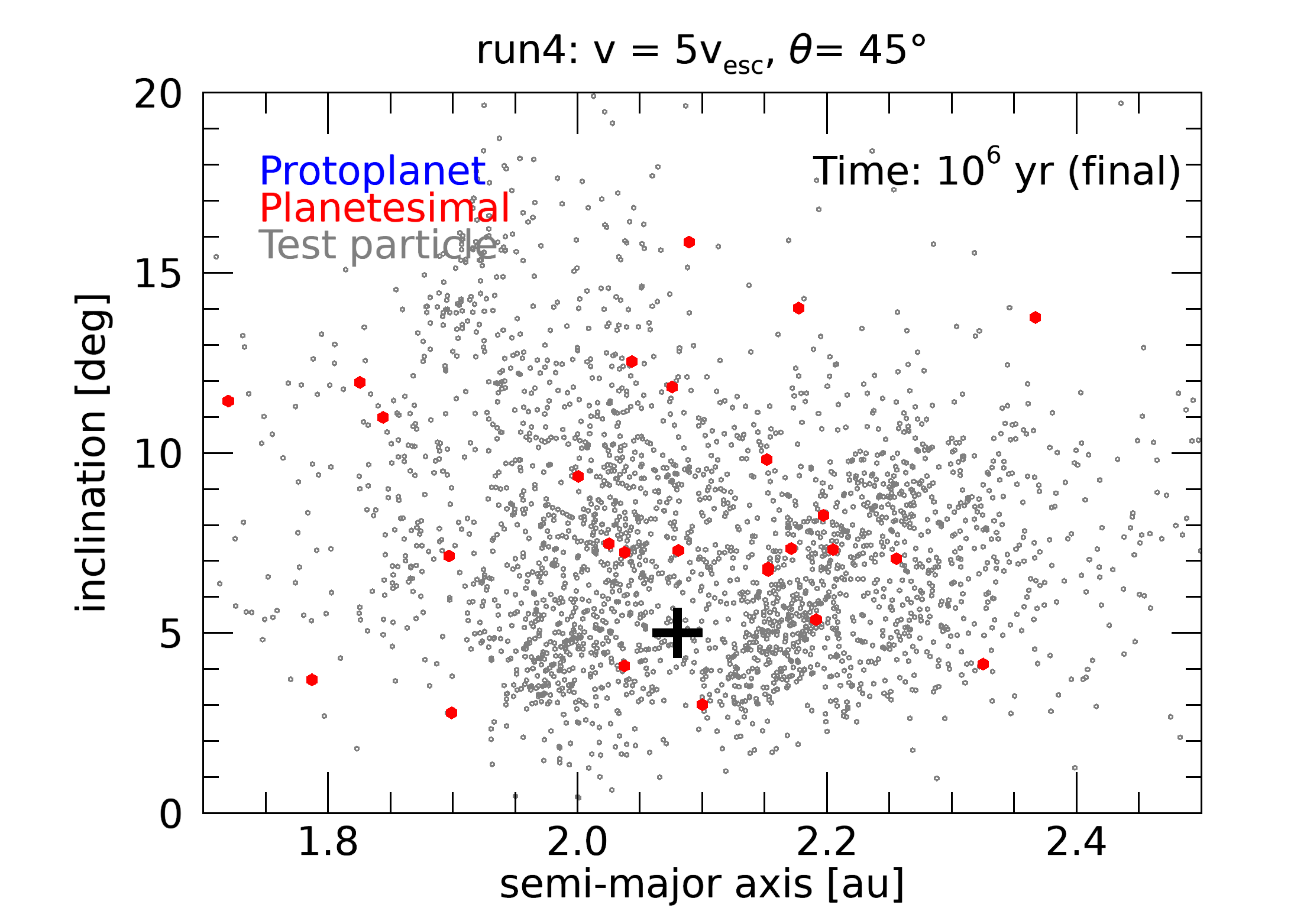}
\includegraphics[width=0.27\paperwidth,keepaspectratio=true]{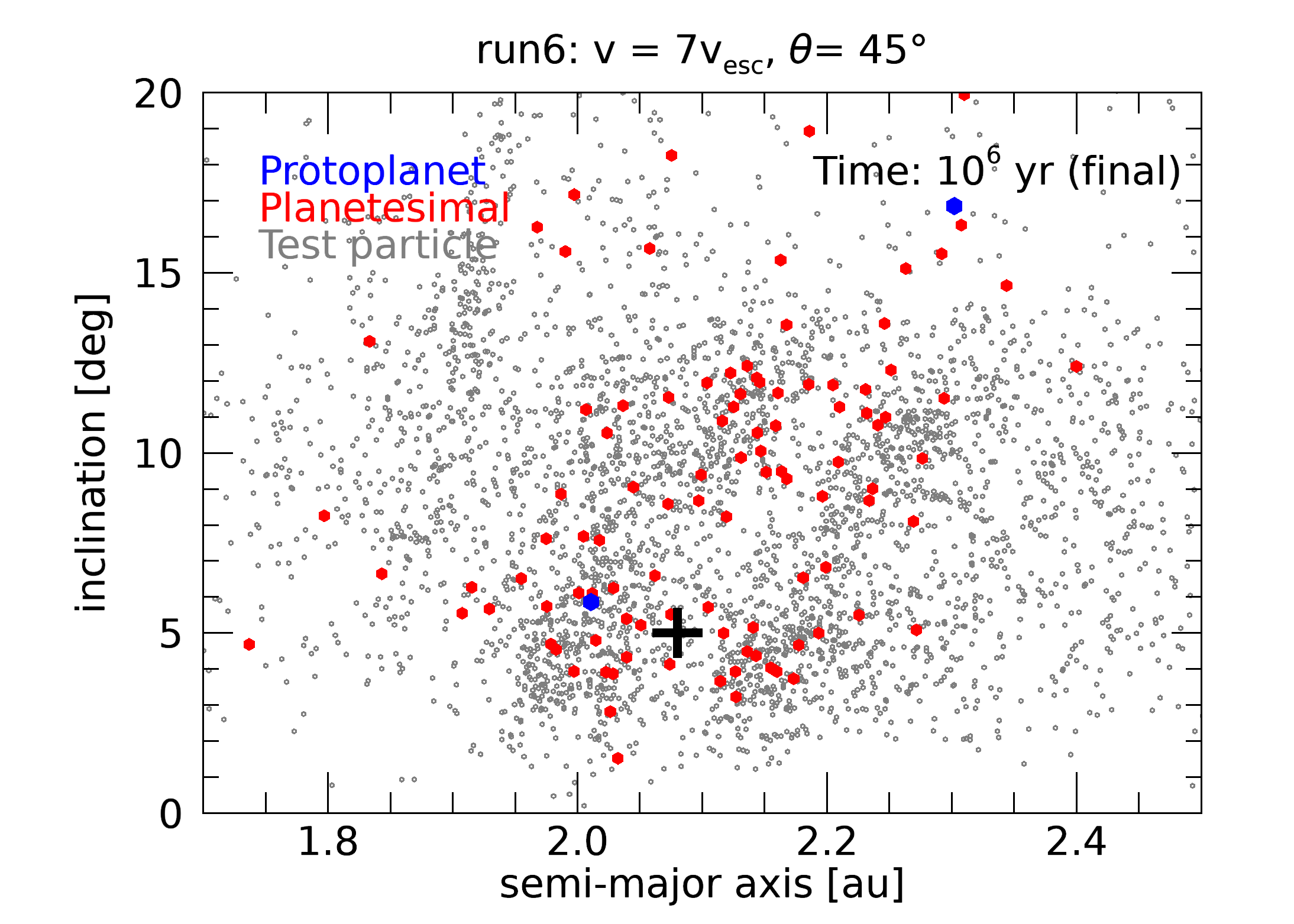} 
\caption{Same as in Fig.~\ref{Fig_06}, but in the ($a-i$) plane.}
\label{Fig_07}
\end{figure*}

From the barycentric position and velocity vectors of the fragments at 
$t = 8\cdot 10^4$ seconds, the orbital elements were computed and the  
eccentricities $e$ and inclinations $i$ versus the semi-major axis $a$ are plotted 
in Figs.~\ref{Fig_06} and \ref{Fig_07}, respectively for all six runs. The large black 
plus sign indicates the location of the collision at $a=2.08$ au, $e=0.05$, and $i=5^\circ$
(the values of the other orbital elements were set to zero, c.f. the data for the 
Target in Table \ref{tab_01}). The blue-filled circles denote the protoplanets, 
the smaller red circles are the planetesimals, while the small grey dots are 
the test particles.

\par
In the $(a-e)$ planes, an interesting, checkmark-shaped structure forms in all 
cases for $t=0$. Both the massive and the massless bodies are confined to this shape. 
On the contrary, no well-defined shape appears on the ($a-i$) plane, where the fragments 
are scattered around the impact point. However, as velocity increases, distinct arcs 
become visible, particularly noticeable in the planetesimal distribution. The mechanism 
and theory behind the formation of the checkmark-shaped structure will be addressed in 
another article, it is out of the scope of the present paper. 
We also note that the semi-major axes of the smaller fragments span a wide range, 
e.g. for run1 it is approximately $a \in [1.75, 2.5]$ au, which is a difference of 
$\Delta a \approx 0.75$ au, and about $1000$ times larger than in the 
$\mathrm{BC}xyz$ space. The higher the collision velocity, the wider the range over which the 
debris scatters. This fact plays a crucial role in the frequency of collisions with Earth, 
as discussed in Section~\ref{section_results} and demonstrated in Table~\ref{tab_03}.
For all six runs the initial $e$ falls in the interval of $[0, 0.25]$ while the $i$ is 
in the range of $[0, 20]$ degree, c.f. Figs.~\ref{Fig_06} and \ref{Fig_07}. With an increase 
in $v$, the upper limits of the initial $e$ and $i$ grows.

\par
The simulation parameters and initial orbital elements for the backward integrations
are provided in Table \ref{tab_01}. 
The planets were initially placed in their current orbits.
Now we turn our focus to the determination of the 
fraction of collisions with an inflated Earth. The collision parameters 
$v,\,\theta$ were selected based on the results of \cite{Suli2021}, where the author 
performed 2D $N$-body simulations of planetary accretion. From the compiled statistics, 
the impact parameter $b = \sin\theta$ has a uniform distribution and the minimum of 
the impact speed is 1. Based on preliminary results from 3D planetary accretion simulations, 
the collision angle has a maximum of around 40 degrees, and the collision velocity reaches 
its maximum between 3 and 10. In light of these results, we selected the collision angles 
of 30 and 45 degrees, as well as the collision velocities of 3, 5, and 7, see Table 
\ref{tab_01} for details.

\par
At the beginning, both the target and the projectile are initially in the $\nu_6$ 
resonance, raising the question of why they have not been ejected from there, given the 
relatively short timescale required for a significant increase in eccentricity 
\citep{Morbidelli1994,Smallwood2018}. On the other hand, \cite{Yoshikawa1987} and \cite{Knezevic1991},
have shown that the $\nu_6$ resonance is an efficient collector 
capable of capturing objects from the region between 2.4 and 2.7 au with moderate inclination
($15^\circ$ to $20^\circ)$. Therefore, we may  
assume that the target and the projectile were formed further away, and when they reached 
the size of Ceres, they entered the $\nu_6$ resonance and eventually collided.

\section{Results}
\label{section_results}

We numerically integrate the trajectory and simulate the outcomes of fragments. Each 
simulation incorporates the presence of the three terrestrial planets Venus, Earth and Mars
and two giants Jupiter and Saturn. The initial distribution of the eccentricities
and inclinations are shown in the first and third rows of Figs.~\ref{Fig_06} and 
\ref{Fig_07}, respectively. The large 
black plus marks the impact location, blue circles indicate protoplanets, red ones 
represent planetesimals, and grey dots correspond to test particles. 
In the ($a-e$) plane, a distinct checkmark shape is clearly visible in the initial distribution 
of the generated debris. Additionally, it can be observed that collisions at higher velocities 
scatter the debris over a wider range along the $e$ axis, widening the arms of the checkmark. 
Increasing the collision angle from 30 to 45 degrees roughly halves the number of fragments. 
It's also worth noting that the distribution of planetesimals concentrates on the right arm 
of the checkmark, and as the velocity ($v$) increases, it clusters towards the lower and
upper boundaries of the checkmark shape. In general, after each collision, two protoplanets 
remain, with the exception of run5, where collision parameters are such that both the target
and the projectile completely disintegrate into smaller fragments, leaving no protoplanets.

\par
Fig.~\ref{Fig_07} similarly illustrates the initial and final states of collision 
remnants but in the ($a-i$) plane. With increasing velocity, arcs become apparent 
in the distribution of planetesimals.

\par
At the end of the simulation ($t=10^6$ years), we also depicted the distribution of 
debris in both the ($a-e$) plane, (Fig.~\ref{Fig_06}) and the ($a-i$) plane (Fig.~\ref{Fig_07}). 
For better comparison, the displayed range is the same as the one used in the initial 
state. Consequently, some of the debris is not visible in the figures, even if it 
did not collide. The most prominent feature in the final state depicted in the 
figures is the depletion of the region around the $\nu_6$ secular resonance, 
between $\approx 2.05$ and $\approx 2.15$ au. This phenomenon is most pronounced in 
the ($a-e$) plane, as it is more sensitive to secular perturbations (see, for example, 
Figure 11 in \cite{Forgacs2022}) than in the ($a-i$) plane. Generally, dynamical maps 
are usually constructed using the maximum eccentricity method. In the ($a-i$) plane, 
the $\nu_6$ secular frequency is also beginning to manifest itself and is expected 
to become more pronounced with longer integration times.

\begin{figure*}
\includegraphics[width=0.27\paperwidth,keepaspectratio=true]{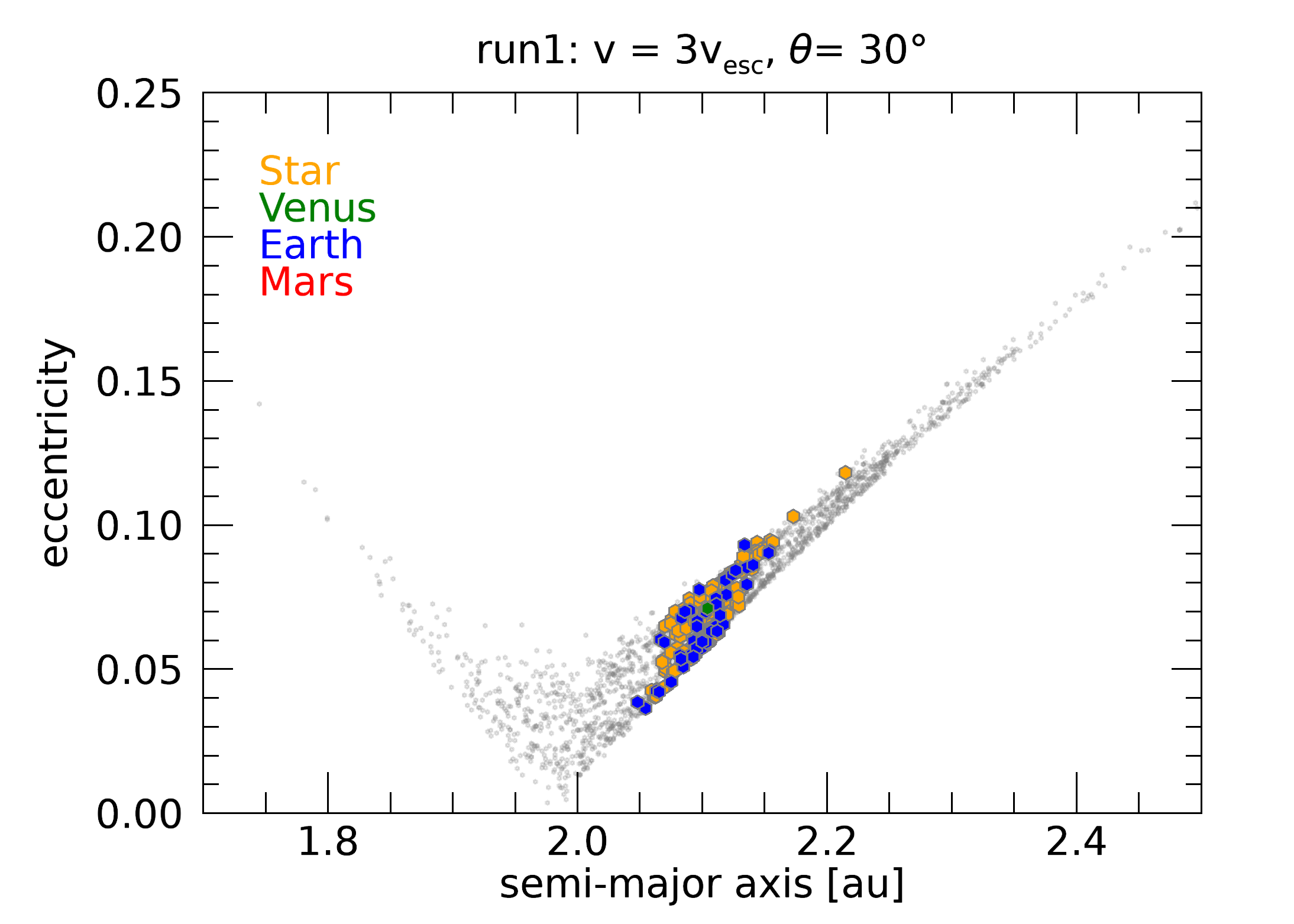}
\includegraphics[width=0.27\paperwidth,keepaspectratio=true]{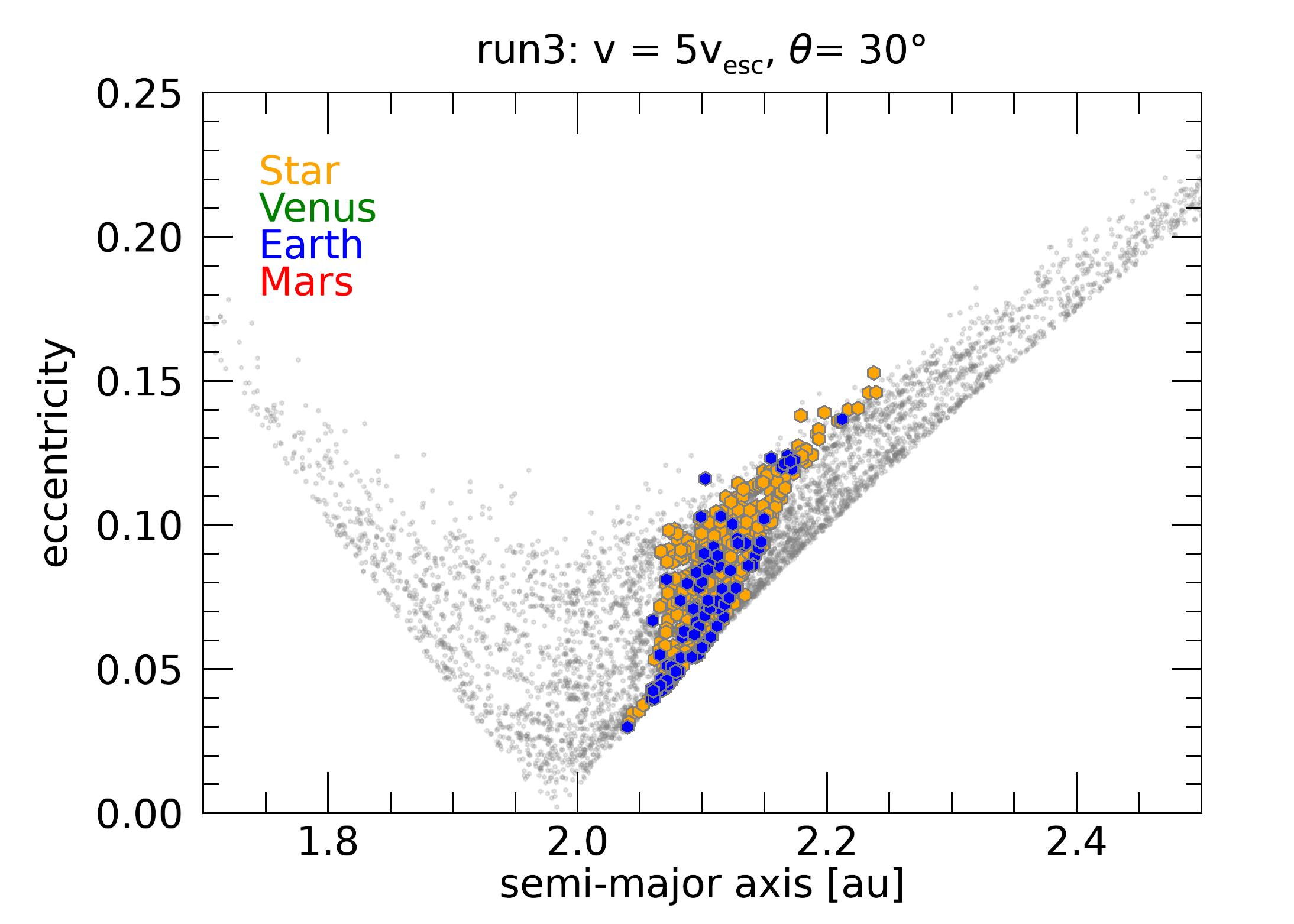}
\includegraphics[width=0.27\paperwidth,keepaspectratio=true]{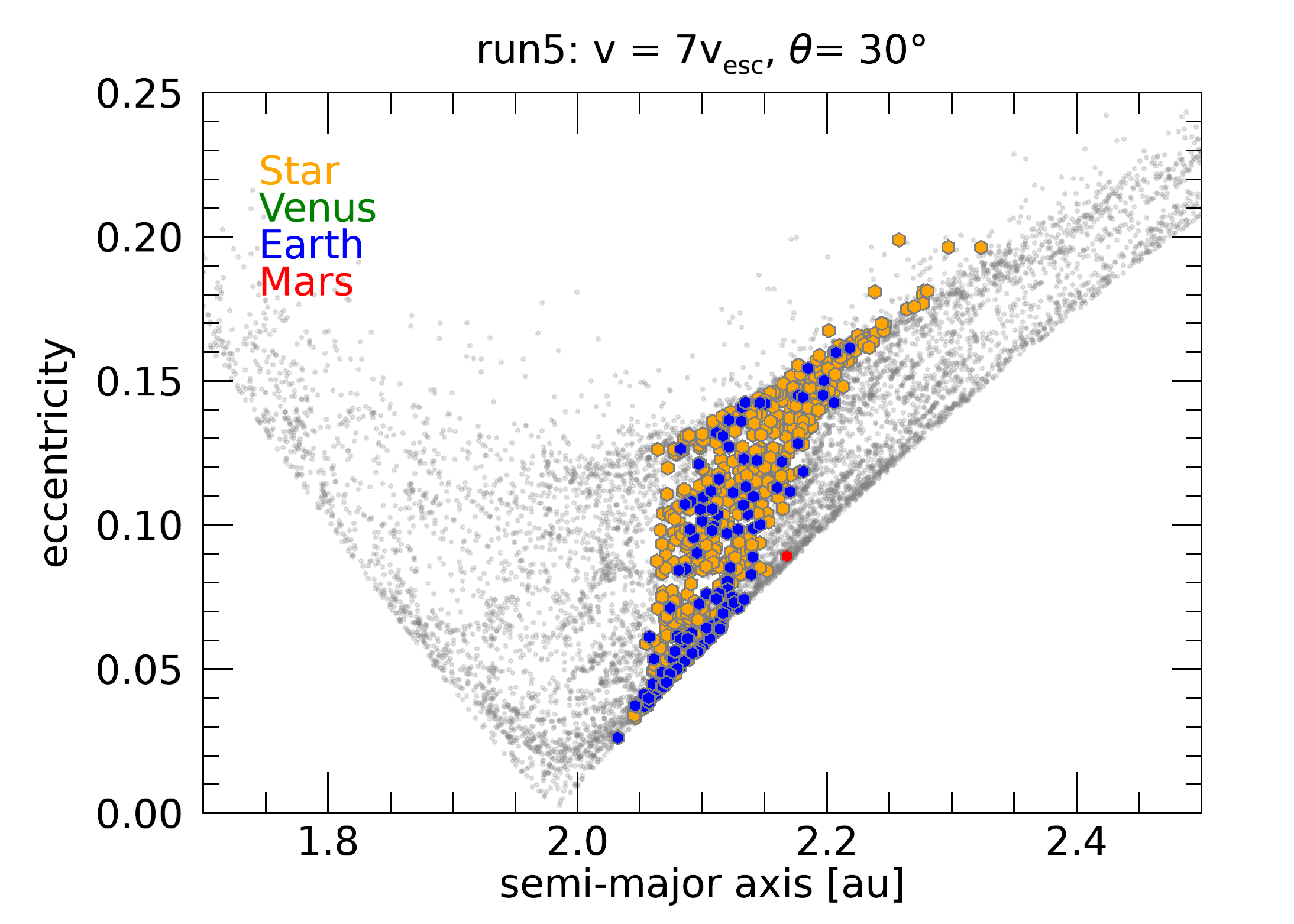} \\
\includegraphics[width=0.27\paperwidth,keepaspectratio=true]{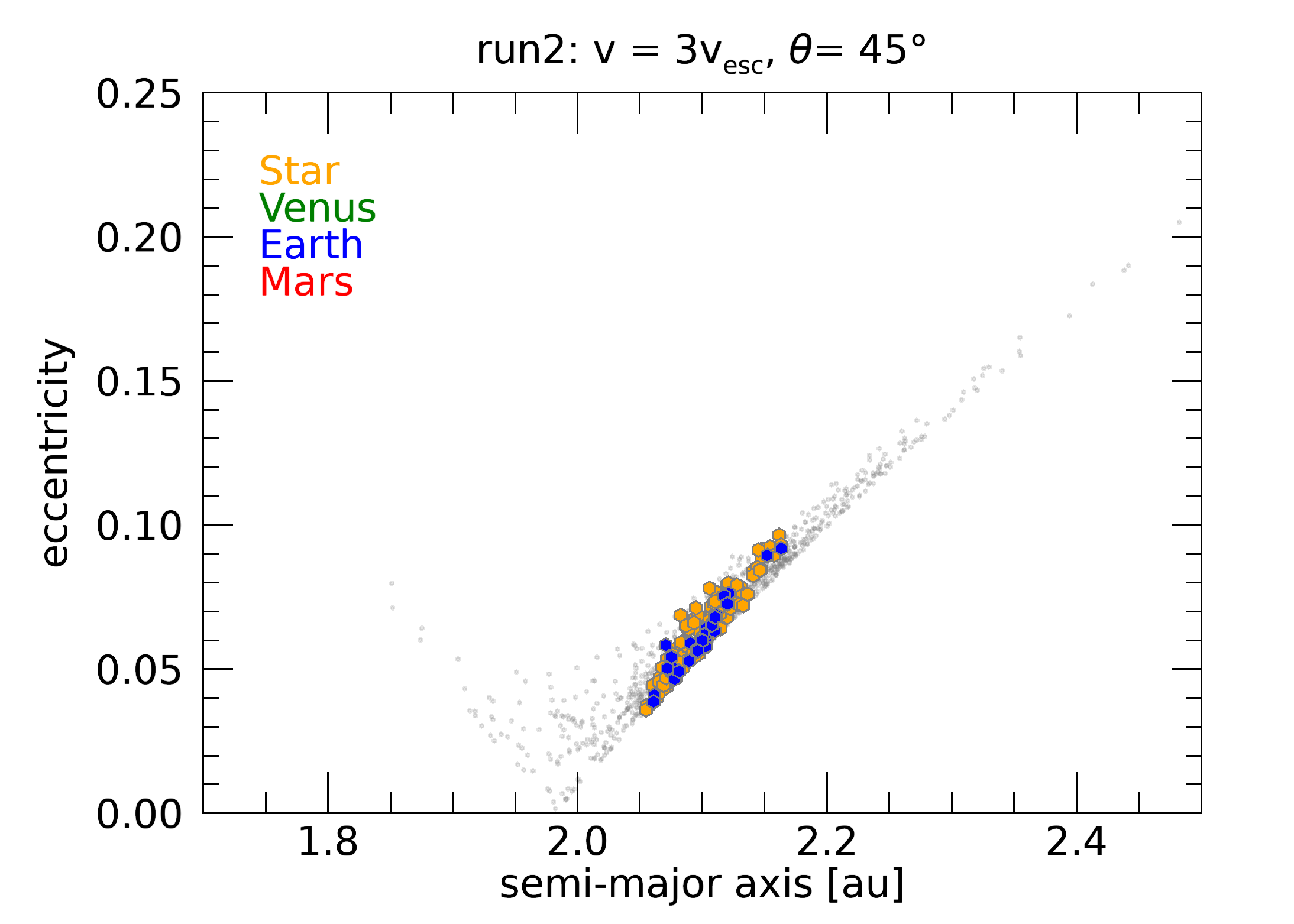}
\includegraphics[width=0.27\paperwidth,keepaspectratio=true]{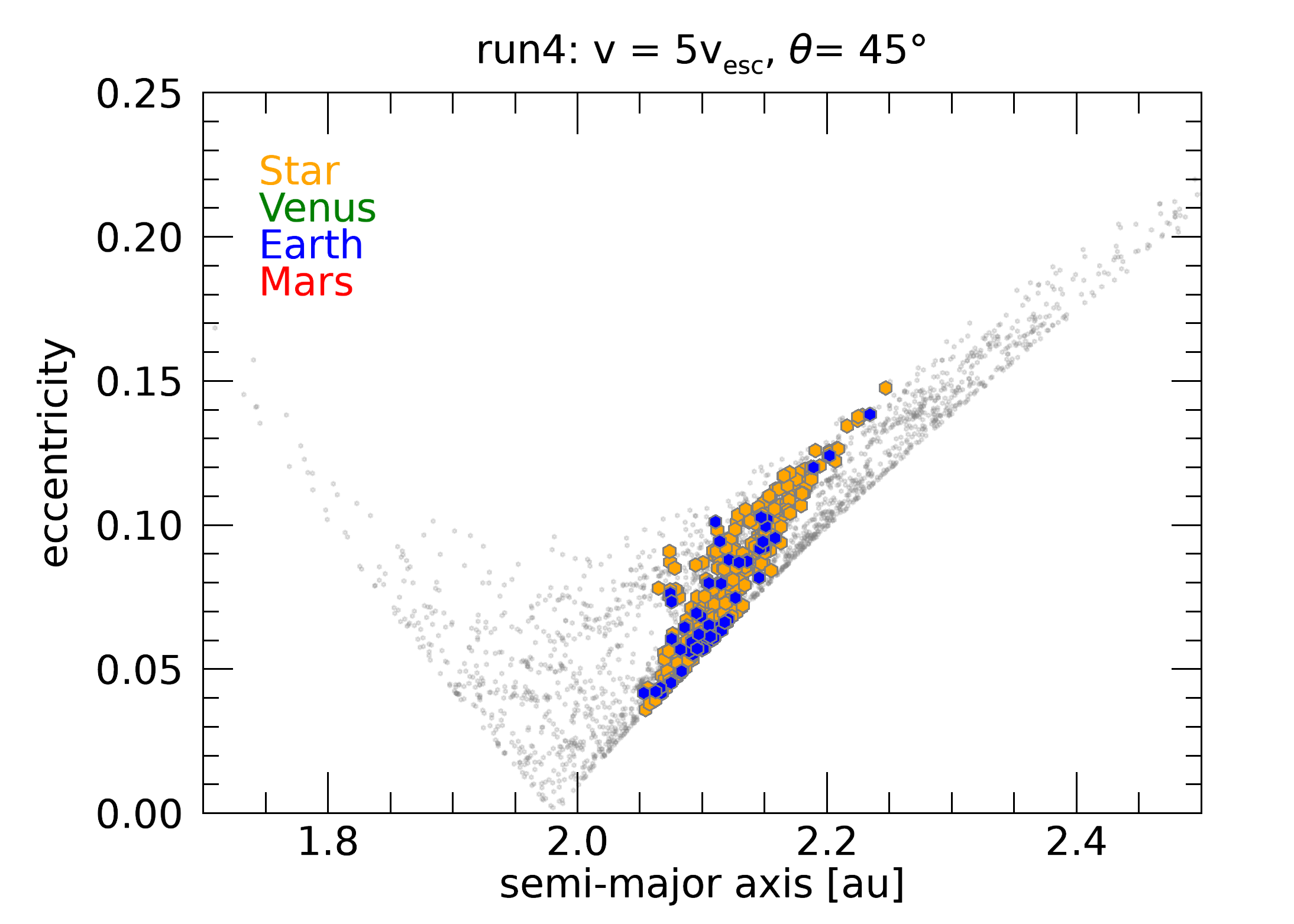}
\includegraphics[width=0.27\paperwidth,keepaspectratio=true]{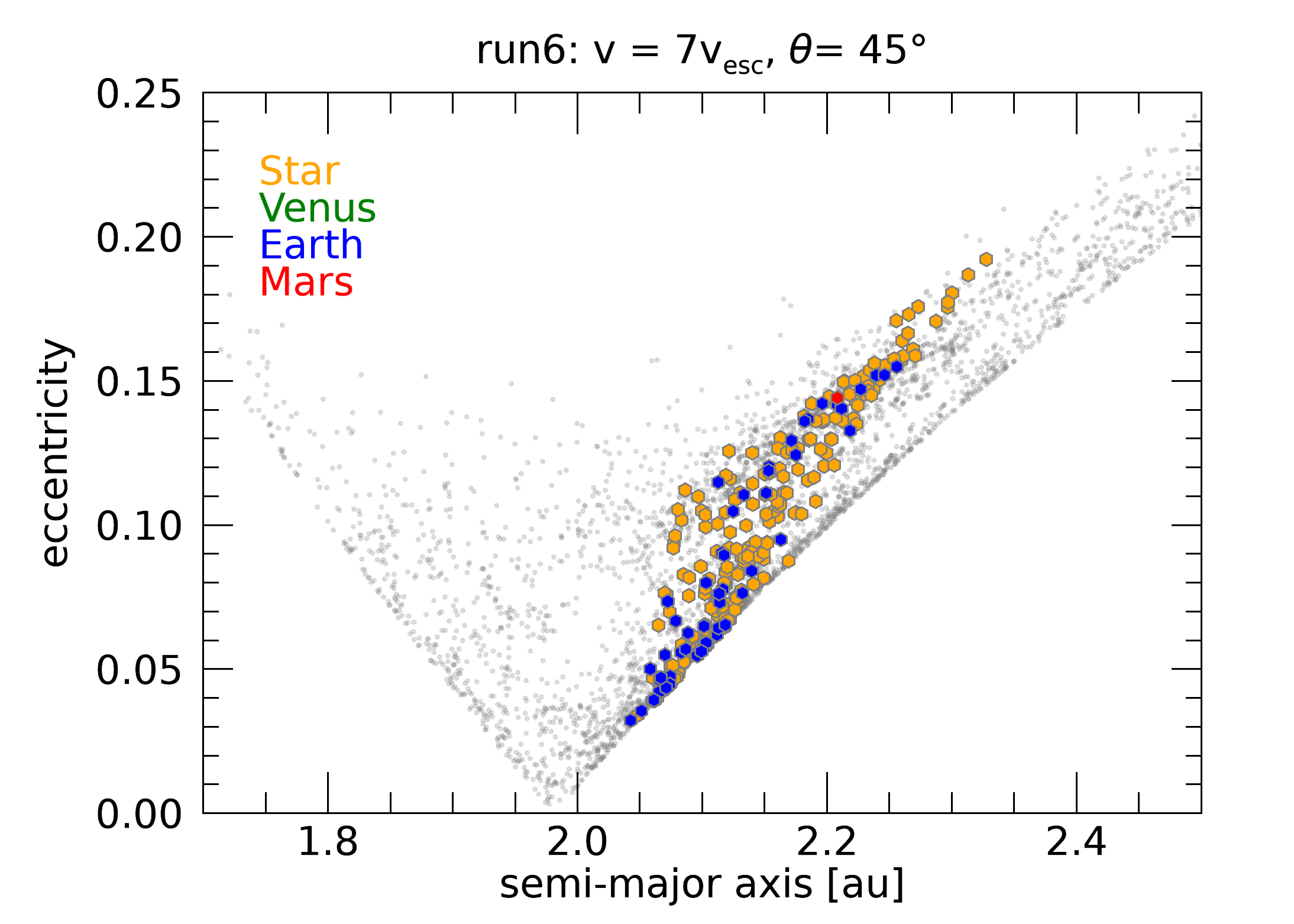} 
\caption{
The positions of the fragments in the ($a-e$) plane that collided with the Sun or one of 
the rocky planets for different impact velocities and angles (see Table \ref{tab_02}). The upper 
panels depict simulations with $\theta = 30^\circ$ while the lower panels for $\theta = 45^\circ$. 
The colors indicate the fragments 
colliding with the different celestial bodies, while the initial position of all bodies involved 
in the $N$-body simulation is indicated in grey in the background.}
\label{Fig_08}
\end{figure*}

\begin{figure*}
\includegraphics[width=0.27\paperwidth,keepaspectratio=true]{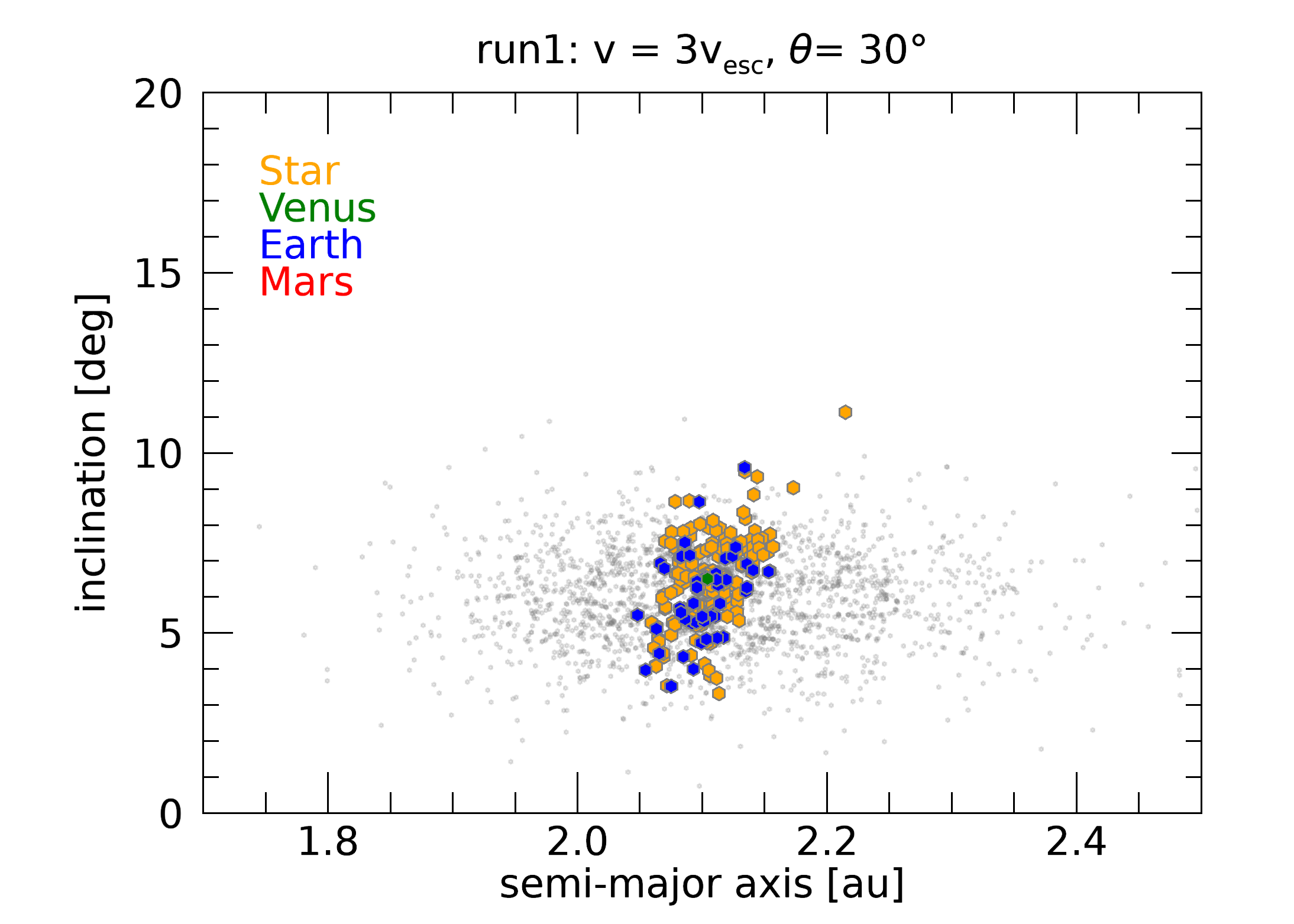}
\includegraphics[width=0.27\paperwidth,keepaspectratio=true]{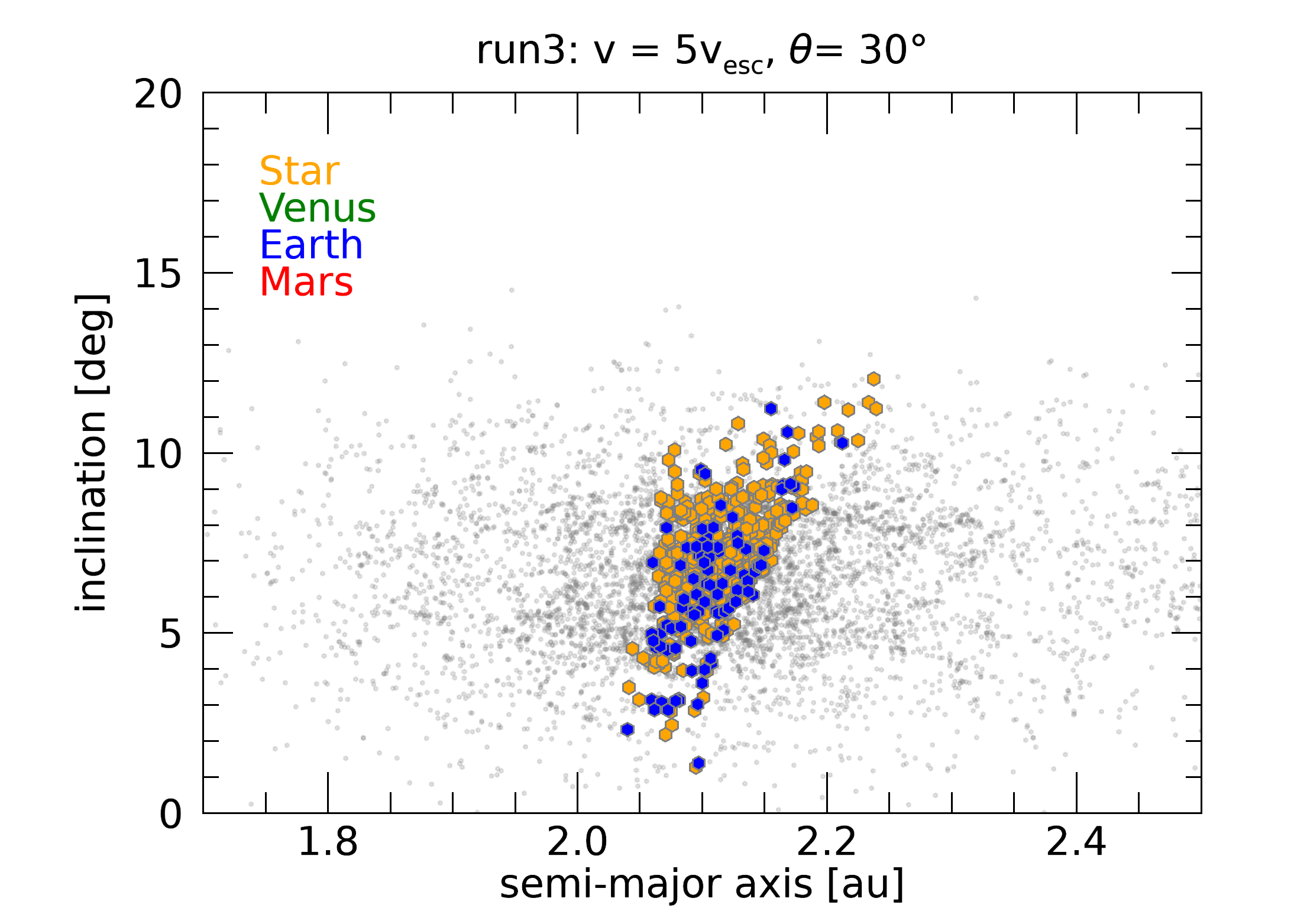}
\includegraphics[width=0.27\paperwidth,keepaspectratio=true]{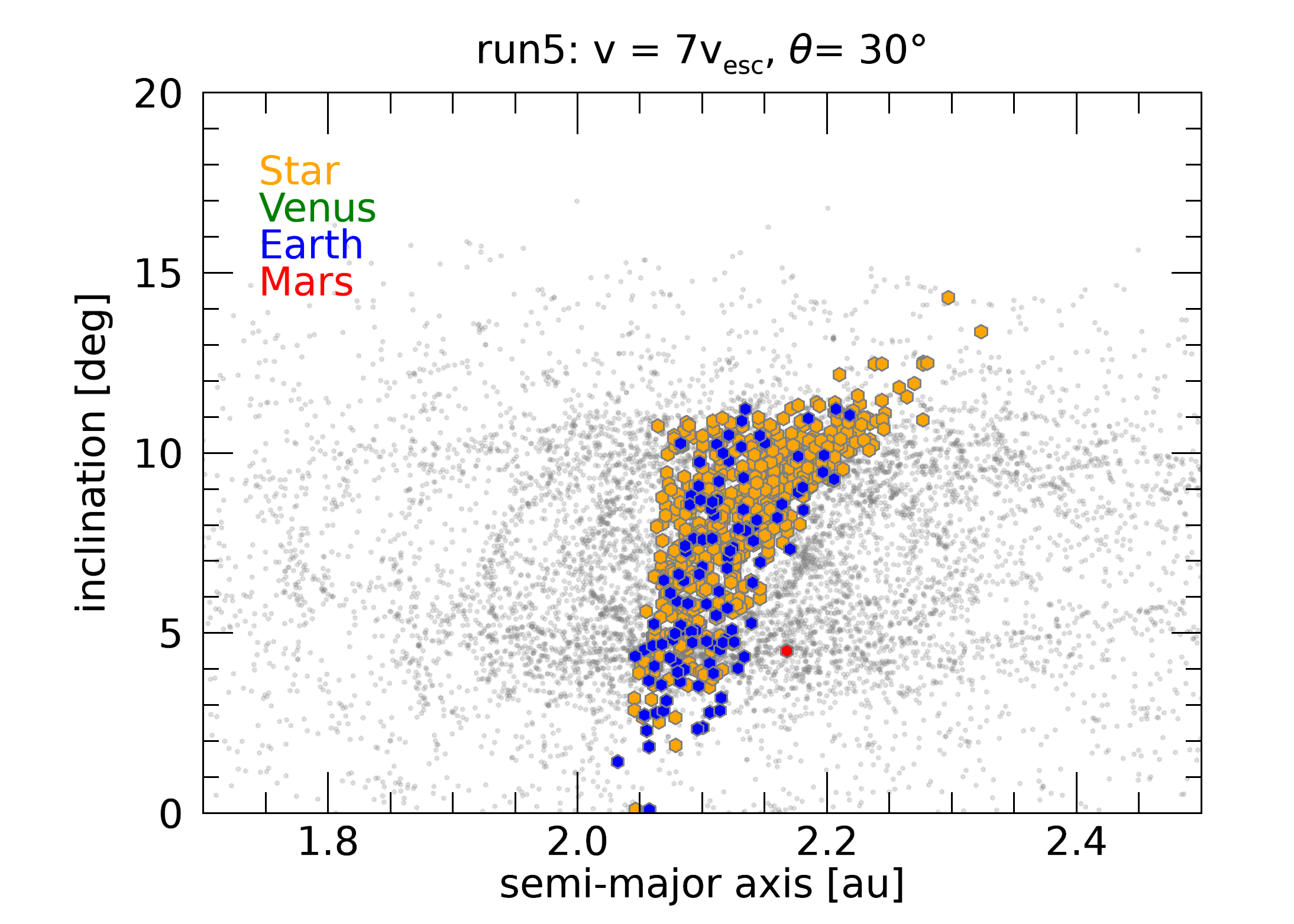} \\
\includegraphics[width=0.27\paperwidth,keepaspectratio=true]{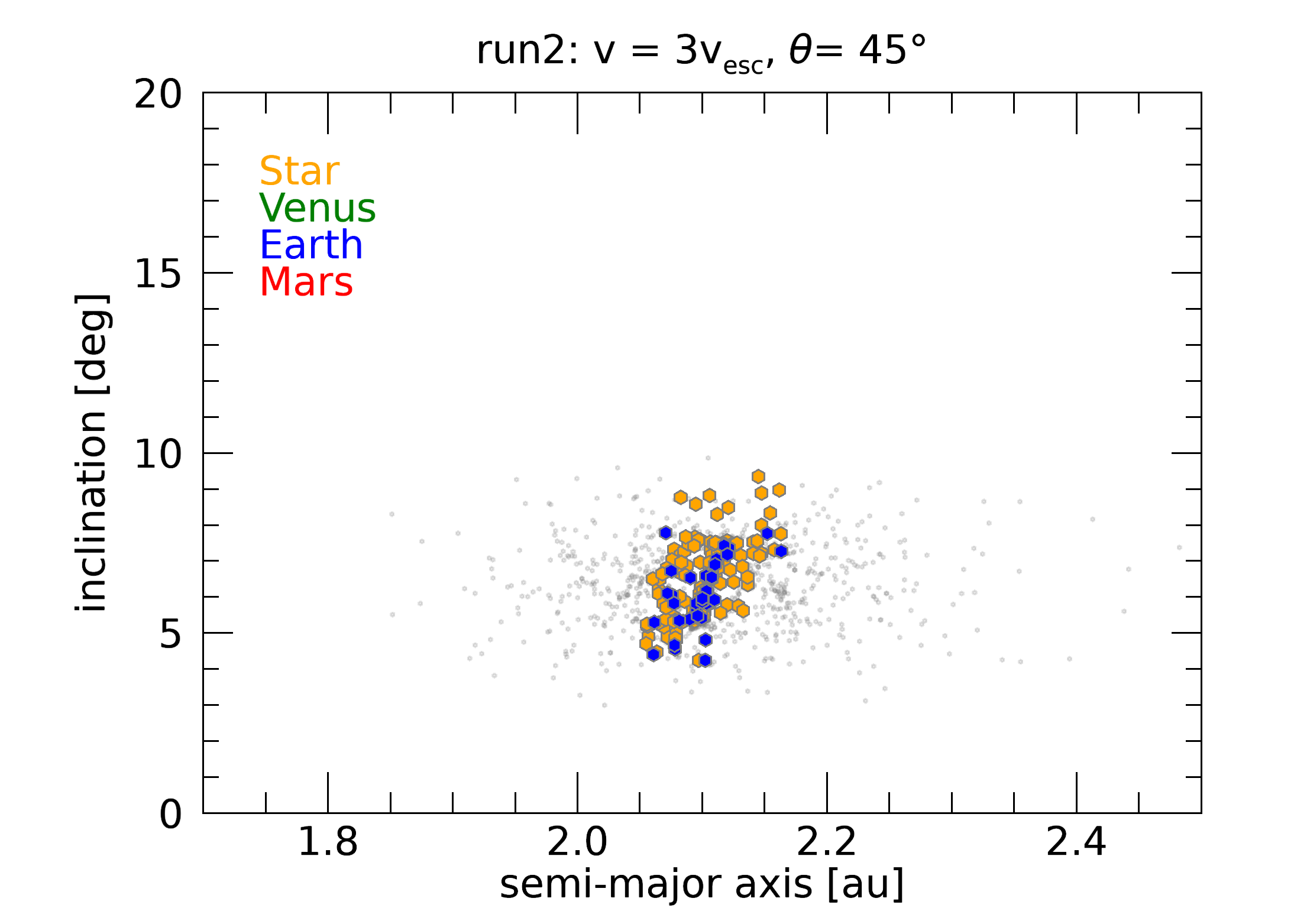}
\includegraphics[width=0.27\paperwidth,keepaspectratio=true]{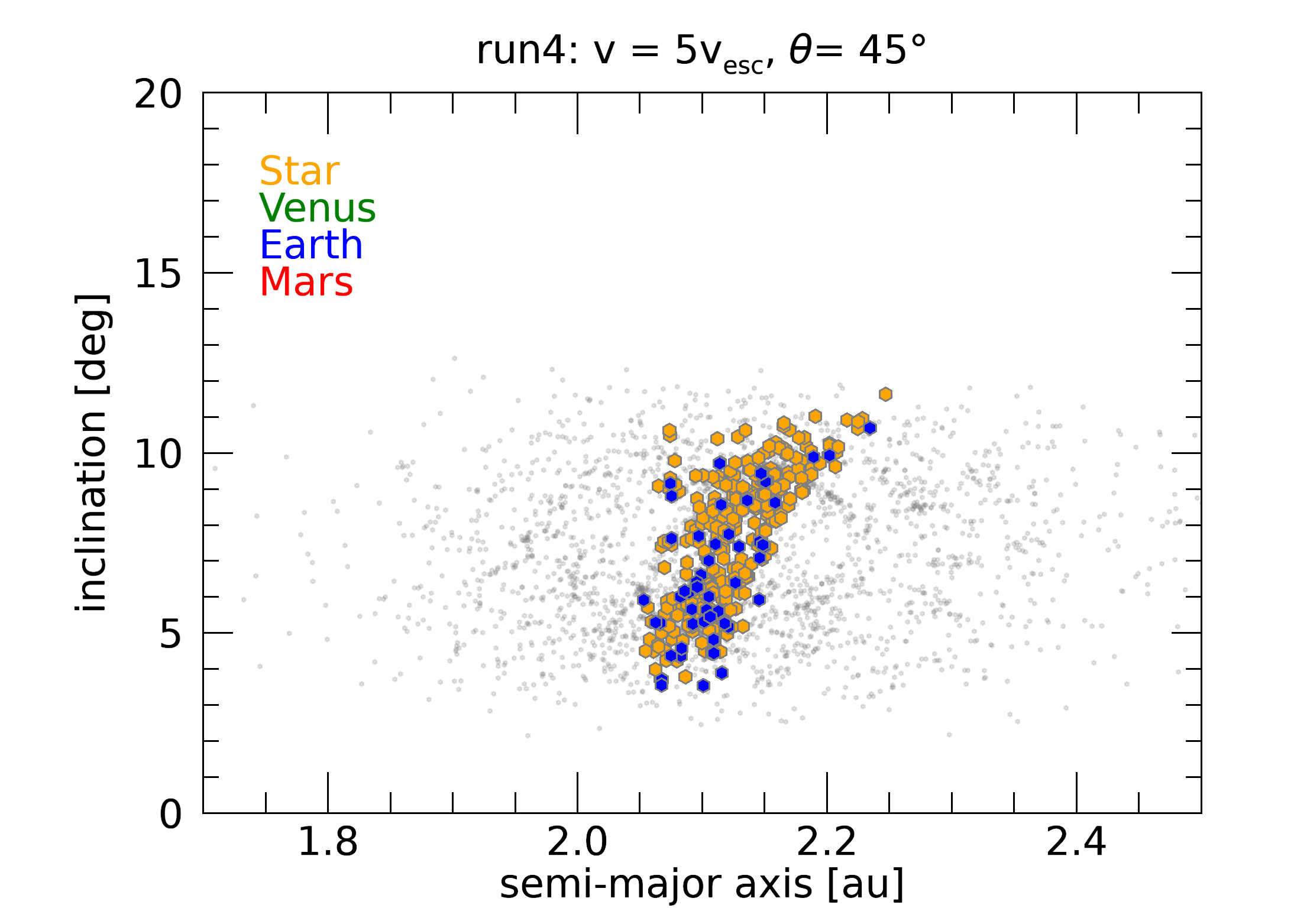}
\includegraphics[width=0.27\paperwidth,keepaspectratio=true]{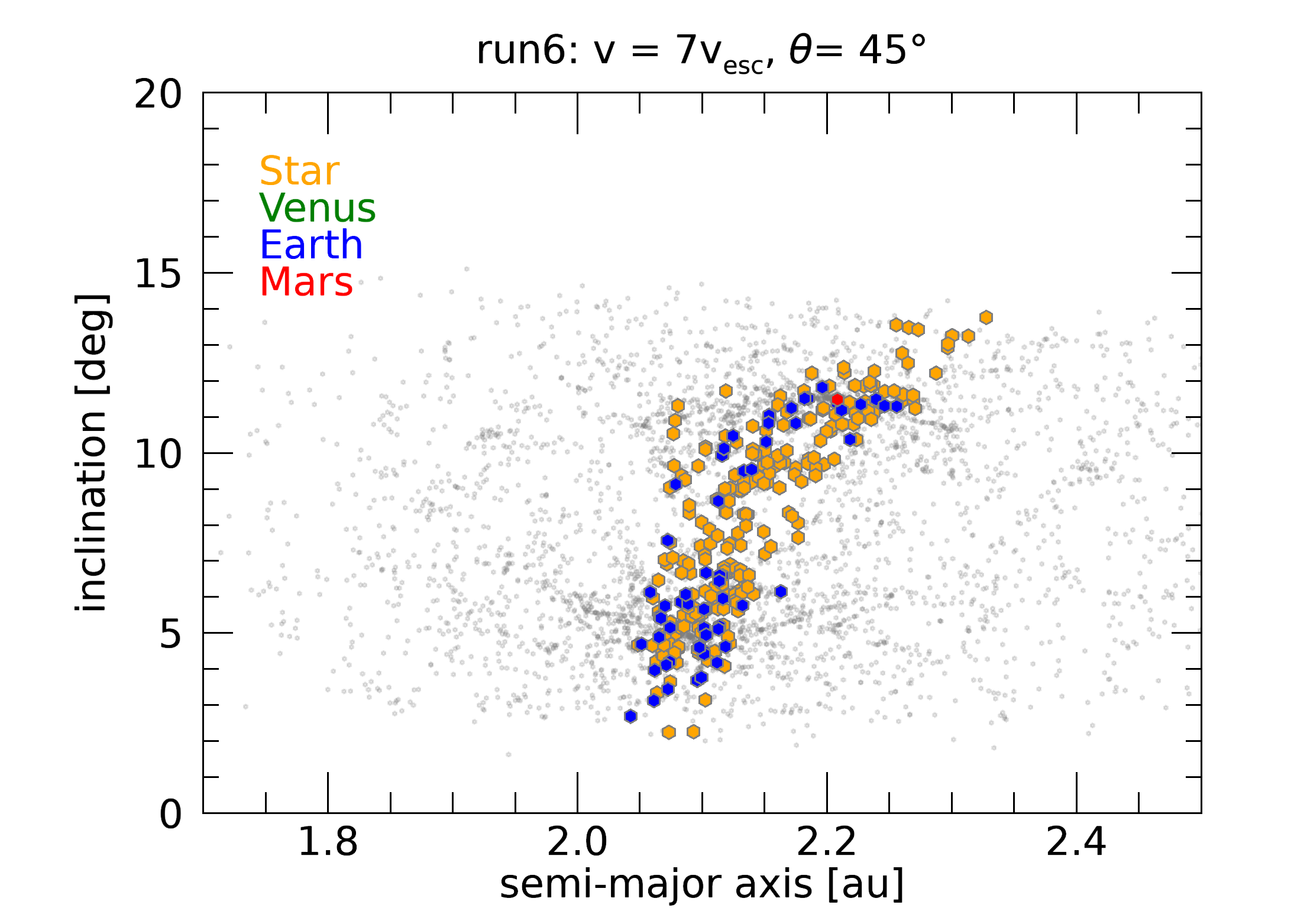} 
\caption{Same as in Fig.~\ref{Fig_08}, but in the ($a-i$) plane.}
\label{Fig_09}
\end{figure*}

\par
In Figs.~\ref{Fig_08} and \ref{Fig_09} you can see the initial positions of celestial bodies 
that collided with the Sun or one of the rocky planets. These figures clearly illustrate 
the extent and shape of the $\nu_6$ secular resonance. The $\nu_6$ resonance becomes 
more prominent as the initial fragment distribution covers a larger area in the ($a-e$) plane.
This inspires us to conduct further research in which the entire 
area will be mapped with test particles to precisely determine the shape and extent 
of the $\nu_6$ resonance in both the ($a-e$) and ($a-i$) planes.
In Fig.~\ref{Fig_09} the colored region delineated by colliding bodies extends further
to larger $a$-values as inclination increases. This observation aligns with 
Figure 1a in the study by \cite{Froeschle1986}, although in our figure, the vertical 
axis represents the osculating inclination while in \cite{Froeschle1986}'s figure, it 
represents the proper inclination.

According to the results presented in ML, there were no collisions observed when considering 
the actual radius of the Earth ($R_\oplus = 6371$ km). Therefore, based on their results, 
we used a 10 times larger radius for the Earth $R_\mathrm{Earth} = 10R_\oplus$ in the 
simulations to ensure that collisions occur. The radii of the other planets were taken 
as their actual values. We considered a particle to hit the Sun if its astrocentric 
distance became smaller than 0.3 au.

\par
We will use the estimation employed by ML to determine the number of objects colliding 
with a real-sized Earth based on the number of objects colliding with an inflated Earth 
of size $10R_\oplus$. The estimation resulted in a factor $f \approx 20$ for 
the position of the $\nu_6$ secular resonance.

\par
According to Table \ref{tab_02} it can be observed that at $\theta = 30^\circ$, approximately 
twice as many fragments are produced compared to the case of $\theta = 45^\circ$ collision
for all three impact velocities. A significant portion of the debris collides with 
larger fragments, primarily with the remnants of 
the collided bodies, which are the two largest objects in the debris cloud. The sum of these 
collisions can be found in the table under the $N_\mathrm{TP}$ header 
(the subscript TP stands for target and projectile). This process, known as 
reaccretion, returns a substantial fraction of the mass back to the target and projectile, the 
specific mass can be found in Table \ref{tab_05}. The reaccretion occurs within a few years 
following the collision, and after that, debris rarely impacts on the remnants of the impacted 
bodies. However, in the following years collisions still occur between smaller fragments and 
test particles, but the rate of these collisions decline with time.

\par
Overall, based on the table, it can be concluded that a smaller impact angle, representing a 
more frontal collision, generates more debris than a larger impact angle, which corresponds 
to a more tangential collision. Of course, for both impact angles, higher velocities further 
increase the amount of debris, and they scatter more widely in both the ($a-e$) and ($a-i$) 
planes. The wider the range in which the fragments are scattered along the semi-major axis, 
the proportionally fewer of them end up in the $\nu_6$ resonance. Therefore, in the case of 
higher-velocity collisions, fewer fragments enter the $\nu_6$ resonance region.

\par
The outcomes for $v=3$, $\theta=30^\circ$ and $\theta=45^\circ$ are shown in the left panels of 
Fig.~\ref{Fig_11} and the data are listed in the first two rows of Table \ref{tab_02}. 
It is clear that the majority of collisions occur with the Sun and with Earth. 
The number of collisions with the Sun significantly exceeds the number of impacts with Earth.
Within the simulation time span of 1 Myr, there were a total of 47 impacts on the Earth, 
for $\theta=30^\circ$, while 33 impacts on Earth for $\theta=45^\circ$. In the case of $v=5$ 
there were 95 and 52 Earth impacts for $30^\circ$ and $45^\circ$ impact angles, respectively.
For the case of $v=7$, the measured values are 116 and 58, respectively. 
For the Sun, these numbers are as follows: in the case of run1 and 2, 178 and 181; 
in the case of run3 and 4, 427 and 307; while in the last two runs, 600 and 245.
The observed trend is that 
with more oblique impacts, a larger number of fragments are produced and as a result of this the 
number of impacts also increases. The number of impacts into the giant planets are in 
close agreement with the results of ML and are in the order of unity.

\par
We have calculated two ratios using two different benchmarks to measure the chance of 
collision with Earth. In the first case $P_{10} = N_\mathrm{E}/(N_\mathrm{pp} + N_\mathrm{pl} + N_\mathrm{tp})$, 
while in the second one $P_{10}^\star = N_\mathrm{E}/N_\mathrm{event}$ In order to directly 
compare the results with those of ML, the latter definition matches what was used in the ML paper. 
Both $P_{10}$ and $P_{10}^\star$ characterize the likelihood of objects impacting the 
inflated Earth, thus these ratios can be considered as a loosely defined probability 
of impact with Earth. The calculated values of $P_{10}$ and $P_{10}^\star$ are listed in 
Table \ref{tab_03}. The measured $P_{10}$ decreases with increasing $v$, the maximum is 
$\approx 0.0258$ ($v = 3,\,\theta = 45^\circ$), the minimum value is $\approx 0.0111$ 
($v = 7,\,\theta = 30^\circ$), thus both the minimum and the maximum fall into the same 
order of magnitude. The decreasing trend is a consequence of the fact that at higher 
velocities, the fragments making up the debris scatter over a larger range along the semi-major 
axis (as mentioned above). Consequently, proportionally fewer fragments enter the $\nu_6$ 
resonance, resulting in fewer fragments on Earth-crossing orbits.

\par
The ratio $P_{10}^\star$ does not follow a similar trend, it fluctuates around a mean value
of 0.0876 with a standard deviation of 0.0112. To adjust for the inflated Earth, these values 
need to be divided by a factor of $f = 20$, as described above. The computed ratios $P$ and 
$P^\star$ are given in the last two columns of Table~\ref{tab_03}. Since $P$ and $P^\star$ 
differs by a factor of $f$ they follow the same pattern as $P_{10}$ and $P_{10}^\star$.

\par
In general, the histograms in Fig.~\ref{Fig_11} are similar. In the time following the first 
collision, approximately equal amounts of debris collide with both the Earth and the Sun. 
However, over a period of a few hundred thousand years, this ratio significantly shifts in 
favour of collisions with the Sun. The earliest event occurs between $10^5$ and $2\cdot 10^5$ 
years, indicating that it takes roughly this amount of time for the first debris fragments 
from the $\nu_6$ resonance to reach the inner Solar System. Interestingly, in the run1 
simulation, the first collision occurs with Venus at $t \approx 1.107 \cdot 10^5$ years, 
followed by a longer period without any event until around $t \approx 4 \cdot 10^5$ years. 
From this point on, debris fragments collide with both the Earth and the Sun.

\par
We note that a few impacts on the other rocky planets were also detected, 1 impact into 
Venus (green bar) for run1 and 2 collisions with Mars both in run5 and run6 (red bar). 
Mars collision occurs only for the $v=7$ cases. 

\begin{figure}
\centering
\includegraphics[width=0.95\columnwidth,keepaspectratio=true]{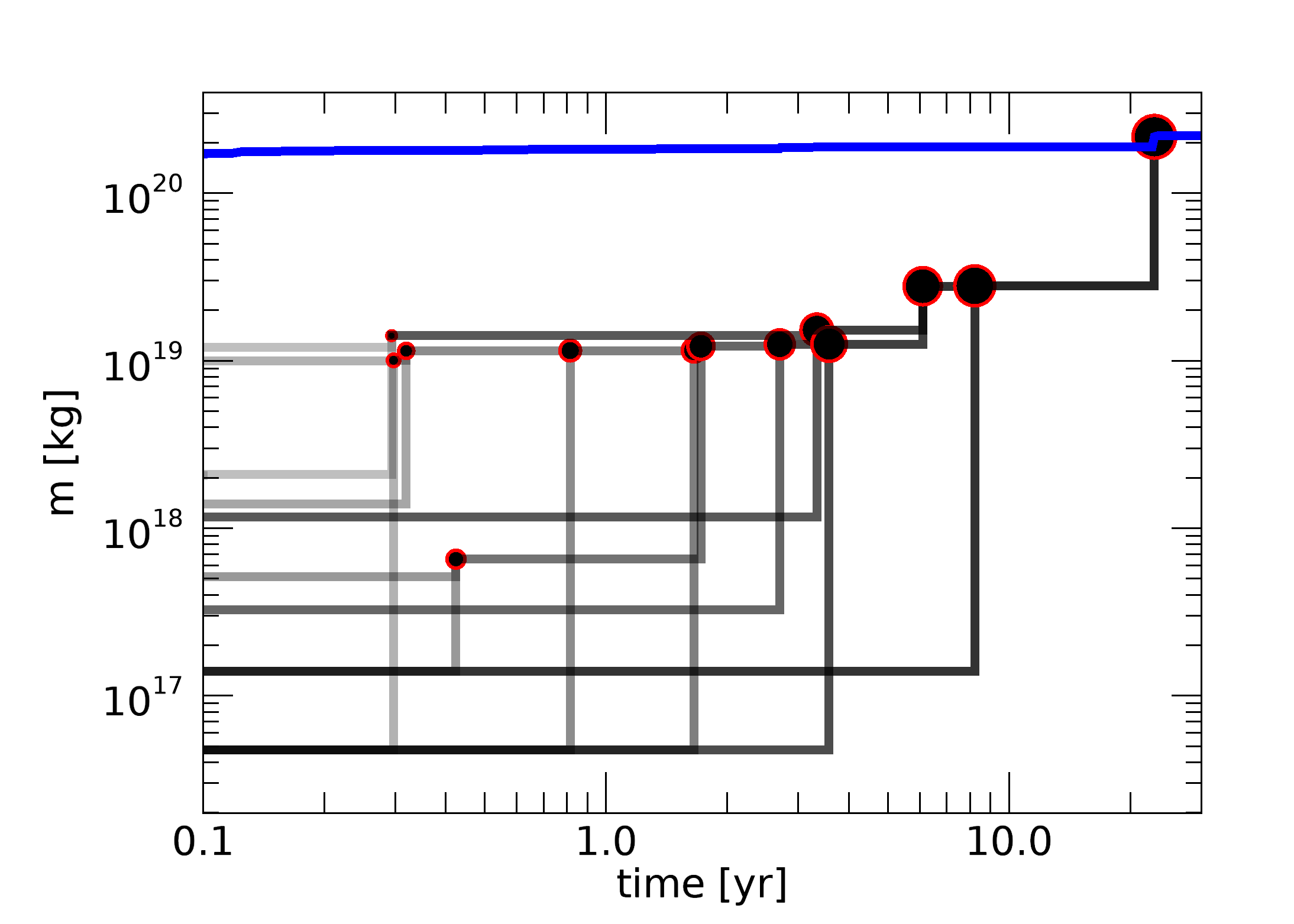}
\caption{An example of a collision tree, showing the mass of the fragments 
(gray-black) and of the remnants of the target (solid blue line) vs. time
in run5. After a collision is simulated with SPH, the orbital evolution of the 
fragments is tracked using our $N$-body integrator and if a collision is detected 
then the two bodies merge. This figure shows only a part of the whole 
collision tree, where the mass evolution of some fragments that 
eventually impact the target are depicted by gray-black lines. 
Every jump in the lines correspond to a collision.}
\label{Fig_10}
\end{figure}

In Fig.~\ref{Fig_10} a part of the collision history of an individual 
protoplanet is visualised using a collision tree. The collision events are
plotted only for $t \le 30$ years. The gray-black lines show the mass of 
fragments as a function of time, while the blue line shows 
the increase in mass of the target body left after the SPH simulation. 
In the figure, for clarity, only one collision tree has been drawn. The 
smaller fragments merge together during the first 10 years, forming a 
larger body with mass $\approx 3\cdot 10^{19}$ kg, which hits the target at 
$t \approx 23$ years. Notice that initially several fragments have the same
mass, thus gray-black lines are overlap in the beginning of the simulation.

\begin{table*} 
\centering
\caption{The simulation outcomes with $R_\mathrm{Earth} = 10 R_\oplus$ for $T=10^6$ years. The first
column contains the names of the simulations. Columns 2 and 3 show the impact velocity $v$ 
and angle $\theta$, respectively. Column 4 represents the total number of bodies (including 
the Sun and 5 planets) $N$. Columns 5,6 and 7 display the number of protoplanets 
$N_\mathrm{pp}$, planetesimals $N_\mathrm{pl}$ and test particles $N_\mathrm{tp}$ respectively. 
Column 8 - 13 contains the number of events (outcomes) $N_\mathrm{event}$, fragments that 
collided with the Sun $N_\mathrm{Sun}$, Venus $N_\mathrm{V}$, Earth $N_\mathrm{E}$, 
Mars $N_\mathrm{M}$ and Jupiter or Saturn $N_\mathrm{JS}$, respectively. Column 13 
lists the number of collisions $N_\mathrm{TP}$ with the remnants of the target and 
projectile.}
\label{tab_02}
\begin{tabular}{ccc|rrrr|rrrrrrr}
\hline
Sim. & $v [v_\mathrm{esc}]$ & $\theta$ [deg]   & $N$ & $N_\mathrm{pp}$ & $N_\mathrm{pl}$ & $N_\mathrm{tp}$      & $N_\mathrm{event}$ & $N_\mathrm{Sun}$ &
$N_\mathrm{V}$  & $N_\mathrm{E}$       & $N_\mathrm{M}$   & $N_\mathrm{JS}$ & $N_\mathrm{TP}$ \\
\hline
run1 & 3.0 & 30 &  2662 & 2 & 102 & 2552 &  612 & 178  & 1 &  47 & 0 & 0 & 210 \\
run2 & 3.0 & 45 &  1283 & 2 &  34 & 1241 &  383 & 181  & 0 &  33 & 0 & 0 & 101 \\
run3 & 5.0 & 30 &  6847 & 2 &  44 & 3125 & 1067 & 427  & 0 &  95 & 0 & 0 & 208 \\
run4 & 5.0 & 45 &  3176 & 2 & 518 & 6321 &  526 & 307  & 0 &  52 & 0 & 1 & 156 \\
run5 & 7.0 & 30 & 10502 & 0 & 927 & 9569 & 1576 & 600  & 0 & 116 & 2 & 2 & 134 \\
run6 & 7.0 & 45 &  4442 & 2 & 155 & 4279 &  573 & 245  & 0 &  58 & 2 & 0 & 199 \\
\hline
\end{tabular}
\end{table*}

\begin{table} 
\centering
\caption{The simulation results with $R_\mathrm{Earth} = 10 R_\oplus$ for $T=10^6$ years. The first
column contains the names of the simulations. Columns 2 and 3 show the collision probability 
with an inflated Earth $P_{10} = N_\mathrm{E}/(N_\mathrm{pp} + N_\mathrm{pl} + N_\mathrm{tp})$ and 
$P_{10}^\star = N_\mathrm{E}/N_\mathrm{event}$. Columns 4 and 5 list the approximate rate with actual 
size Earth $P = P_{10} / f$ and $P^\star = P_{10}^\star / f$, where $f=20$.}
\label{tab_03}
\begin{tabular}{c|rrrr}
\hline
Sim. & $P_{10}$ & $P_{10}^\star$ & $P$ & $P^\star$ \\
\hline
run1 & 0.0177 & 0.0768 & 0.00088 & 0.00384 \\ 
run2 & 0.0258 & 0.0862 & 0.00129 & 0.00431 \\ 
run3 & 0.0139 & 0.0890 & 0.00069 & 0.00445 \\ 
run4 & 0.0164 & 0.0989 & 0.00082 & 0.00494 \\ 
run5 & 0.0111 & 0.0736 & 0.00055 & 0.00368 \\ 
run6 & 0.0131 & 0.1012 & 0.00065 & 0.00506 \\ 
\hline
\end{tabular}
\end{table}

\begin{figure*}
\includegraphics[width=0.27\paperwidth,keepaspectratio=true]{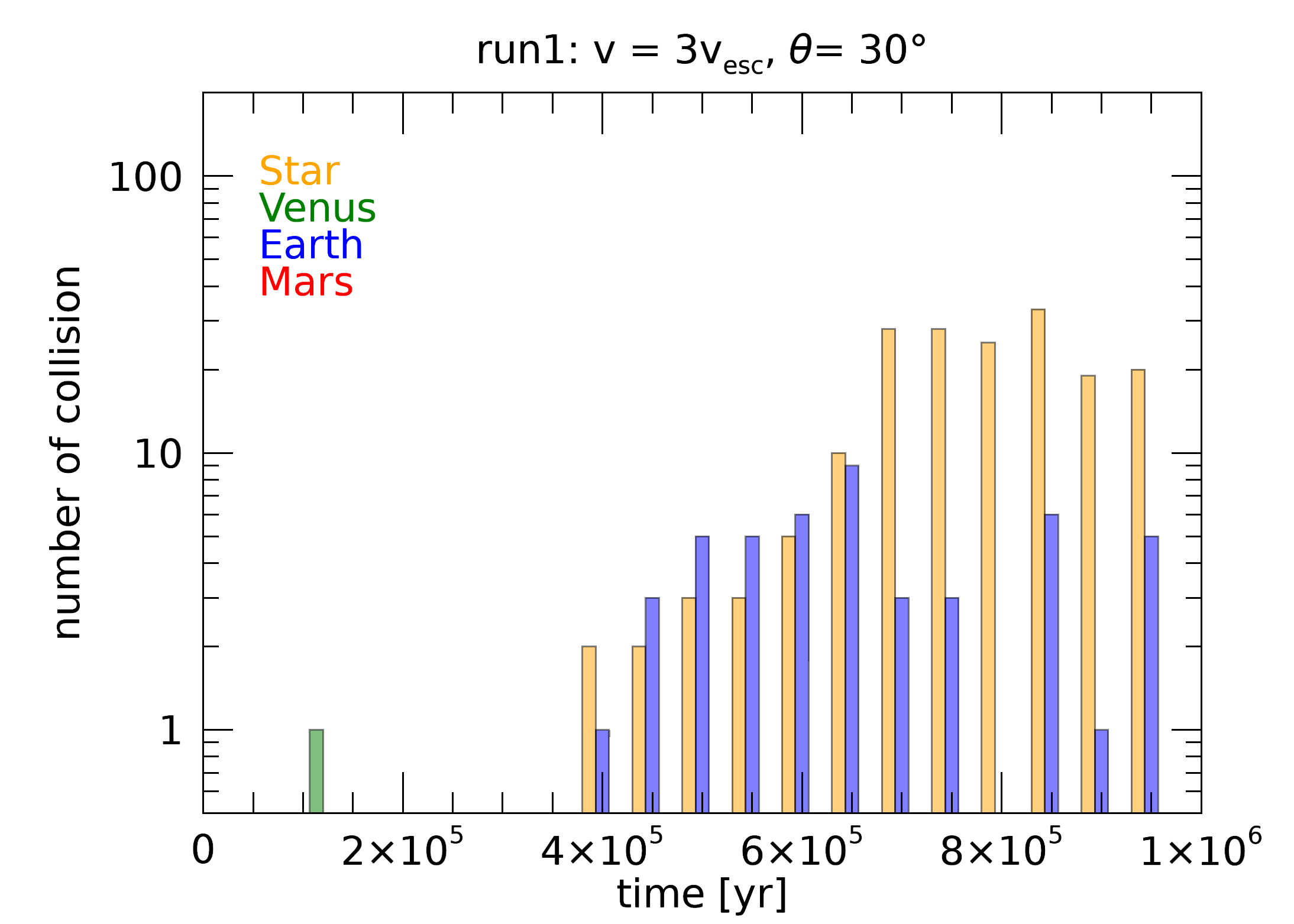}
\includegraphics[width=0.27\paperwidth,keepaspectratio=true]{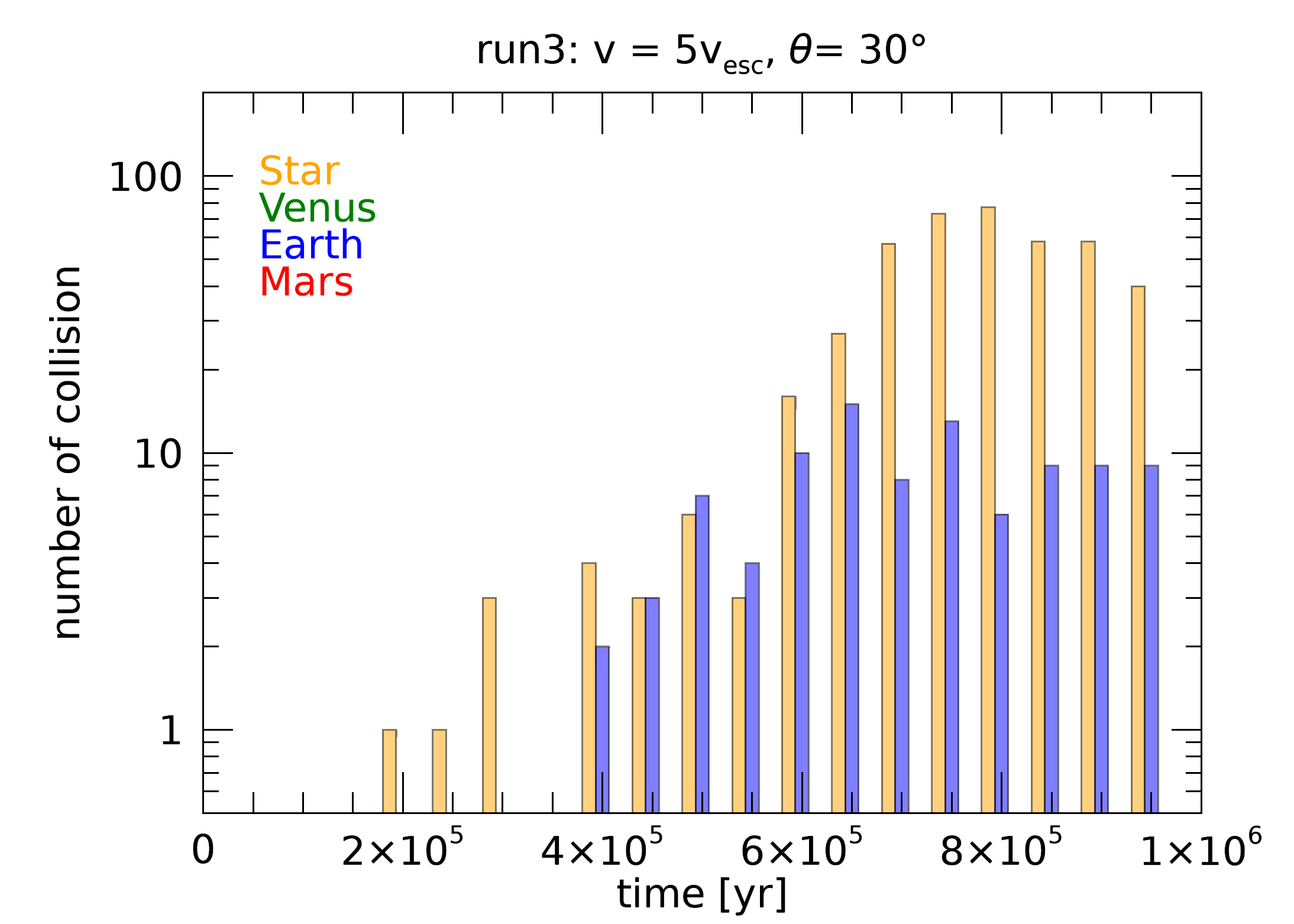}
\includegraphics[width=0.27\paperwidth,keepaspectratio=true]{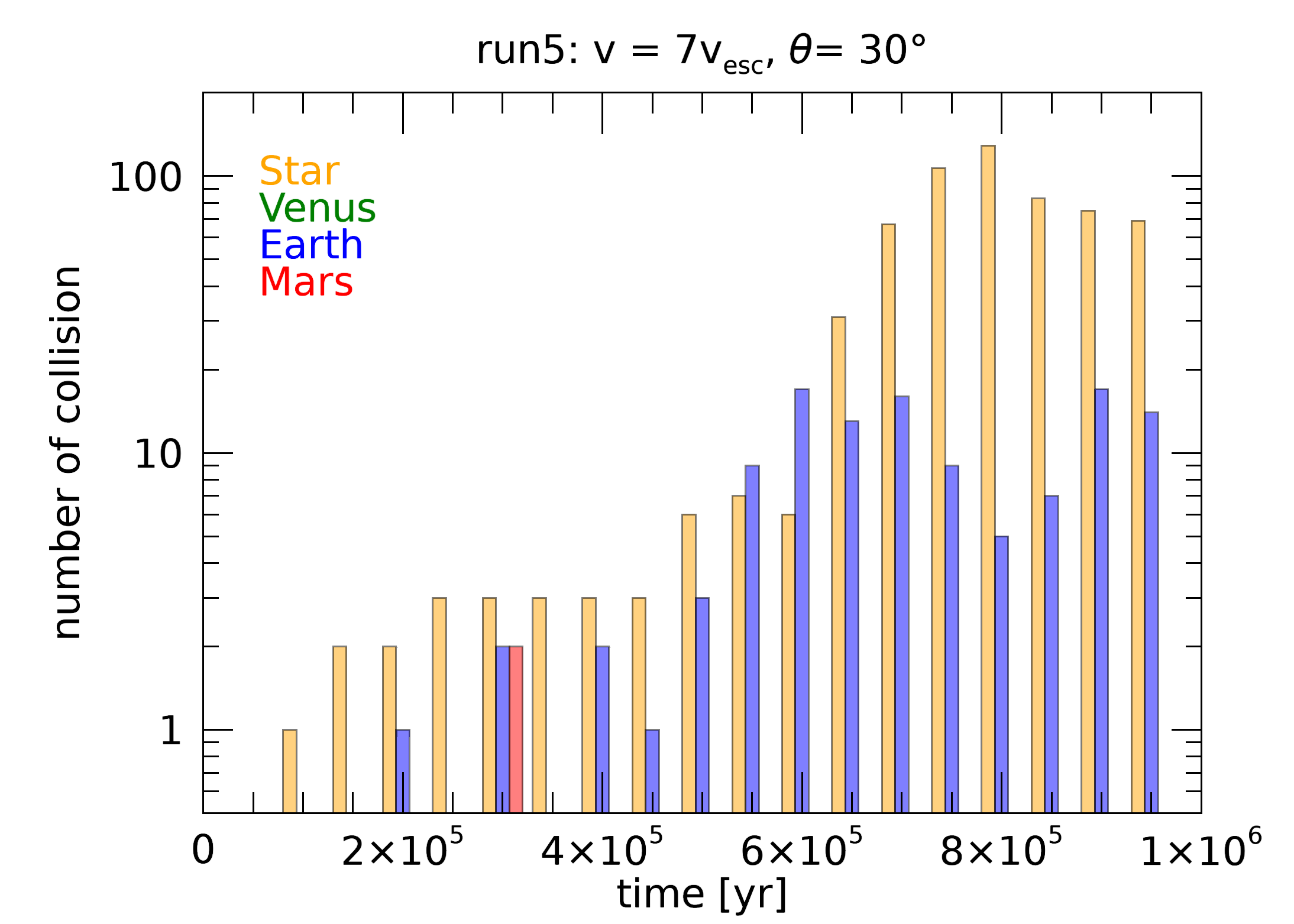} 
\\
\includegraphics[width=0.27\paperwidth,keepaspectratio=true]{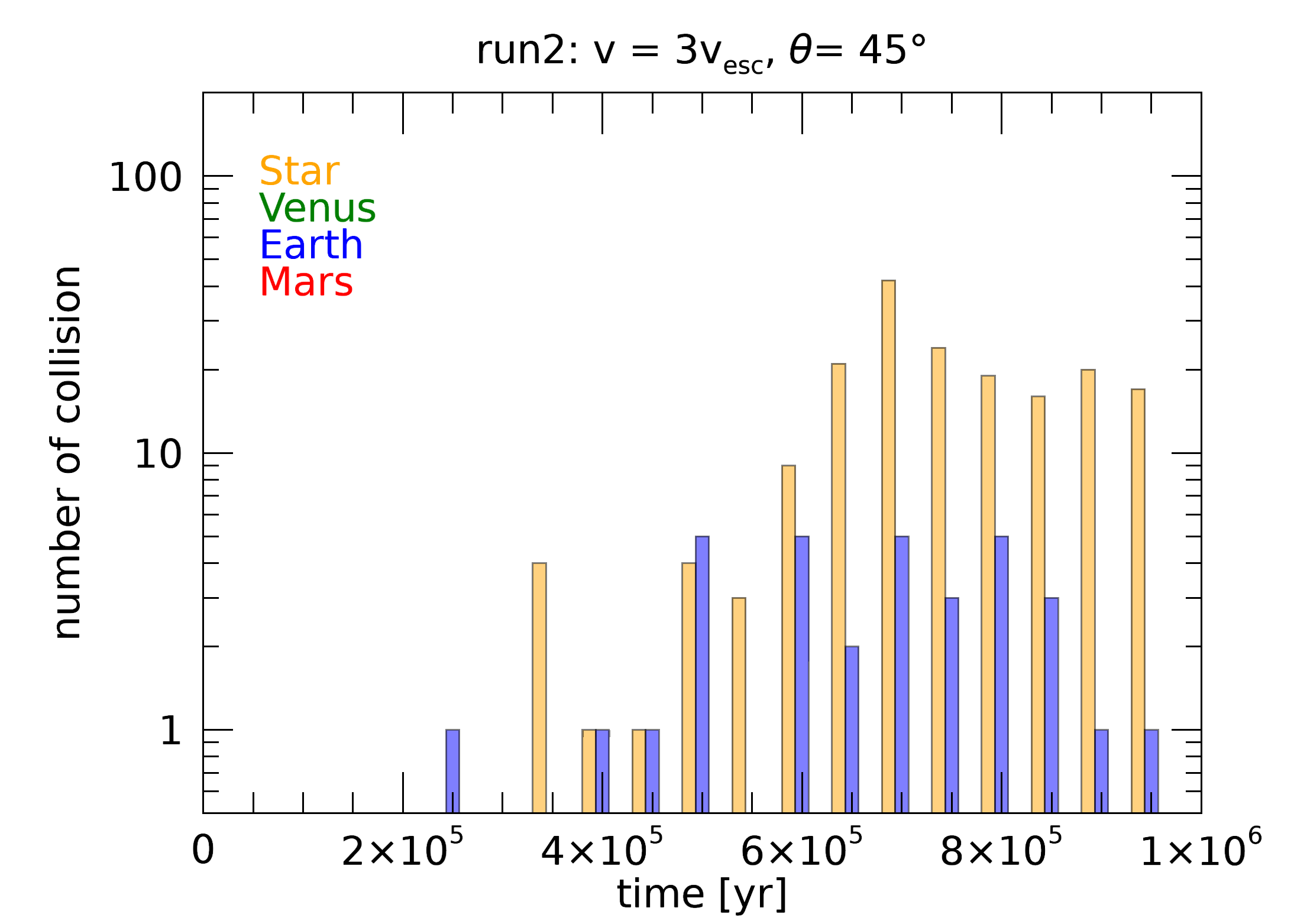}
\includegraphics[width=0.27\paperwidth,keepaspectratio=true]{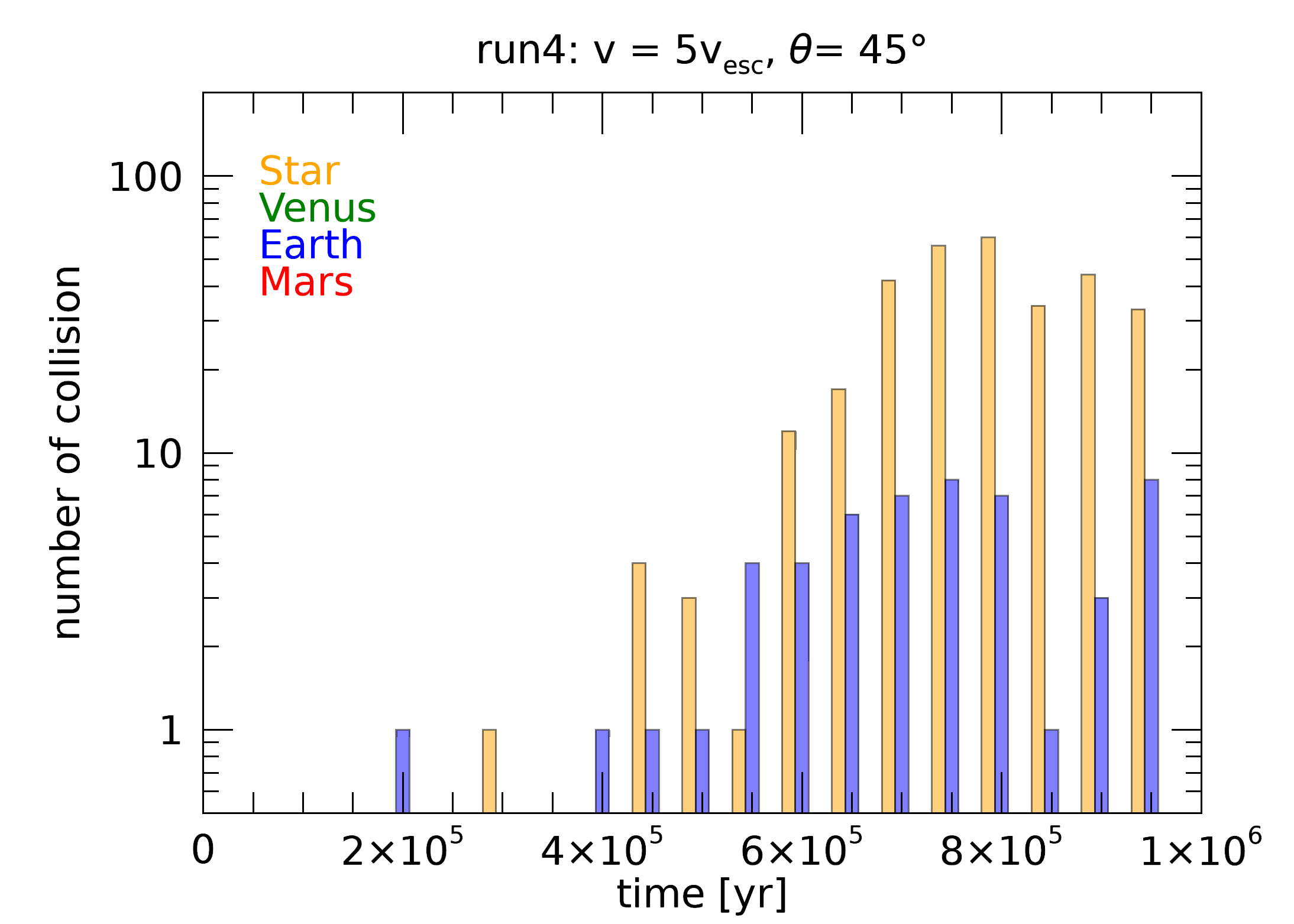}
\includegraphics[width=0.27\paperwidth,keepaspectratio=true]{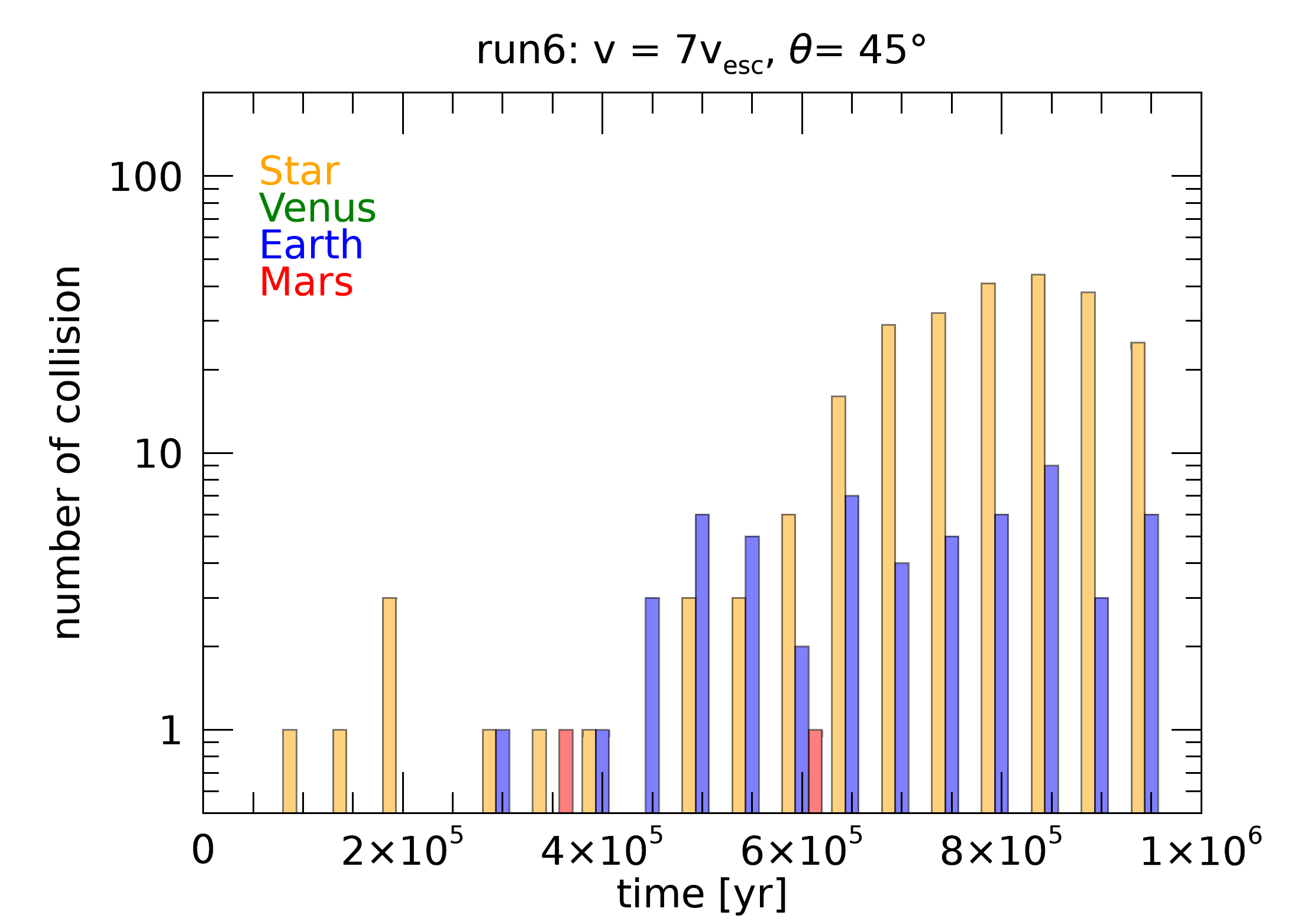} 
\caption{The event history of the $N$-body simulations are binned into 50 kyr time intervals for the 
different impact speed and angle, see Table \ref{tab_02}. The upper panels depict simulations 
with $\theta = 30^\circ$ while the lower panels for $\theta = 45^\circ$. The Earth radius 
is $10R_\oplus$ for all cases, the other planets have their actual radius. In the plots, 
the yellow bars show the number of fragments that hit the Sun, while the green, 
blue and red bars represent asteroids colliding with Venus, Earth and Mars, respectively.}
\label{Fig_11}
\end{figure*}

\subsection{Water delivery by fragments}

As previously stated, the inner region of the original planetesimal disk is dry, therefore
several explanations for the presence of water on Earth were devised. According to a group 
of models explaining the origin of Earth's water, the source is located in the outer
regions of the Solar System. Among these models, the external pollution model is currently 
widely accepted, which successfully accounts for the quantity and chemical composition
of water on Earth and is consistent with various theories of Solar System formation.
The present study seamlessly fits into this and the other models too, assuming that Earth's
water originate from the outer Solar System, as collisions between protoplanets are 
inevitable and frequent during the early stages of planetary evolution.

\par
The composition of the bodies in the asteroid belt depends on the radial distance, 
as indicated by studies e.g. \cite{Kerridge1985,Gradie1982, DeMeo2014}. The snowline 
$r_\mathrm{SL}$, which is approximately 2.7 au from the Sun is a boundary beyond which 
volatile substances, such as water can condense and freeze. The inner regions, 
$r < r_\mathrm{SL}$ au are predominantly populated by S-type asteroids, their fraction of water 
is typically quite low, with estimates ranging from less than 0.1\% to 1\% by mass. 
In contrast, beyond $r_\mathrm{SL}$ the asteroid belt are dominated by C-type asteroids, 
which generally have a water fraction $f_\mathrm{water,C}$ of approximately 10\% by 
mass \citep{Kerridge1985}. The current mass of water in Earth's oceans is 
$2.5 \cdot 10^{-4} \mathrm{M_\oplus}$, while the mass of the asteroid belt is only 
$4.5 \cdot 10^{-4} \mathrm{M_\oplus}$ despite covering a vast region from 2 to 4 au 
\citep{Krasinsky2002,Kuchynka2013}. This mass is orders of magnitude smaller than 
what would be predicted by models of planet-forming disks, such as the minimum-mass 
solar nebula model (MMSN) \citep{Weidenschilling1977,Hayashi1981}. 
We note that actually the MMSN is not a nebula, but represents a protoplanetary 
disc with solar composition containing the requisite amount of metals for the formation 
of the eight planets within the solar system, including the asteroid belts.
In order to determine the original total mass of solids in the asteroid belt the 
MMSN were used in which the surface density of solids is
\begin{equation}
\Sigma_{\mathrm{solid}} = f_\mathrm{n}f_\mathrm{ice}\Sigma_1 r^{-\frac{3}{2}},
\label{Eq_surfacedensity}
\end{equation}
where $f_\mathrm{n}$ is a nebular mass scaling factor, 
$f_\mathrm{ice}$ is the ice condensation factor, $\Sigma_1$ is the surface density of 
solids at 1 au and $r$ is the distance from the Sun. According to
\cite{Hayashi1981} $\Sigma_1 = 7.1\, \mathrm{g cm^{-2}}$, $f_\mathrm{ice} = 1$ for 
$r \le r_\mathrm{SL}$ and $f_\mathrm{ice} = 4.22$ for $r > r_\mathrm{SL}$. 
By integrating Eq. (\ref{Eq_surfacedensity}) from 1.55 (3 times the Mars' Hill 
radius) to 4.1 au (3 Hill radii from Jupiter) we obtain the mass of solid material
\begin{equation}
M_\mathrm{solid} = f_\mathrm{n} \Sigma_1 \int_{0}^{2\pi} d\theta 
\int_{1.55}^{4.1} f_\mathrm{ice} r^{-\frac{3}{2}} \,dr = 6.69 f_\mathrm{n},
\label{Eq_mmsn_mass}
\end{equation}
in $\mathrm{M_\oplus}$ units.
Despite its widespread use, the MMSN model has several limitations. It assumes planets 
gathered all nearby solids, but only some of the solid material were in planetesimals, 
while others were in the form of dust which was later lost as the solar nebula's dissipated. 
Adjusting for this lost mass relies on the 
fraction of solids in planetesimals. However, the model assumes that all planetesimals 
were accreted, which may not hold true for terrestrial planets. In the region between 0.4 and 4 
au, where terrestrial planets were forged, simulations show significant, $\ge 50\%$ mass 
loss and redistribution during the final assembly stage \citep{Raymond2005}. 
The asteroid belt and the region near Mercury have also been depleted of solids at some 
point \citep{Weidenschilling1977}. This suggests that the MMSN model doesn't fully 
capture the complexities of planet formation in these areas.

\par
Thus, for the MMSN $f_\mathrm{n} = 1$, so Eq. (\ref{Eq_mmsn_mass}) results in
6.69 $\mathrm{M_\oplus}$ which is the minimum mass required to forge the terrestrial
planets. In light of the above, it is evident that $f_\mathrm{n} > 1$, and simulations 
examining the formation of the solar system suggest that $2 \le f_\mathrm{n} \le 3$ 
\citep[see e.g.~][]{Fogg2005,Fogg2007,Raymond2005,Raymond2017}. Since S-type asteroids are very dry,
we are interested only in C-type asteroids which are primarily located beyond the snowline.
The mass $M_\mathrm{C}$ contained in $r_\mathrm{SL} < r < 4.1$ au is $M_\mathrm{C} = 5.38 
f_\mathrm{n} \, \mathrm{M_\oplus}$ and it is composed predominantly of C-type asteroids.

\par
If we focus only on C-type asteroids and assume that they entered the $\nu_6$ 
resonance via some mechanisms (e.g. chaotic diffusion, gravitational scattering, etc.), 
the estimation for the maximum amount of water that could be delivered to Earth can 
be given by the following formula (see Eq. (2) of ML)
\begin{equation}
    M_\mathrm{wd} = P\cdot f_\mathrm{water,C}\cdot M_\mathrm{C}.
    \label{Eq_M_wd}
\end{equation}
The computed results for $M_\mathrm{wd}$ are listed in Table \ref{tab_04}.
The results range between $\sim 1.19$ and $\sim 2.78$ 'ocean' quantities, which is 
in terms of magnitude consistent with the previous value of
ML's estimation of 8 $\mathrm{M_\mathrm{o}}$. 
If we take $f_\mathrm{n} = 3$, we obtain that $M_\mathrm{wd}$ falls within 
the $[3.57,\,8.34]$ $\mathrm{M_\mathrm{o}}$ interval.
It should be noted that these values were determined without considering any "delivery" 
and/or "impact" losses. It was assumed that the the total water inventory of the debris 
was deposited on Earth. Clearly, this provides a generous upper bound estimate on Earth's 
water inventory. These results are in good agreement with estimates for the total water 
inventory on Earth may range up to 10 $M_\mathrm{o}$ mantel water
\citep[see e.g.][]{Abe2000,Hirschmann2006,Marty2012}.
The value of $M_\mathrm{wd}^\star$ listed in Table \ref{tab_04} was calculated similarly 
to $M_\mathrm{wd}$, but using $P^\star$. The values of $M_\mathrm{wd}^\star$ with 
$f_\mathrm{n} = 3$ range from 23.76 to 32.67, which are significantly larger than 
$M_\mathrm{wd}$. These results greatly exceed the estimate for the Earth's total 
water supply.

\begin{figure*}
\includegraphics[width=0.27\paperwidth,keepaspectratio=true]{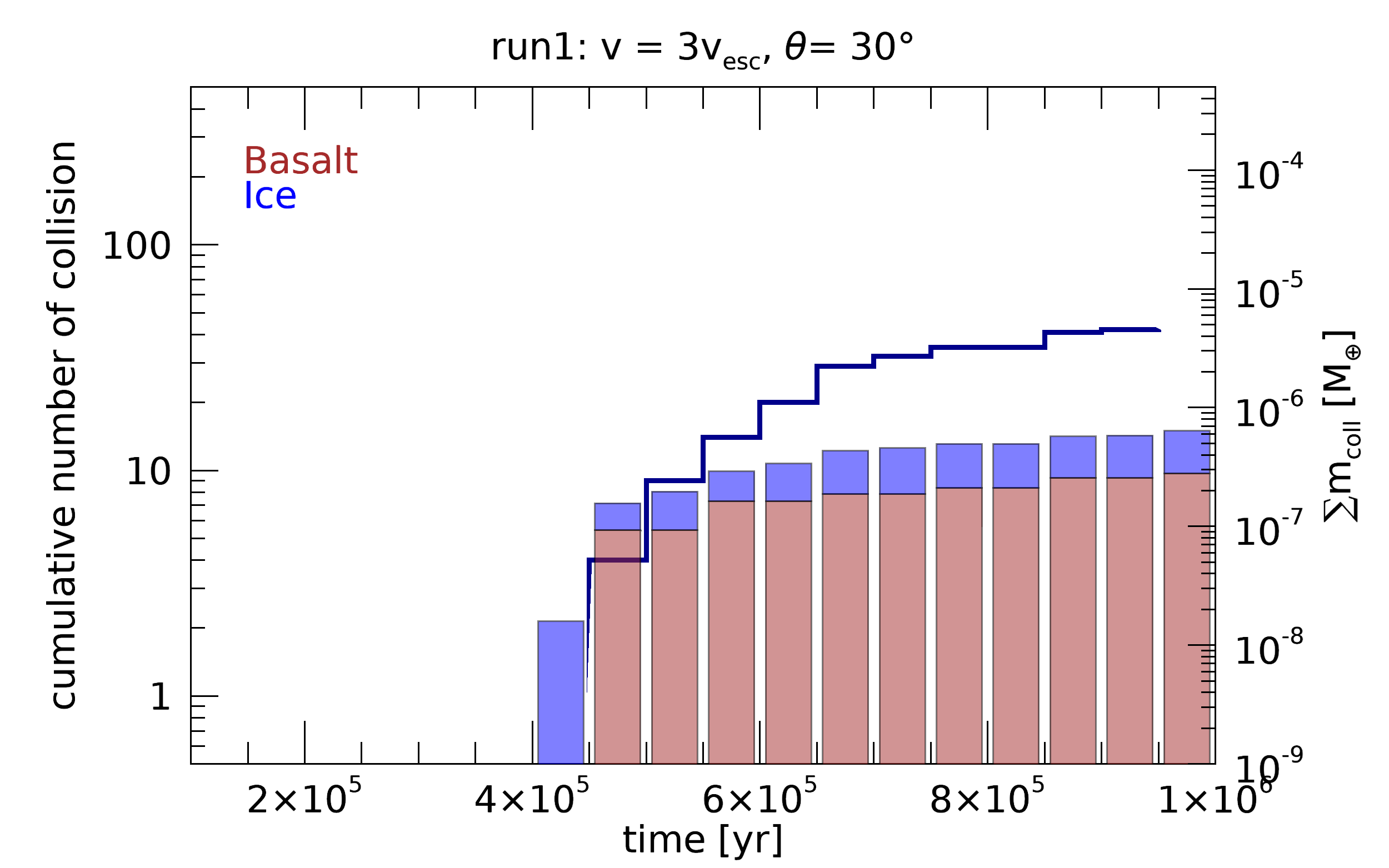}
\includegraphics[width=0.27\paperwidth,keepaspectratio=true]{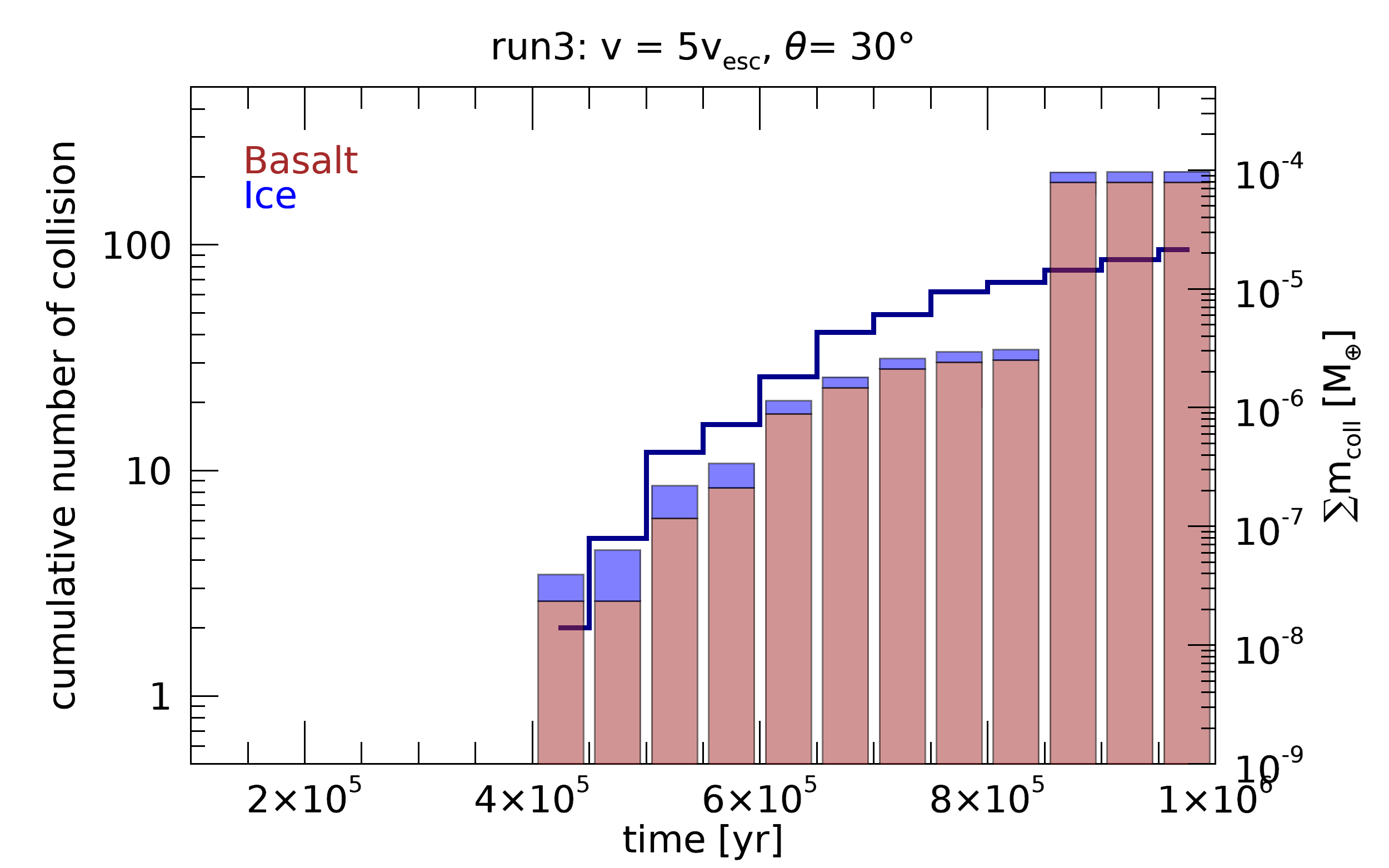}
\includegraphics[width=0.27\paperwidth,keepaspectratio=true]{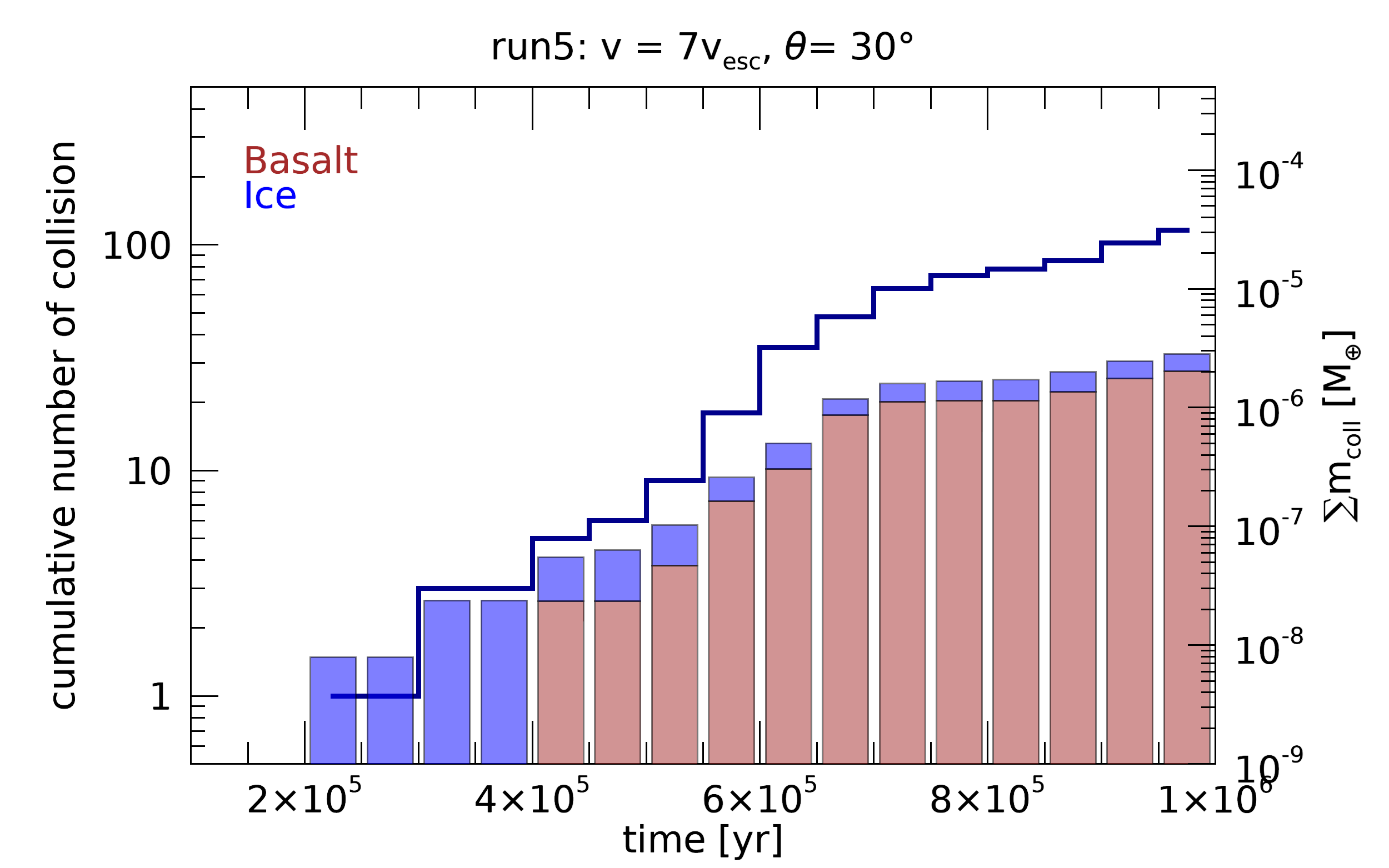} 
\\
\includegraphics[width=0.27\paperwidth,keepaspectratio=true]{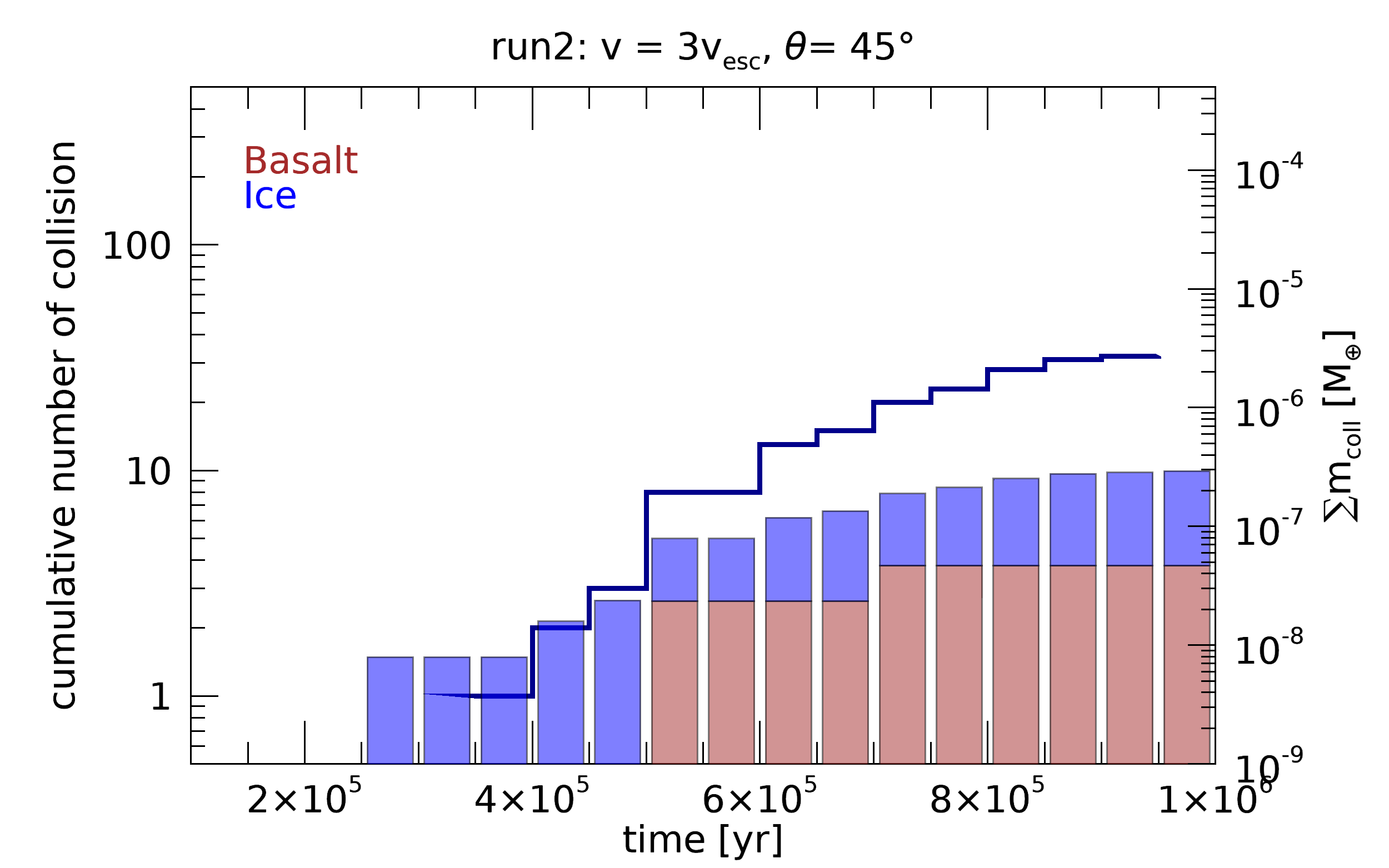}
\includegraphics[width=0.27\paperwidth,keepaspectratio=true]{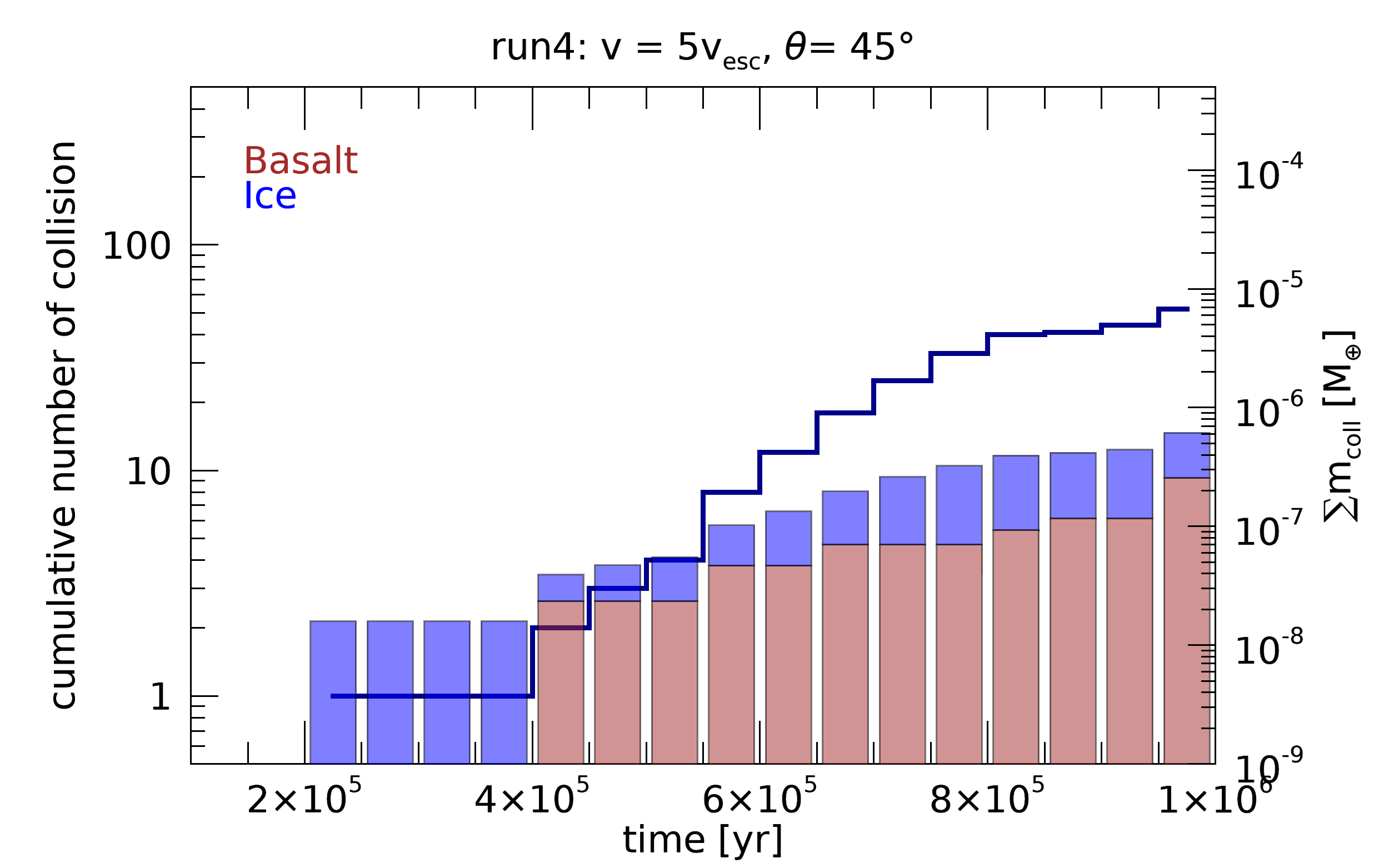}
\includegraphics[width=0.27\paperwidth,keepaspectratio=true]{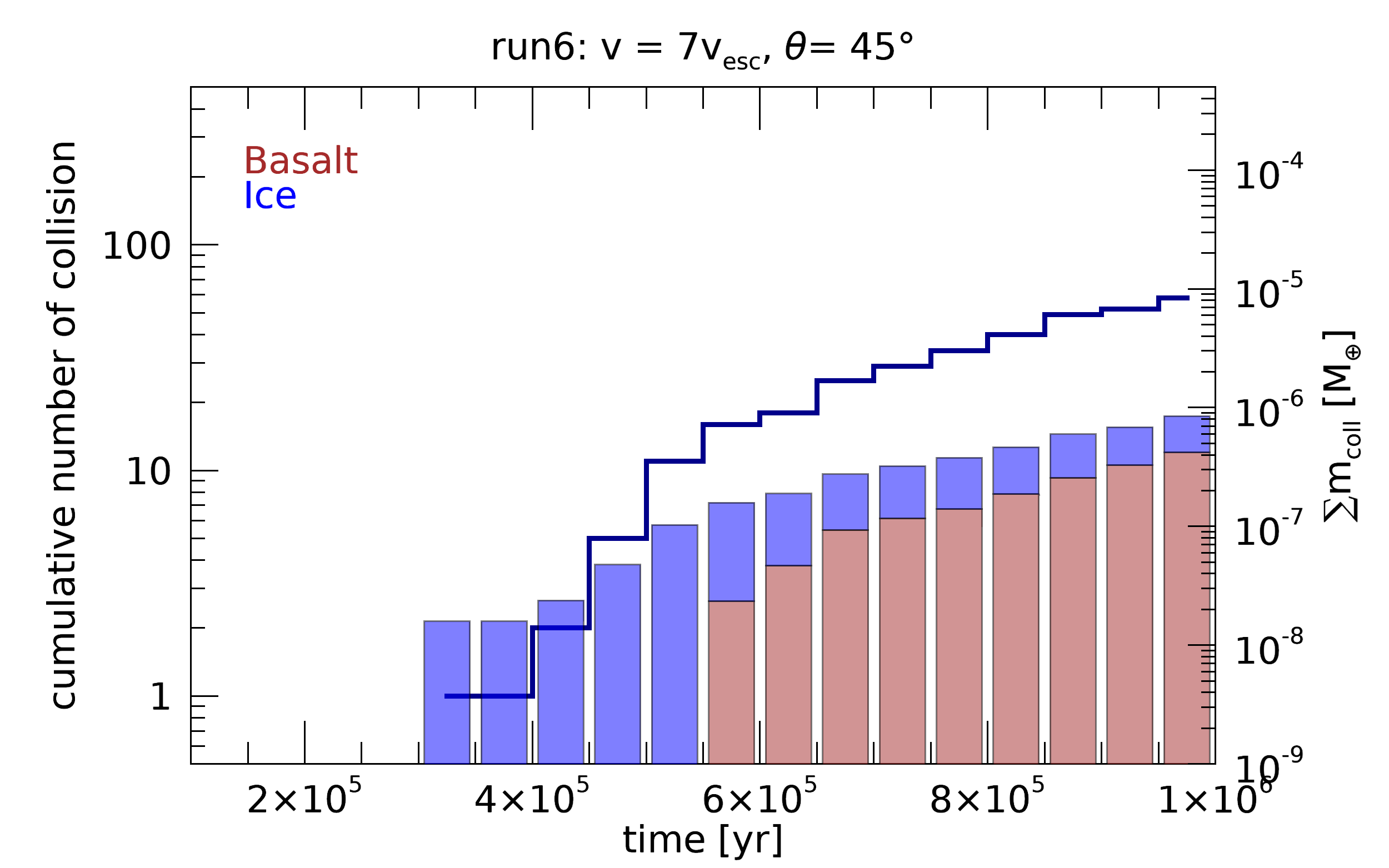} 
\caption{The event history of bodies colliding with the Earth. On the left axis, the cumulative 
number of bodies that collided with the Earth (blue line). On the right axis, the cumulative
mass of bodies that collided with the Earth (brown and blue bars) in units of Earth-mass. 
The brown colour represents the amount of basalt in the total mass, while the blue colour 
represents water ice. Note that the horizontal axis begins at $t = 10^5$ yr.}
\label{Fig_12}
\end{figure*}

In Figure \ref{Fig_12} the chronological record of collisions between fragments and Earth
is shown. On the left hand the vertical axes shows the cumulative count of these collisions
(blue line), while on the right hand axis the cumulative mass of the fragments are plotted. 
From the figure it is evident that initially smaller debris composed solely of pure water 
ice collide with the Earth. This typically begins around $\sim 2\cdot10^5 - 4\cdot10^5$ 
years after the impact event. Larger debris carrying both basalt and water ice, on the 
other hand, collide with the Earth later, around $\sim 4\cdot10^5 - 6\cdot10^5$ years 
after the impact.

\par

\begin{table}
\centering
\caption{Maximum amount of water that may deliver to Earth from the asteroid belt. 
Columns 2 and 3 list $M_\mathrm{wd}$
as computed from Eq. (\ref{Eq_M_wd}) using $P$ from Table \ref{tab_03} for $f_\mathrm{n} = 1$ 
and 3, respectively. Columns 4 and 5 are the same as Columns 2 and 3 but for $P^\star$.
Column 6 presents $\Delta m_\mathrm{ice}$ as determined from impacts of
fragments into the Earth using the results of the simulations (cf. Table \ref{tab_05}).}
\label{tab_04}
\begin{tabular}{crrrrr}
\hline
Sim. & \multicolumn{2}{|c|}{$M_\mathrm{wd}\,[\mathrm{M_\mathrm{o}}]$} & \multicolumn{2}{c}
{$M_\mathrm{wd}^\star\,[\mathrm{M_\mathrm{o}}]$} & $\Delta m_\mathrm{ice}\,[\mathrm{M_\mathrm{o}}]$ \\
 & \multicolumn{1}{c}{$f_\mathrm{n}=1$} & \multicolumn{1}{c}{$f_\mathrm{n}=3$} & \multicolumn{1}{c}{$f_\mathrm{n}=1$} & \multicolumn{1}{c}{$f_\mathrm{n}=3$} &  \\
\hline
run1 & 1.90 & 5.71 &  8.26 & 24.79 & 0.00142 \\
run2 & 2.78 & 8.34 &  9.27 & 27.81 & 0.00098 \\
run3 & 1.49 & 4.48 &  9.58 & 28.74 & 0.07093 \\
run4 & 1.76 & 5.29 & 10.64 & 31.91 & 0.00142 \\
run5 & 1.19 & 3.57 &  7.92 & 23.76 & 0.00322 \\
run6 & 1.41 & 4.22 & 10.89 & 32.67 & 0.0017  \\
\hline
\end{tabular}
\end{table}

\begin{figure}
\centering
\includegraphics[width=0.95\columnwidth,keepaspectratio=true]{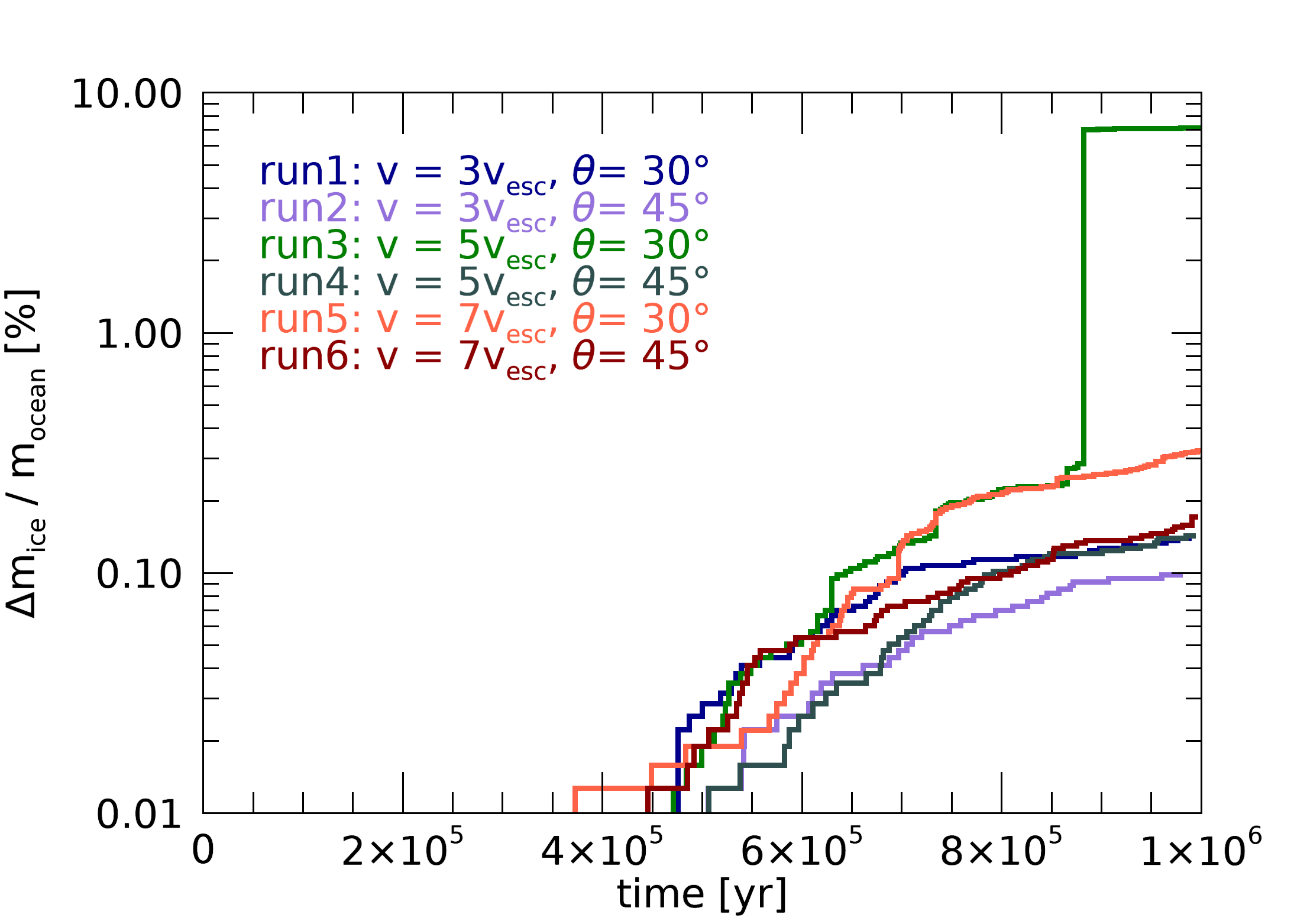}
\caption{The amount of water delivered to Earth in ocean units 
over time in the 6 simulations. The run3 is a special case where 
the water is primarily delivered by the remnants of a single object, 
the remnant of the projectile, which collided with Earth at 
$t\approx 8.7\cdot10^5$ years.}
\label{Fig_13}
\end{figure}

According to Table \ref{tab_04} in most cases, the quantity of water delivered 
to Earth by the debris fluctuates around 0.001 ocean units, except for the run3 
where more than 0.07 'ocean' of water is delivered. This means that as a result of
a single distant collision, over 7\% of the surface water is delivered to Earth. 
In this particular case, the remnants of the projectile collided with Earth, 
bringing along a significant amount of water (initially the projectile 
contained $\sim 0.16$ 'ocean' of water). 
According to \cite{Morbidelli2000}, the majority of the water currently present on 
Earth was delivered by a small number of planetary embryos that were initially formed 
in the outer asteroid belt and subsequently accreted by the Earth during its final 
stage of formation. The results of run3 are consistent with this finding.
The special case of run3 is very well visible in Fig.~\ref{Fig_13} (green solid
line), where the quantity of water transported to Earth, measured in terms of ocean 
units versus time is plotted for the 6 simulations. 
In general, the amount of water delivered to Earth initially increases rapidly, then the
growth rate slows down and stabilizes at around 0.1\%. An exception to this trend is 
observed in run3, as discussed earlier, when the remnants of the projectile hits the Earth at
about $t\approx 8.7\cdot10^5$ years, and depositing large quantity of water ice.

\par
In Table \ref{tab_05} we have summarized the total mass delivered by impacts in all 
six simulations, along with its distribution specifically between basalt and water ice. 
Only the bodies that have participated in at least one collision are listed in the table. 
The second column provides information about the mass increase resulting from impacts. 
We can see that different bodies experience varying degrees of mass increase, 
with values spanning several orders of magnitude. Among the terrestrial planets
Earth receives the highest amount of material, while Venus and Mars acquire
negligible mass. The higher mass increase on the Earth could be attributed to 
various factors, predominantly its inflated size and gravitational attraction.

\par
In all six runs the mass increase for the Sun is significantly higher compared 
to other bodies, ranging from $2.252\cdot 10^{19}$ kg to $1.897\cdot 10^{21}$ kg, 
depending on the specific run.
Roughly half of the mass absorbed by the Sun consists of basalt, with the remaining 
half being water ice, but in all cases, the mass of basalt is greater. The number 
of collisions involving the Sun varies from 178 to 600, indicating a relatively high 
frequency of impacts compared to other bodies.

\par
Generally the mass increase resulting from impacts is roughly the same for 
the Target and the Projectile in all scenarios, except two cases:
In run1, the target accretes five times more mass than the projectile, while 
in run5, the post-projectile (pl7) accumulates three times more mass than the target.
The number of collisions varies between the Target and Projectile. 
In some scenarios, the Target experiences a higher number of collisions, 
while in others, the Projectile has more collisions. 

\par
Overall, the data in the table provides insights into the effects of impacts on 
mass increase and collision occurrences for various bodies. 

\begin{table}
\centering
\caption{Summary of the collisions with $R_\mathrm{Earth} = 10 R_\oplus$ for $T=10^6$ years
for all six giant impacts. A specific object is only listed if it has suffered from a 
collision. The first column contains the names of the bodies. In the second column, the 
mass increase resulting from impacts is displayed in kg. In the third and fourth columns, 
this mass increase is specified for different materials, namely basalt and water ice. The 
fifth column shows the number of collisions (cf. it with the data in Table \ref{tab_02}).
In run5 both the target and the projectile disintegrate into smaller fragments, 
leaving no protoplanets but planetesimals pl6 and pl7.}
\label{tab_05}
\begin{tabular}{lrrrrr}
Name       &  $\Delta m$ [kg] &  $\Delta m_\mathrm{Bas}$ [kg] &  $\Delta m_\mathrm{ice}$ [kg] & $N_\mathrm{coll}$ \\
\hline
\multicolumn{5}{c} {run1 $\left( v = 3,\, \theta = 30^\circ \right)$} \\
\hline
Sun       &    5.525E19     &     3.921E19    &      1.604E19    &    178  \\
Venus      &    1.395E17     &     1.395E17    &             0    &      1  \\ 
Earth      &    3.802E18     &     1.673E18    &      2.129E18    &     47  \\
Target     &    9.570E18     &     4.601E18    &      4.969E18    &     96  \\
Projectile &    5.214E19     &     3.879E19    &      1.335E19    &    114  \\
\hline
\multicolumn{5}{c} {run2 $\left( v = 3,\, \theta = 45^\circ \right)$} \\
\hline
Sun       &    9.146E20        &  6.982E20        &  2.165E20     &   181  \\
Earth      &    1.746E18        &  2.790E17        &  1.467E18     &    33  \\
Target     &    5.259E18        &  2.230E18        &  3.028E18     &    52  \\
Projectile &    2.833E18        &  2.787E17        &  2.555E18     &    49  \\
\hline
\multicolumn{5}{c} {run3 $\left( v = 5,\, \theta = 30^\circ \right)$} \\
\hline
Sun       &    6.352E20        &  5.137E20        &  1.215E20     &   427  \\
Earth      &    5.790E20        &  4.725E20        &  1.064E20     &    95  \\
Target     &    3.899E19        &  3.085E19        &  8.140E18     &    93  \\
Projectile &    2.380E19        &  1.633E19        &  7.477E18     &   115  \\
\hline
\multicolumn{5}{c} {run4 $\left( v = 5,\, \theta = 45^\circ \right)$} \\
\hline
Sun       &    1.897E21        &  1.407E21        &  4.900E20     &   307  \\ 
Earth      &    3.664E18        &  1.534E18        &  2.130E18     &    52  \\ 
Jupiter    &    1.393E17        &  1.393E17        &         0     &     1  \\ 
Target     &    4.931E18        &  2.092E18        &  2.839E18     &    66  \\ 
Projectile &    6.398E18        &  2.091E18        &  4.307E18     &    90  \\ 
\hline
\multicolumn{5}{c} {run5 $\left( v = 7,\, \theta = 30^\circ \right)$} \\
\hline
Sun       &  1.279E+20       &   1.005E+20      &  2.735E+19     &    600  \\               
Earth      &  1.697E+19       &   1.214E+19      &  4.827E+18     &    116  \\               
Mars       &  1.866E+17       &   1.393E+17      &  4.729E+16     &    2    \\               
Jupiter    &  9.456E+16       &           0      &  9.456E+16     &    2    \\               
pl6        &  5.207E+19       &   4.303E+19      &  9.039E+18     &    71   \\               
pl7        &  1.422E+20       &   1.169E+20      &  2.527E+19     &    63   \\               
\hline
\multicolumn{5}{c} {run6 $\left( v = 7,\, \theta = 45^\circ \right)$} \\
\hline
Sun       &   2.252E19       &   1.311E19       &   9.417E18     &     245 \\
Earth      &   5.066E18       &   2.510E18       &   2.556E18     &      58 \\
Mars       &   9.458E16       &          0       &   9.458E16     &       2 \\
Target     &   1.188E19       &   7.810E18       &   4.069E18     &     102 \\
Projectile &   1.659E19       &   1.158E19       &   5.016E18     &      97 \\
\hline
\end{tabular}
\end{table}

\section{Conclusions}

The focus of this study is to examine the fate of debris ejected during 
collisions among celestial bodies in the asteroid belt and to study their 
role in the delivery of water to the Earth. In the applied model all major 
planets in the Solar System have already formed and are revolving around 
the Sun on their current orbits. To achieve our objective, we utilized 
SPH and $N$-body simulations to quantify the water transported to Earth 
through debris ejected from collisions between two Ceres-sized bodies near 
the $\nu_6$ secular resonance. We successfully combined the SPH code with 
the $N$-body simulator and provided a detailed description of generating initial 
conditions that lead to collisions at the desired location and velocity.

\par
As a result of $N$-body simulations in the ($a-e$) planes (see Fig.~\ref{Fig_06}), 
an intriguing checkmark-shaped pattern emerges at $t=0$ in all instances, encompassing both 
massive and massless bodies. In contract, in the distribution of bodies on the ($a-i$) 
plane (see Fig.~\ref{Fig_07}) such a distinct structure is not as apparent. However, 
at higher collision velocities, two arc-like structures become evident in the distribution 
of planetesimals. The other type of fragments are dispersed around the impact site. 
It is clear that in the vicinity of the secular resonance, the ($a-e$) plane clears out, 
and the bodies end up on highly eccentric orbits (which is not visible in the figures 
because $e < 0.25$). The $\nu_6$ secular resonance pumps up the eccentricities, and the 
highly elongated orbits intersect with the orbits of Mars and even Earth, allowing for 
impacts on these rocky planets. In Figs.~\ref{Fig_08} and \ref{Fig_09}, the extent and 
shape of the $\nu_6$ secular resonance are clearly visible from the distribution of 
objects that collided with Earth during the simulations.

\par
After the collision, the created the check-mark-shaped pattern's extension 
along the semi-major axis, increases with velocity $v$. At $v=3$, it spans more than 
0.5 au, while at $v=7$, it covers nearly 1 au in width. Consequently, if a collision 
deviates from the location of the $\nu_6$ secular resonance by no more than 0.5 au, due 
to the scattering effect, debris will enter the resonance. The quantity of debris 
that enters depends on the distance and velocity of the collision. Since collisions 
are frequent in the early stages of planetary formation, it's likely that the $\nu_6$ 
resonance receives a continuous supply, which it then "transmits" to the inner regions 
through the mechanism outlined above. From the perspective of the inner planets, this 
represents a continuous water delivery system that persists as long as there are 
collisions and debris enters the resonance.

\par
Based on the data from the $N$-body simulations, we calculated the probability of the ejected 
debris impacting Earth. Using this information, we provided an upper estimate for the 
amount of water that can be transported by the debris. Considering the significant 
uncertainties and strong model dependencies, our estimate is in reasonable agreement with
previous results, such as ML and \cite{Morbidelli2000}. According to our findings, a minimum 
of $\approx 1.2$ and a maximum of $\approx 2.8$ ocean equivalents of water can be delivered 
if the MMSN model is considered. 
However, if we assume an equivalent mass of 3 $\times$ MMSN of solids and 
calculate the water quantity, the range increases to 3.6 to 8.3 ocean equivalents, 
indicating the potential delivery of larger amounts of water to Earth.
Therefore, if the total water inventory on Earth is estimated to be between 1 and 10 ocean 
equivalents, the external pollution model can provide an explanation for Earth's total 
water inventory. If this is true, it would not require assuming local sources to explain 
the origin of water, although they cannot be ruled out either.

\par
During the $N$-body simulations, we were able to directly measure the amount of water brought 
to Earth by the ejected debris during individual collisions. Typically, this amounted 
to 0.001 ocean equivalents of water. An interesting exception was observed when a single 
large body, the remnants of the projectile, collided with Earth and deposited approximately 
0.07 ocean equivalents of water on the Earth's surface. 
The results of this study support the widely accepted external pollution model.
The maximum estimated amount of water that was transported to the Earth via fragments
varies between 1.2 and 8.3 ocean's worth, depending on the initial mass of the protoplanetary disk.

\section*{Acknowledgements}

We thank an anonymous referee for valuable comments that improved
the quality of our manuscript.
This work was supported by the János Bolyai Research Scholarship of the Hungarian 
Academy of Sciences (ÁS). ÁS also acknowledge the support of the Hungarian OTKA Grant 
No. 119993. On behalf of the \textit{Simulation of the formation of the "434 Hungaria" family} 
project we are grateful for the possibility to use ELKH 
Cloud\footnote{https://science-cloud.hu/} \citep{Heder2022} which helped us achieve the results 
published in this paper. We acknowledge the computational resources of the GPU 
Laboratory of the Wigner Research Centre for Physics. This project has received funding from the HUN-REN Hungarian Research Network.

\section*{Data Availability}

The data underlying this paper will be shared on reasonable request
to the authors.



\bibliographystyle{mnras}
\bibliography{MN-23-2054-MJ} 





\bsp	
\label{lastpage}
\end{document}